\documentclass[12pt,twoside]{article}
\usepackage{axodraw}
\usepackage{colordvi}
\usepackage{latexsym} \usepackage{epsfig}
 
\begin{document}
\begin{titlepage}
\begin{center}
{\Large  \bf JOINT INSTITUTE FOR NUCLEAR RESEARCH\\[0.3cm]
Bogoliubov Laboratory of Theoretical Physics} %\vspace{2cm}
\vspace*{-2cm}

\hspace*{-1cm}\includegraphics[height=.55\textheight]{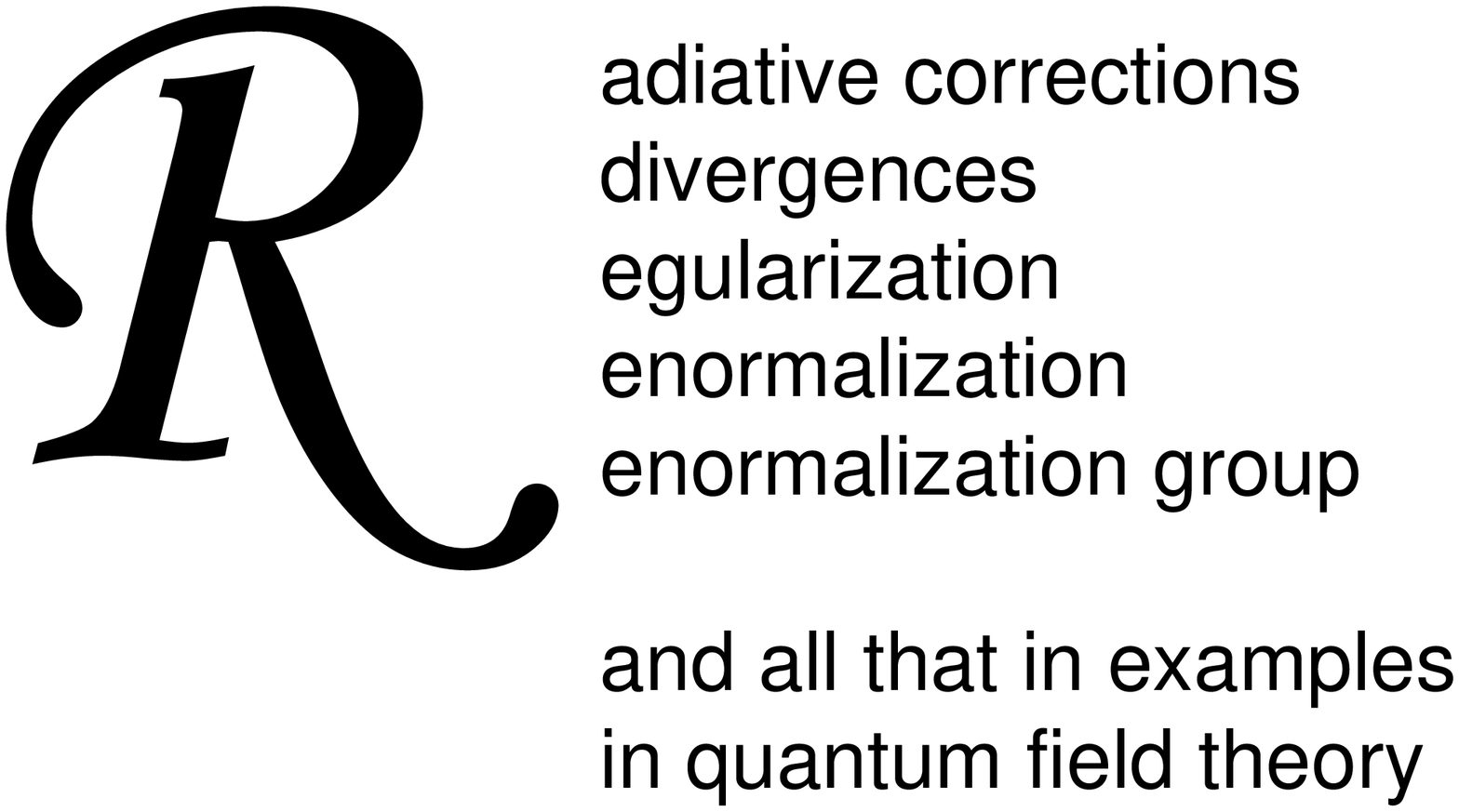}

%\vspace{-2cm}

{\Large \bf D.I.KAZAKOV}

\vspace{7cm}

{\large \bf DUBNA\\[0.3cm] 2008}
\end{center}
\end{titlepage}

\thispagestyle{empty}

\noindent {\bf Kazakov D.I.}\\

 \noindent{\bf \small RADIATIVE CORRECTIONS, DIVERGENCES, REGULARIZATION,
RENORMALIZATION, RENORMALIZATION GROUP AND ALL THAT IN EXAMPLES IN QUANTUM FIELD
THEORY}\\

{\small The present lectures are a practical guide to the calculation of radiative
corrections to the Green functions in quantum field theory. The appearance of ultraviolet
divergences is explained, their classification is given, the renormalization procedure
which allows one to get the finite results is described, and  the basis of the
renormalization group in QFT is presented. Numerous examples of calculations in scalar
and gauge theories are given. Quantum anomalies are discussed. In conclusion the
procedure which allows one to get rid of infrared divergences in S-matrix elements is
described. The lectures are based on the standard quantum field theory textbooks, the
list of which is given at the end of the text.%\vspace{-0.5cm}

 These lectures were given to the 4-th year students of the Department of General and
Applied Physics of the Moscow Institute of Physics and Technology (Technical
University).}\vspace{0.2cm}

\noindent Figs.-- 42, Refs.-- 13.\vspace{0.5cm}

\noindent {\bf Казаков Д.И.}\\

\noindent {\bf \small РАДИАЦИОННЫЕ ПОПРАВКИ, РАСХОДИМОСТИ, РЕГУЛЯРИЗАЦИЯ, РЕНОРМИРОВКА,
РЕНОРМГРУППА И ВСЁ ТАКОЕ В ПРИМЕРАХ В КВАНТОВОЙ ТЕОРИИ ПОЛЯ}\\

{\small Настоящие лекции являются практическим руководством по вычислению радиационных
поправок к функциям Грина в квантовой теории поля. Объясняется как возникают
ультрафиолетовые расходимости, даётся их классификация, описана процедура перенормировки,
позволяющая получать конечные ответы, излагаются основы группы перенормировок в КТП.
Приведены многочисленные примеры вычислений в скалярных и калибровочных теориях с
использованием размерной регуляризации. Обсуждаются квантовые аномалии. В заключении
описана процедура избавления от инфракрасных расходимостей, возникающих при вычислении
элементов матрицы рассеяния. Изложение основывается на стандартных учебниках по квантовой
теории поля, список которых приведён в конце текста.%\vspace{-0.5cm}

 Лекции были прочитаны студентам 4-го курса кафедры "Физика элементарных частиц" факультета
Общей и прикладной физики  МФТИ.}

 \vfill \copyright\ Kazakov D.I., 2008

%\noindent
%-----------------------------------------------------------
\pagebreak \thispagestyle{empty} %\normalsize
\vspace*{1.0cm} \tableofcontents
 \vglue 0.34cm {\bf References} \hfill {\bf 91}
\newpage

\renewcommand{\thesubsection}{\thesection.\arabic{subsection}}
\renewcommand{\thesubsubsection}{\thesubsection.\arabic{subsubsection}}
\renewcommand{\theequation}{\thesection.\arabic{equation}}
\renewcommand{\thefigure}{\arabic{figure}}
\setcounter{section}{-1}\vspace*{1cm}

\section{Preface}

Today there exist many excellent textbooks on quantum field theory. The most popular ones
are listed in the bibliography to the present lectures. Nevertheless, everyone who gives
lectures on quantum field theory faces the problem of selection of material and writing
the lecture notes for students. The present text is just the lecture notes devoted to the
radiative corrections in QFT. On this way, one encounters two problems, namely, the
ultraviolet and the infrared divergences.  Our task is to demonstrate how one can get rid
of these divergences and obtain finite corrections to the cross-sections of elementary
processes. During the course we describe the methods of Feynman diagram evaluation and
regularization of divergences. In more detail, we consider the renormalization theory and
elimination of ultraviolet divergencies in the Green functions off mass shell, as
exemplified by  scalar and gauge theories. In connection with the renormalization
procedure we describe also the renormalization group formalism in QFT. As for the
infrared divergences, in the literature one can find mainly the discussion of the IR
divergencies in quantum electrodynamics. In non-Abelian theories as well as in QED with
massless particles the situation is much more involved as there arise collinear
divergences as well. In the last lecture, we show how one can get rid of these
divergences using the methods developed in quantum chromodynamics. One more topic also
related to divergences is the so-called anomalies. They also lead to unwanted ultraviolet
divergent contributions. Therefore, a separate lecture is dedicated to the axial and
conformal anomalies.

The presented text overlaps with many textbooks and is partly borrowed from there.
However, the composition of the material and most of the calculations belong to the
author, so we omit the direct references to any textbooks. It should be admitted that the
style of presentation in different textbooks varies very much and the reader can choose
the book according to his preferences. We mostly used the classical monograph by
N.Bogoliubov and D.Shirkov when describing the renormalization theory and more modern
book by M.Peskin and D.Schreder which we followed when discussing the infrared
divergences.

Our experience in giving lectures on quantum field theory, the renormalization theory and
the renormalization group tells us that this material is still complicated for perception
and is not always presented clearly enough. One often meets with the lack of
understanding of the complicated structure of the field theory which manifests itself in
renormalization theory. Sometimes the nonrenormalizable theories are simplistically
treated as the field theories with a dimensional coupling constant which otherwise have
no difference  from the renormalizable ones. The collinear divergences arising in
theories with massless particles, despite a long history, have not also become the
well-known part of the QFT course. Here we make an attempt of a simplified presentation
of this complicated material. Of course, this means that one has to sacrifice some
rigorousness and completeness. We hope that together with the existing literature the
present lectures will serve the goal of clarification and mastering of quantum field
theory and its applications to particle physics.

\newpage
\setcounter{section}{0}\vspace*{1cm}
\section{Lecture I: Radiative corrections. General analysis of divergent integrals}
\setcounter{equation}{0}
\subsection{Radiative corrections}

\begin{center}
 \begin{figure}[ht]\hspace*{0.5cm}
\includegraphics[width=.9\textwidth]{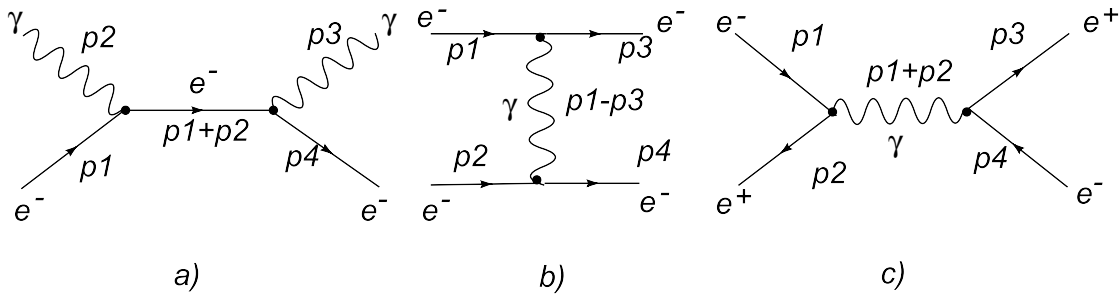}
 \caption{The examples of tree diagrams of different processes in QED: а) the Compton scattering,
 b) the Mueller scattering, c) the annihilation of the particle-antiparticle pair. Shown are the momenta
 of external (real) and internal (virtual) particles
 \label{tree}}
 \end{figure}
\end{center}

The formalism of quantum field theory, being the generalization of quantum mechanics to
the case of an infinite number of degrees of freedom with nonconservation of the number
of particles, allows one to describe the processes of scattering, annihilation, creation
and decay of particles with the help of the set of well-defined rules. As in quantum
mechanics the cross-section of any process is given by the square of the modulus of the
probability amplitude calculated according to the Feynman rules for the corresponding
Lagrangian integrated over the phase space. Since the exact calculations of the
probability amplitudes seem to be impossible, one is bound to use the perturbation theory
with a small parameter - the coupling constant - and get the result in the form of a
power series. The leading terms of this series can be presented by Feynman diagrams
without loops, the so-called tree diagrams. The examples of such diagrams for some
typical processes in QED are shown in Fig.\ref{tree}.
\vspace{-1cm}

All the diagrams shown in Fig.\ref{tree} are proportional to the square of the coupling
constant $e^2$. They are constructed according to the well-known Feynman rules and do not
contain any integration over momenta (when working in momentum representation) since due
to the conservation of four-momentum all momenta are defined uniquely.

\begin{figure}[ht]\hspace*{0.5cm}
\includegraphics[width=0.9\textwidth,height=0.3\textheight]{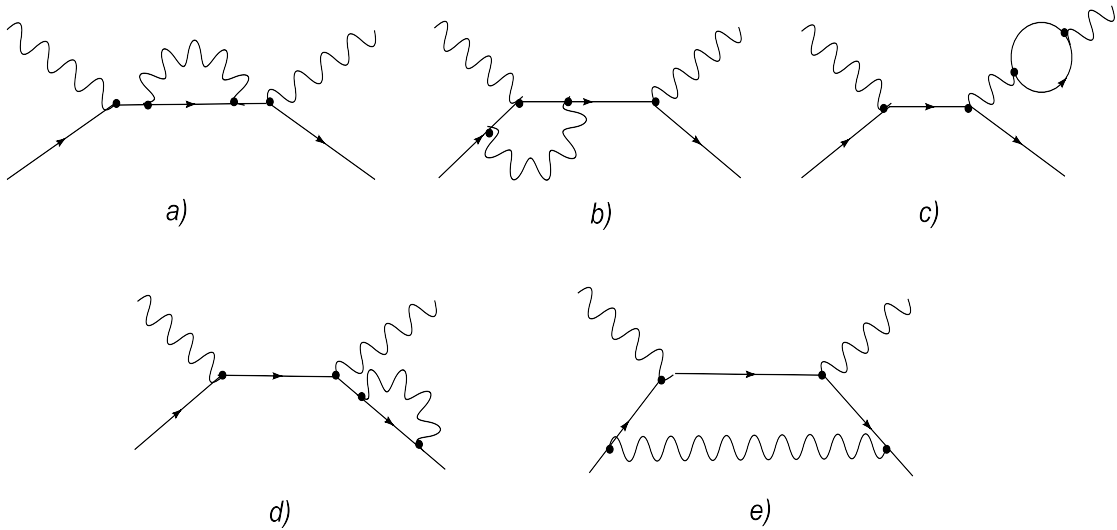}
 \caption{The one-loop diagrams for the process of the Compton scattering\label{loop}}
 \end{figure}

The situation changes when considering the next order of perturbation theory. As an
example, in Fig.\ref{loop} we show the corresponding diagrams for the Compton scattering.

They got the name of {\it radiative corrections} since in electrodynamics they correspond
to the emission and absorption of photons. This name is also accepted in  other theories
for perturbative corrections. All these diagrams are proportional to the fourth power of
the coupling constant $e^4$ and, hence, are the next order perturbations with respect to
the tree diagrams. However, contrary to the tree diagrams, they contain a closed loop
which requires the integration over the four-momenta running through the loop. Any loop
corresponds to the bifurcation of momenta similarly to the bifurcation of the electric
current, according to the Kirchhoff rules, so that the total momentum is conserved but
the momentum running along each line is arbitrary. Therefore, one has to integrate over
it.

\subsection{Divergence of integrals}

Prior to calculating the radiative corrections let us consider the behaviour of the
integrand and the integral as a whole. As an example we take the diagrams of the Compton
scattering shown in Fig.\ref{loop}. The integral corresponding to the diagram shown in
Fig.\ref{loop}.a) has the form
\begin{equation}\label{i1}
\int d^4k \frac{\gamma^\mu
(\hat{p}-\hat{k}+m)\gamma^\mu}{[k^2+i\varepsilon][(p-k)^2-m^2+i\varepsilon]},
\end{equation}
where the photon propagator is written in Feynman gauge and the integration takes place
in Minkowskian space. We shall not calculate explicitly this integral now (we shall do it
later) but consider the integrand from the point of view of the presence of singularities
as well as the behaviour at small and large momenta.

The presence of poles in the propagators for momentum equal to the mass squared does not
create any problem for the integration since according to the Feynman rules the
denominator contains the infinitesimal imaginary term   $\sim\varepsilon \to 0$, which
defines the way to bypass the pole. The choice accepted in (\ref{i1}) corresponds to the
causal Green function.

Consider now the case of  $k_\mu \to 0$, the so-called  {\it infrared} behaviour. Despite
the presence of $k^2$ in the denominator, the singularity is absent due to the measure of
the 4-dimensional integration which is also proportional to $k^4$. This is true for all
such integrals. The singularities appear only for certain external momenta which are on
mass shell and have a physical reason. Off shell the singularities are absent.  For this
reason we shall not discuss the infrared behaviour of the integrals so far.

Consider at last the case of $k_\mu \to \infty$, the so-called  {\it ultraviolet}
behaviour. Notice that in the denominator one has 4 powers of momenta, while in the
numerator one has 1 plus 4 powers in the measure of integration. Hence one has 5-4=1,
i.e. the integral is linearly divergent as $k_\mu \to \infty$. Is it the property of a
particular integral or is it a general situation? What happens with the other diagrams?

Consider the integral corresponding to the diagram shown in Fig.\ref{loop}.б). One has,
according to the Feynman rules
\begin{equation}\label{2}
\int d^4k \frac{\gamma^\mu
(\hat{p_1}-\hat{k})+m)\gamma^\nu(\hat{p_2}-\hat{k})+m)\gamma^\mu}{k^2
[(p_1-k)^2-m^2][(p_2-k)^2-m^2]}.
\end{equation}
We are again interested in the behaviour for $k_\mu \to \infty$. The counting of the
powers of momenta in the numerator and the denominator gives: 6 in the denominator and 2
in the numerator plus 4 in the integration measure. Altogether one has   6-6=0, i.e., the
integral is logarithmically divergent as $k_\mu \to \infty$.

Here we met the difficulty called the ultraviolet divergence of the integrals for the
radiative corrections. The  examples considered above are not exceptional but the usual
ones. The corrections are infinite, which makes perturbation theory over a small
parameter meaningless. The way out of this trouble was found with the help of the
renormalization theory which will be considered later and now we  try to estimate the
divergence of the integrals in a theory with an arbitrary Lagrangian.

\subsection{General analysis of ultraviolet divergences}

\begin{figure}[ht]\hspace*{1.5cm}
\includegraphics[width=0.9\textwidth]{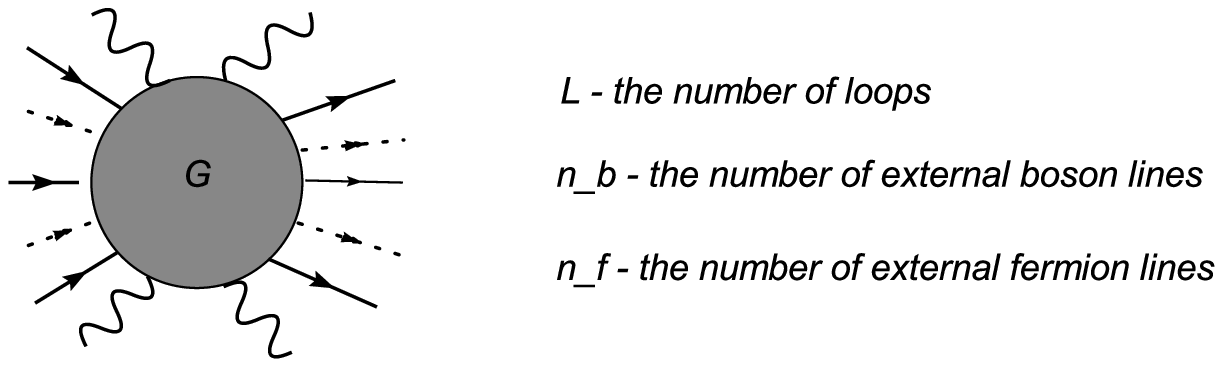}
  \caption{An arbitrary diagram containing L integrations\label{graph}}
\end{figure}

Consider an arbitrary Feynman diagram  $G$ shown in Fig.\ref{graph}.
and try to find out whether it is ultraviolet  divergent  or not. For this purpose we
have to calculate the number of powers of momenta in the integrand: each internal loop
leads to integration  $d^4p$ that gives 4 powers of momenta; each derivative in the
vertex gives the momentum in p-space, i.e., $1$; each internal line gives a propagator
which behaves as $p^{r_l}/p^2$, i.e., $r_l-2$ powers of momenta, where $r_l=0,1,2$ for
various fields. Combining all these powers together we get the quantity called
 {\it the index of divergence of the diagram} (UV)
\begin{equation}\label{ind}
\omega (G)=4L+\sum\limits_{vertices}\delta _v+\sum\limits_{internal\ lines}(r_l-2),
\end{equation}
where $L$ is the number of loops  and $\delta _v$ is the number of derivatives in a
vertex $v$.

The absence of the ultraviolet divergences means that $\omega (G)<0.$ However, one has to
be careful, there might be subdivergences in some subgraphs. Hence, the necessary
condition for finiteness is
$$
\mbox{The finiteness condition (UV):}\ \ \omega (\gamma _i)<0,\quad \forall \gamma
_i\subset G,
$$
where $\gamma _i$ are all possible subgraphs of the graph $G$ including the graph $G$
itself.

There exists, however, a simpler way to answer the same question which does not need to
analyse all the diagrams. One can see it directly from the form of the Lagrangian. To see
this, let us introduce the quantity called    {\it the index of the vertex} (UV)
\begin{equation}\label{indv}
\omega _v=\delta _v+b_v+\frac 32f_v-4,
\end{equation}
where $\delta _v,b_v$ and $f_v$ are the number of derivatives, internal boson and fermion
lines, respectively. Then the index of a diagram (\ref{ind}) can be written as
\begin{equation}\label{ind2}
\omega (G)=\sum\limits_{vertices}\omega _v^{max}+4-n_b-\frac 32 n_f,
\end{equation}
where $\omega _v^{max}$ corresponds to the vertex where all the lines are internal, $n_b$
and $n_f$ are the number of external boson and fermion lines, and we have used the fact
that usually $r_l(boson)=0$ and $r_l(fermion)=1.$

Equation (\ref{ind2}) tells us that the finiteness ($\omega (G)<0$) can take place if
$\omega _v\leq0$ and the number of external lines is big enough. Prior to the formulation
of conditions when it happens, let us consider some examples.\vspace{0.3cm}

\underline{Example 1}: The scalar theory ${\cal L}_{int}=-\lambda \varphi ^4.$\\[-0.3cm]

\begin{figure}[ht]\vspace{0.2cm}\hspace*{0.5cm}
\includegraphics[width=0.9\textwidth]{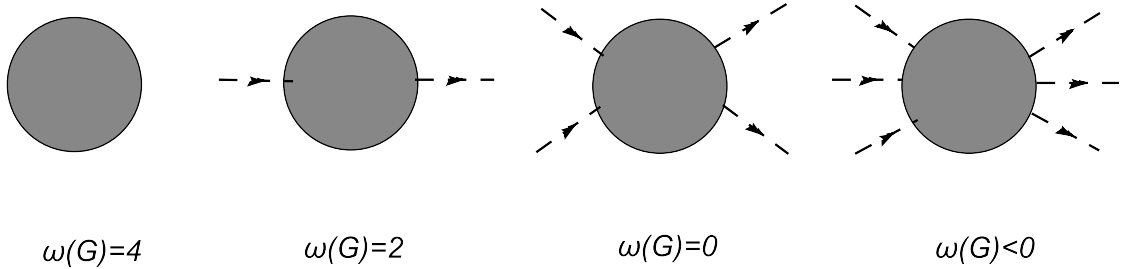}
  \caption{The indices of divergence of the diagrams in the scalar theory\label{scal}}
\end{figure}

In this case  $\delta _v=0,\ f_v=0, \ b_v=4$ and, hence,  $\omega _v^{max}=0$. Thus,
according to (\ref{ind2}), $\omega(G)=4-n_b-\frac 32 n_f$ and everything is defined by
the number of external lines. The situation is illustrated in Fig.\ref{scal}.

We see that there exists a limited number of divergent structures in the  $\varphi ^4$
theory. These are the vacuum graphs, the two- and four-point functions. All the other
diagrams having more than 4 external lines are convergent (though may have divergent
subgraphs).\\

\underline{Example 2}: Quantum Electrodynamics ${\cal L}_{int}=е\bar\psi \hat A \psi
.$\\[-0.3cm]

In this case  $\delta _v=0,f_v=2, b_v=1, \ \omega _v^{max}=0$. Hence,
$\omega(G)=4-n_b-\frac 32 n_f$ and the situation is similar to the previous example,
everything is defined by external lines. Divergent are the vacuum diagrams
$(\omega(G)=4)$, the photon propagator $(\omega(G)=2)$, the electron propagator
$(\omega(G)=1)$ and the triple vertex $(\omega(G)=0)$.
All the other diagrams are convergent.\\

\underline{Example 3}: Four-fermion interaction ${\cal L}_{int}=G\bar\psi \psi \bar\psi
\psi.$\\[-0.3cm]

Here $\delta _v=0,f_v=4, b_v=0,\  \omega _v^{max}=2$ and, hence,
$\omega(G)=2N_{вершин}-\frac32 n_f$. Therefore, increasing the number of  vertices we get
new divergent diagrams independently of the number of external lines. The number of
divergent structures happens to be  infinite.

Thus, the key role is played by the maximal index of the vertex. All the theories may be
classified according to the value of  $\omega _v^{\max }:$
 \begin{equation}\label{class} \omega _v^{\max
}=\left\{
\begin{array}{cl}
<0 & Finite\ number \ of\ divergent\ diagrams, \\ 0 &
Finite\  number \ of\ divergent\ structures, \\
>0 &  Infinite\  number \ of\ divergent\ structures.
\end{array}
\right.
\end{equation}
Below we show that for the first two types of theories we can handle the ultraviolet
divergences with the help of the renormalization procedure. The theories with $\omega
_v^{\max}=0$ are called  {\it renormalizable}, the theories with $\omega _v^{\max}>0$ are
called {\it nonrenormalizable}, and the theories with $\omega _v^{\max}<0$ are called
{\it superrenormalizable}.

\subsection{The analysis of dimensions}

The property of a theory with respect to ultraviolet divergences can be reformulated in
terms of dimensions. Consider for this purpose an arbitrary term of the interaction
Lagrangian which is the product of the field operators and their derivatives
\begin{equation}\label{lag}
{\cal L}_I(x)=g\prod\limits_{i,j}\varphi _i(x)\partial \varphi _j(x).
\end{equation}
Consider the action which is the four-dimensional integral of the Lagrangian density
\begin{equation}
A=\int d^4x{\cal L}(x),
\end{equation}
and find the dimensions of parameters in eq.(\ref{lag}). As a unit of measure we take the
dimension of a mass equal to 1. Then the dimension of length $[L]=-1$, the dimension of
time is also $[T]=-1$, the dimension of derivative $[\partial_\mu]=1$, the dimension of
momenta $[p_\mu]=1$. Since the action is dimensionless (we use the natural units where
$\hbar =c=1$)
$$
[A]=0,
$$
the dimension of the Lagrangian is
$$[{\cal L}]=4, \ \ (D - \mbox{in D dimensional space.})$$
This gives us the dimensions of the fields. Indeed, from the kinetic term for the scalar
field one finds
$$ [(\partial \phi)^2]=4 \to [\phi]=1, \ \ \
(\frac{D-2}{2}\  \mbox{in D dimensional space}), $$ for the spinor field
$$ [\bar\psi\hat\partial \psi]=4 \to [\psi]=\frac 32, \ \ \
(\frac{D-1}{2}\  \mbox{in D dimensional space}), $$ for the vector field
$$ [(\partial_\mu A_\nu-\partial_\nu A_\mu)^2]=4 \to [A_\mu]=1, \ \ \
(\frac{D-2}{2}\  \mbox{in D dimensional space}). $$ This allows one to find the dimension
of the coupling constant in (\ref{lag})
\begin{equation} [g]=4-\delta_v-b_v-\frac
32f_v=-\omega_v^{max}.
\end{equation}
Then the classification of interactions (\ref{class}) can be written as
\begin{equation}
[g]=\left\{
\begin{array}{cl}
> 0 & Superrenormalizable \ theories, \\ 0 &
 Renormalizable\ theories, \\
< 0 &  Nonrenormalizable \ thoeries.
\end{array}
\right.
\end{equation}

Consider which category various theories belong to. For this purpose we have to calculate
the dimensions of the couplings.

\underline{Illustration}
$$
\begin{array}{llll}
{\cal L}_{\varphi ^3}=-\lambda \varphi ^3 & \Rightarrow &
[\lambda ]=1, & SuperRen. \\
{\cal L}_{\varphi ^4}=-\lambda \varphi ^4 & \Rightarrow &
[\lambda ]=0, & Ren. \\
{\cal L}_{QED}=e\overline{\psi }\gamma ^\mu A_\mu \psi &
\Rightarrow & [e]=0, & Ren. \\
{\cal L}_{gauge}=-\frac 14F_{\mu \nu }^2=-\frac 14\left[ \partial _\mu A_\nu ^a-\partial
_\nu A_\mu ^a+gf^{abc}A_\mu ^bA_\nu ^c\right] ^2 & \Rightarrow & [g]=0, & Ren. \\
{\cal L}_{Yukawa}=y\overline{\psi }\varphi \psi & \Rightarrow & [y]=0. & Ren.
\end{array} $$
Thus, all these models are renormalizable.
$$ \begin{array}{llll} {\cal L}=-h\varphi ^6 &\Rightarrow & [h]
=-2, & Nonren. \\
{\cal L}=G\overline{\psi }\psi \overline{\psi}\psi &\Rightarrow& [G]=-2 & Nonren. \\
{\cal L}=\kappa \overline{\psi }\partial_\mu V_\mu \psi &\Rightarrow &[\kappa ]=-1 & Nonren. \\
{\cal L}=\gamma \overline{\psi }\partial _\mu \varphi \gamma ^\mu \psi & \Rightarrow &
[\gamma ]=-1. & Nonren.
\end{array}
$$
All these models on the contrary are nonrenormalizable. Notice that they include the
four-fermion or current-current interaction which was previously  used in the theory of
weak interactions.

\begin{figure}[ht]\hspace*{2.5cm}
\includegraphics[width=0.6\textwidth]{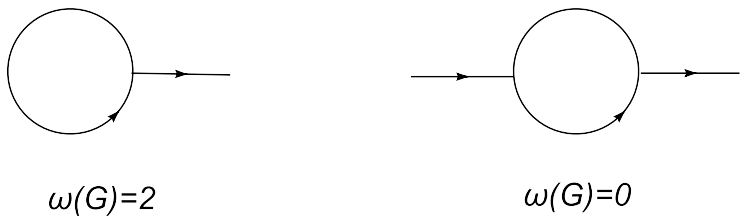}
  \caption{The only divergent diagrams in the  $\phi^3$ theory \label{tri}}
\end{figure}

Hence, we come to the following conclusion: the only renormalizable interactions in four
dimensions  are:

i) the  $\varphi ^4$ interaction;

ii) the Yukawa interaction;

iii) the gauge interaction;

iv)  the theory $\varphi ^3$ is superrenormalizable. It contains only two divergent
diagrams shown in Fig.\ref{tri}.

If one looks at the spins of particles involved in the interactions, one finds out that
they are strongly restricted. The renormalizable  interactions contain only the fields
with spins 0, 1/2 and 1. All the models with spins 3/2, 2, etc. are nonrenormalizable.
The latter include also gravity. Indeed, the coupling constant in this case is the Newton
constant which has dimension equal to $[G]=-2$, i.e., quantum gravity is
nonrenormalizable.

Since we do not know how to handle the nonrenormalizable interactions because the
ultraviolet divergences are out of control, there are only three types of interactions
which are used in the construction of the Standard Model of fundamental interactions,
namely the $\varphi ^4$, the Yukawa and the gauge interactions with the scalar, spinor
and vector particles.

Here one has to make a comment concerning the vector fields with $M\neq 0.$  Remind the
form of the propagator of the massive vector field
$$
\overline{V_\mu V_\nu }=i\frac{g_{\mu \nu }-k_\mu k_\nu /M^2}{ M^2-k^2-i\epsilon }.
$$
It gives  $r_l=2,$ which leads to some modification of the formulas used above and
finally to the nonrenormalizability of the theory. The only known way to avoid this
difficulty is the spontaneous breaking of symmetry. In this case,
$$
\overline{V_\mu V_\nu }=i\frac{g_{\mu \nu }-k_\mu k_\nu /k^2}{ M^2-k^2-i\epsilon },
$$
that gives  $r_l=0$ and the theory happens to be renormalizable. This mechanism is used
in the Standard Model to give masses to the intermediate weak bosons without breaking the
renormalizability of the theory.

\newpage
\vspace*{1cm}
\section{Lecture II: Regularization}
\setcounter{equation}{0} The divergences which appear in radiative corrections are not
yet a catastrophe for a theory (remind, for example, the infinite self-energy of an
electric charge in its own Coulomb field) but require a quantitative description. To get
a finite difference of the two infinite quantities, one has to give them some meaning.
This can be achieved by introducing a kind of regularization of divergent integrals. The
most natural way of regularization is to cut off the integral on the upper or lower bound
of integration. There are also different ways of regularization  based on a modification
of the integrand or of the measure of integration. Below we consider  three  most popular
kinds of regularization: the ultraviolet cutoff in Euclidean space
($\Lambda$-regularization), the Pauli-Villars regularization, and the dimensional
regularization.

\subsection{Euclidean integral and the ultraviolet cutoff}
\begin{figure}[ht]
\begin{center} \epsfxsize=5cm
 \epsffile{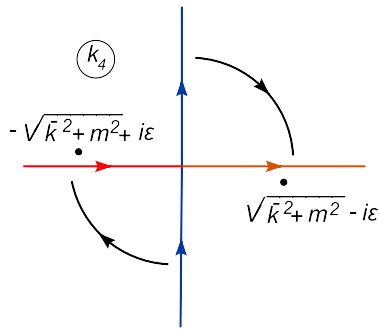}
  \caption{The Wick rotation of the integration contour \label{wik}}
  \end{center}
\end{figure}

All the integrals in quantum field theory are written in Minkowski space; however, the
ultraviolet divergence appears for large values of modulus of momentum and it is useful
to regularize it in Euclidean space. Transition to Euclidean space can be achieved by
replacing the zeroth component of momentum  $k_0 \to ik_4$, so that the squares of all
momenta and the scalar products change the sign $k^2=k_0^2-\vec k^2 \to -k_4^2-\vec
k^2=-k_E^2$ and the measure of integration becomes equal to $d^4k \to id^4k_E$, where the
integration over the fourth component of momenta goes along the imaginary axis. To go to
the integration along the real axis, one has to perform the (Wick) rotation of the
integration contour by  $90^o$ (see. Fig.\ref{wik}). This is possible since the integral
over the big circle vanishes and during the transformation of the contour it does not
cross the poles.

\begin{figure}[ht]
\begin{center} \epsfxsize=4cm
 \epsffile{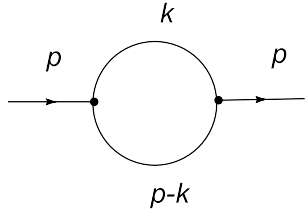}
  \caption{The simplest divergent diagram in a scalar theory \label{exm}}
  \end{center}
\end{figure}

When transferring to Euclidean space the poles in all the propagators disappear. Now the
integral in 4-dimensional Euclidean space can be evaluated in spherical coordinates and
the integral over the  modulus  can be cut on the upper limit. Let us demonstrate how
this method works in the case of the simplest scalar diagram shown in Fig.\ref{exm}.
The corresponding pseudo-Euclidean integral has the form
 \begin{equation}\label{integ1}
 I(p^2)=\frac{1}{(2\pi)^4}\int \frac{d^4k}{[k^2-m^2][(p-k)^2-m^2]}.
 \end{equation}
Transforming it to Euclidean space one gets
\begin{equation}\label{integ2}
 I(p_E^2)=\frac{i}{(2\pi)^4}\int \frac{d^4k_Е}{[k_E^2+m^2][(p-k)_E^2+m^2]}.
 \end{equation}
 (in what follows the index Е will be omitted.)

For calculation of this kind of integrals we use the following approach. First, we
transform the  product of several brackets in the denominator into the single bracket
with the help of the so-called Feynman parametrization. The following general formula is
valid:
\begin{eqnarray}
\frac{1}{A_1^{\alpha_1}A_2^{\alpha_2}\cdots A_n^{\alpha_n}}&=&
\frac{\Gamma(\alpha_1+\alpha_2+\cdots+\alpha_n)}{\Gamma(\alpha_1) \Gamma(\alpha_2)
\cdots\Gamma(\alpha_n)}\int^1_0dx_1dx_2\cdots dx_n \nonumber \\
&.&\frac{\delta(1-x_1-x_2-\cdots-x_n)x_1^{\alpha_1-1}x_2^{\alpha_2-1}\cdots
x_n^{\alpha_n-1}}{[A_1x_1+A_2x_2+\cdots A_nx_n]^{\alpha_1+\alpha_2+\cdots+\alpha_n}}
.\label{feyn}
\end{eqnarray}
Here $\Gamma(\alpha)$ is the Euler $\Gamma$-function which has the following properties:
$$\Gamma(1)=1, \  \Gamma(n+1)=n!, \  x\Gamma(x)=\Gamma(x+1),  \
\Gamma(1+x)=e^{\displaystyle
[-x\gamma_E+\sum\limits^\infty_{n=2}\frac{(-x)^n}{n}\zeta(n)]},$$
 where $\gamma_E$ is the
Euler constant and   $\zeta(n)$ is the Riemann zeta-function. The $\Gamma$-function is
finite for positive values of the argument and has simple poles at negative integer
values and at zero.

In our case,  ($n=2, \alpha_1=\alpha_2=1$) and eq.(\ref{feyn}) has the form:
\begin{eqnarray}
\frac{1}{[k^2+m^2][(p-k)^2+m^2]}&=&
\frac{\Gamma(2)}{\Gamma(1)\Gamma(1)}\int^1_0\frac{dx_1x_2\delta(1-x_1-x_2)}
{\left[[k^2+m^2]x_1+[(p-k)^2+m^2]x_2\right]^2} \nonumber \\
&=& \int^1_0\frac{dx}{[k^2-2pkx+p^2x+m^2]^2} . \label{ffeyn}
\end{eqnarray}
Thus,  integral (\ref{integ2}) can be written as
\begin{equation}\label{inte}
I(p^2)\!=\!\frac{i}{(2\pi)^4}\!\int\limits_0^1\!\! dx\!\! \int\!\!\!
\frac{d^4k}{[k^2\!-\!2kpx\!+\!p^2x\!+\!m^2]^2}\!\! \stackrel{k\to k-px}{=}\!\!
\frac{i}{(2\pi)^4}\!\int\limits_0^1\!\! dx\!\!\int\!\!\!
\frac{d^4k}{[k^2\!+\!p^2x(1\!-\!x)\!+\!m^2]^2}
\end{equation}

Now the integral depends only on the modulus of $k$ and one can use  the spherical
coordinates:
\begin{equation}
I(p^2)= \frac{i}{(2\pi)^4}\int_0^1 dx\ \Omega_4\int_0^\Lambda
\frac{k^3dk}{[k^2+p^2x(1-x)+m^2]^2},
\end{equation}
where the volume of the 4-dimensional sphere equals $\Omega_4=2\pi^2$ (in general
$\Omega_D=\frac{2\pi^{D/2}}{\Gamma(D/2)}$). The integral over the modulus can be easily
calculated
\begin{equation}
\frac 12\int_0^{\Lambda^2} \frac{k^2dk^2}{[k^2+p^2x(1-x)+m^2]^2}=\frac 12
\log(\frac{\Lambda^2}{p^2x(1-x)+m^2})+1,
\end{equation}
and, as one can see, is logarithmically divergent at the upper limit. The full answer has
the form
\begin{equation}\label{lamr}
I(p^2)=\frac{i}{16\pi^2}\int_0^1 dx \left(
\log(\frac{\Lambda^2}{p^2x(1-x)+m^2})+1\right).
\end{equation}
The last integral over  $x$ can also be evaluated and takes the  simplest form in the
limiting cases for $m=0$ or $p=0$. Now one can go back to Minkowski space $p_E^2=>-p^2$.

The regularization with the ultraviolet cut-off is quite natural and relatively simple.
The drawback of this regularization is Euclidean rather than Lorentzian invariance and
also the absence of the gauge invariance. Therefore, it is not useful in the gauge
theories. However, one should notice that the noninvariance of a regularization is
acceptable since the invariance is restored when removing the regularization .  Still,
this aspect complicates the calculation as one has to take care of the validity of all
the identities.

\subsection{Pauli-Villars Regularization}

Another method of regularization which is called the Pauli-Villars regularization is
based on the introduction of a set of additional heavy fields with a wrong sign of the
kinetic term. These fields are not physical and are introduced essentially with the
purpose of regularization of divergent integrals. The main trick is in the replacement
\begin{equation}\label{PV}
\frac{1}{p^2-m^2} \to \frac{1}{p^2-m^2}-\frac{1}{p^2-M^2},
\end{equation}
where $M\to \infty$ is the mass of the Pauli-Villars fields. As a result, the propagator
for large momenta decreases faster, which ensures the convergence of the integrals. The
divergences manifest themselves as logs and powers of $M^2$ instead of the cutoff
parameter $\Lambda^2$.

 One uses sometimes the modifications of the Pauli-Villars regularization when the
replacement (\ref{PV}) is performed not for each propagator but for the loop as a whole.
This method of regularization is called the regularization over circles. It is used in
Abelian gauge theories for the loops made of the matter fields. This way one can preserve
the gauge invariance. However, in non-Abelian theories we face some problems related to
the loops of the gauge fields which cannot become massive without violating the gauge
invariance. This problem is often solved by introducing an additional regularization for
the vector fields, for example, with the help of higher derivatives. Here we will not
consider this regularization.

The positive property of the Pauli-Villars regularization is the explicit Lorentz and
gauge (in abelian case) invariance, but it requires complicated  calculations since one
has to calculate  massive diagrams, while massless integrals are much simpler.

\subsection{Dimensional Regularization}

The most popular in gauge theories is the so-called dimensional regularization. In this
case, one modifies the integration measure.

The technique of dimensional regularization consists of analytical continuation from an
integer to a noninteger number of dimensions. Basically one goes from  some $D$ to
$D-2\epsilon$, where $\epsilon \to 0$. In  particular,  we will be interested in going
from $4$ to $4-2\epsilon$ dimensions.  In this case, all the ultraviolet and infrared
singularities  manifest themselves as  pole terms in $\epsilon$. To perform this
continuation to non-integer  number of dimensions, one has to define all the objects such
as the metric, the measure of integration, the $\gamma$ matrices,  the propagators, etc.
Though this continuation is not unique, one can define a  self-consistent set of rules,
which allows one to perform the calculations.
\par {\it The metric}: \ \
 $g^{\mu\nu}_4 \to g^{\mu\nu}_{4-2\epsilon}$. Though it is rather tricky to
 define the metric in non-integer dimensions, one usually needs  only one
 relation, namely  \    $g^{\mu\nu}g_{\mu\nu} =
\delta^\mu_\mu=D=4-2\epsilon$.
\par {\it The measure}:\ \  $d^4q \to (\mu^2)^\epsilon
d^{4-2 \epsilon}q $ , where $\mu$ is a parameter of dimensional regularization with
dimension of a mass. The integration with this measure is defined by an analytical
continuation from the integer dimensions.
\par
{\it The $\gamma$ matrices} : The usual anticommutation relation holds
$\{\gamma^\mu,\gamma^\nu\}=2g^{\mu\nu}$; however,  some relations involving the dimension
are modified:
$$\gamma^\mu\gamma_\mu=D=4-2\epsilon; \ \ Tr \gamma^\mu\gamma^\nu =
g^{\mu\nu}Tr 1=g^{\mu\nu}\left\{\begin{array}{c} 2^{[D/2]} \\
4\end{array}\right. .$$ Usually $Tr 1 = 4$ is taken. Then the $\gamma$-algebra is
straightforward:
$$Tr\gamma^\mu\gamma^\nu\gamma^\rho\gamma^\sigma= Tr1[g^{\mu\nu}g^{\rho
\sigma}+g^{\mu\sigma}g^{\nu\rho}-g^{\nu\rho}g^{\mu\sigma}],$$
$$\gamma^\mu\gamma^\nu\gamma^\mu= -\gamma^\mu\gamma^\mu\gamma^\nu+
2g^{\mu\nu}\gamma^\mu=-(4-2\epsilon)\gamma^\nu+2\gamma^\nu=-(2-2\epsilon) \gamma^\nu , \
\ \mbox{etc} .$$ What is not well-defined is the $\gamma^5$ since
$\gamma^5=i\gamma^0\gamma^1\gamma^2 \gamma^3$ and cannot be continued to  an arbitrary
dimension. This creates a problem in dimensional regularization since  there is no
consistent way of definition of $\gamma^5$.
\par
{\it The propagator} : In momentum space the continuation is simple
$$\frac{1}{p^2-m^2} \to \frac{1}{p^2-m^2}.$$ However, in coordinate space
one has: (take $m=0$ for simplicity)
$$\int\frac{d^4p}{p^2}e^{ipx} \sim \frac{1}{x^2} \Rightarrow
\int\frac{d^{4-2\epsilon}p}{p^2}e^{ipx} \sim \frac{1}{[x^2]^{1-\epsilon}}.$$
\par
{\it The basic integrals}: The main idea is to calculate the integral in the space-time
dimension where it is convergent and then analytically continue the answer to the needed
dimension.

Consider the earlier discussed example (\ref{integ1}) and use the Euclidean
representation (\ref{inte}). Let us rewrite it formally in  $D$-dimensional space
\begin{equation}\label{rrr}
  \int \frac{d^Dk}{[k^2+M^2]^2}= \frac{\Omega_D}{2}\ \int_0^\infty
  \frac{(k^2)^{D/2-1}dk^2}{[k^2+M^2]^2}, \ \ \ \  M^2\equiv p^2x(1-x)+m^2.
\end{equation}
The integral over $k^2$ is now the table one
\begin{equation}\label{known}
 \int_0^\infty
  \frac{(k^2)^{D/2-1}dk^2}{[k^2+M^2]^2}\stackrel{k^2\to k^2M^2}{=}
  (M^2)^{\frac D2-2}\int_0^\infty \frac{
  x^{D/2-1}dx}{(x+1)^2}=(M^2)^{\frac D2-2}\ \frac{\Gamma(\frac D2)\Gamma(2-\frac D2)}{\Gamma(2)},
\end{equation}
where we assume that the dimension $D$ is such that the integral exists. In this case
this is 2 and 3. The main formula (\ref{known}) allows one to perform the analytical
continuation over $D$ into the region $D=4-2\varepsilon$. For $\varepsilon =0$, i.e., in
4 dimensions, the integral does not exist since the  $\Gamma$-function has a pole at zero
argument. However, in the vicinity of zero we get a regularized expression.

Collecting all together we get
\begin{equation}\label{result}
  I(p^2)=\frac{i}{(2\pi)^D}\frac{\Omega_D}{2}\int_0^1 \ dx
  \frac{\Gamma(D/2)\Gamma(2-D/2)}{[p^2x(1-x)+m^2]^{2-D/2}}.
\end{equation}
Substituting now $D=4-2\varepsilon$ and transforming back into the pseudo-Euclidean space
one finds
\begin{equation}\label{result2}
  I(p^2)=\frac{i(-\pi)^{2-\varepsilon}}{(2\pi)^{4-2\varepsilon}}\Gamma(\varepsilon)\int_0^1
  \frac{dx (\mu^2)^\varepsilon }{[p^2x(1-x)-m^2]^\varepsilon}
\end{equation}
Expanding the denominator into the series over $\varepsilon$, we finally arrive at
\begin{equation}\label{result3}
  I(p^2)=\frac{i}{16\pi^2}\Gamma(1+\varepsilon)\left(\frac 1\varepsilon-\int_0^1
  dx \log[\frac{p^2x(1-x)-m^2}{-\mu^2}]+\log(4\pi)\right).
\end{equation}

Comparing it with eq.(\ref{lamr}) we see that the ultraviolet divergence now takes the
form of the pole over $\varepsilon$ instead of the logarithm of the cutoff. This is less
visual but much simpler in the calculations and also is automatically gauge invariant.

We present below the main integrals needed for the one-loop calculations. They can be
obtained via the analytical continuation from the integer values of $D$. We will write
them down directly in the pseudo-Euclidean space.
\begin{eqnarray}
\int\frac{d^Dp}{[p^2-2kp+m^2]^\alpha}&=&i\frac{\Gamma(\alpha-D/2)}{\Gamma( \alpha)}
\frac{(-\pi)^{D/2}}{[m^2-k^2]^{\alpha-D/2}} , \label{gen}\\
\int\frac{d^{4-2\epsilon}p}{[p^2-2kp+m^2]^2}&=&i\frac{\Gamma( \epsilon)}{\Gamma(2)}
\frac{(-\pi)^{2-\epsilon}}{[m^2-k^2]^{\epsilon}} , \  \ \ \ \Gamma(\epsilon)
\sim \frac{1}{\epsilon} \to \infty, \nonumber \\
\int\frac{d^{4-2\epsilon}p \ \  p_\mu}{[p^2-2kp+m^2]^2}&=&i\frac{\Gamma(
\epsilon)}{\Gamma(2)}
\frac{(-\pi)^{2-\epsilon}k_\mu}{[m^2-k^2]^{\epsilon}},  \label{m} \\
\int\frac{d^{4-2\epsilon}p \ \ p_\mu p_\nu}{[p^2-2kp+m^2]^2}
&=&i(-\pi)^{2-\epsilon}\!\left[ \frac{\Gamma(\epsilon)}{\Gamma(2)}\frac{k_\mu
k_\nu}{[m^2-k^2]^{\epsilon}}\!+\!\frac{g^{\mu\nu}}{2}\frac{\Gamma(\epsilon-1)}{\Gamma(2)}
\frac{1}{[m^2-k^2]^{\epsilon-1}}\right]\nonumber
\end{eqnarray}
The key formula is (\ref{gen}). All the rest can be obtained from it by the
differentiation. Notice the singularity in the r.h.s. of (\ref{gen}) for $\alpha=D/2-n$,
$n=0,1,..$. These integrals remain non-regularized. However, they usually do not appear
in the real calculations.

Let us mention one important rule used in dimensional regularization and related to the
massless theories. By definition it is accepted that zero to any power is zero. Thus, for
example, the following integral is zero
\begin{equation}\label{zero}
  \int \frac{d^Dk}{(k^2)^\alpha}=0, \ \ \ \ \forall \ \alpha.
\end{equation}
In fact, here we have a cancellation of the ultraviolet and infrared divergences which
both have the form of a pole over $1/\varepsilon$. There is no any inconsistency here and
this way of doing is self-consistent in the calculations of dimensionally regularized
integrals.

This rule leads, in particular, to the vanishing of all the diagrams of the tad-pole type
in the massless case. However, in the massive case they survive and are important for the
restoration of the gauge invariance. As it will be clear later, in the Standard Model the
tad-poles give their contribution to the renormalization of the quark masses and provide
the transversality of the vector propagator in a theory with spontaneous symmetry
breaking.
\newpage
\vspace*{1cm}
\section{Lecture III: Examples of Calculations. One-loop Integrals}
\setcounter{equation}{0} All further calculations will be performed using dimensional
regularization. Below we show how the rules described above can be applied to calculate
in various models of quantum field theory.

\subsection{The scalar theory}

We start with the simplest scalar case and consider the theory described by the
Lagrangian
\begin{equation}\label{scalar}
{\cal L} = \frac{1}{2}(\partial _\mu \varphi )^2-\frac{m^2}{2}\varphi^2-\frac{\lambda
}{4!} \varphi ^4.
\end{equation}
The Feynman rules in this case are:
$$\hspace*{-1cm}=\frac{\displaystyle i}{\displaystyle p^2-m^2}, \hspace{3cm}
= -i\lambda$$
\begin{picture}(30,15)\put(59,38){$\bullet$}
\put(60,41){\line(1,0){30}}\put(89,38){$\bullet$}
\put(200,25){\line(1,1){30}}\put(200,55){\line(1,-1){30}} \put(212,37){$\bullet$}
\end{picture}\vspace{-1cm}

First, we find the one-loop divergent diagrams. As it follows from Fig.\ref{scal},
they are the propagator of the scalar field and the quartic vertex.\\

\begin{figure}[ht]\hspace*{8cm}
\begin{picture}(0,30)(40,40)\put(19,34){$\bullet$}
\put(20,37){\line(1,0){60}}\put(79,34){$\bullet$} \put(50,54){\circle{30}}
\put(48,34){$\bullet$}
\end{picture}
\caption{The one-loop propagator diagram\label{golo}}
\end{figure}
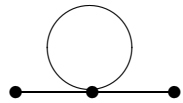

\underline{The propagator:} In the first order there is only one diagram of the tad-pole
type shown in Fig.\ref{golo}.

\noindent The corresponding integral is
\begin{equation}\label{gol}
 J_1(p^2)= \frac{-i\lambda}{(2\pi)^{4-2\varepsilon}}\frac i2 \int \frac{d^{4-2\varepsilon}k
  (\mu^2)^\varepsilon}{k^2-m^2},
\end{equation}
where $1/2$ is the combinatoric factor. Calculating the integral (\ref{gol}), according
to (\ref{m}), we find
\begin{equation}\label{resp}
J_1(p^2)=\frac{-i\lambda}{(4\pi)^{2-\varepsilon}}\frac{\Gamma(-1+\varepsilon)}{2\Gamma(1)}
m^2(\frac{\mu^2}{m^2})^\varepsilon = \frac{i\lambda}{32\pi^2}m^2\left[ \frac
1\varepsilon+1\!-\!\gamma_E\!+\!\log(4\pi)\!-\!\log\frac{m^2}{\mu^2}\right]
\end{equation}
The fact that the integral diverges quadratically manifests itself in the structure of
the multiplier $\Gamma(-1+\varepsilon)$ which has a pole at  $\varepsilon=0$ as well as
at $\varepsilon=1$. However, since we are interested in the limit $\varepsilon\to 0$, we
expand the answer in the Loran series in $\varepsilon$. As one can see, even in the case
of quadratically divergent integrals the divergence takes the form of a simple pole over
$\varepsilon$, but the integral has the dimension equal to two. Notice, however, that for
$m=0$ the integral equals zero in accordance with the properties of dimensional
regularization mentioned above.

\begin{figure}[ht]\hspace*{4cm}
\begin{picture}(200,35)(0,15)
\put(19,20){\line(-1,1){10}}\put(19,20){\line(-1,-1){10}}
\put(16,17){$\bullet$}\put(30,20){\circle{22}}\put(-5,36){$p_1$}\put(-5,1){$p_2$}
\put(55,36){$p_3$}\put(55,1){$p_4$}\put(75,49){$p_1$}\put(75,-11){$p_2$}
\put(99,49){$p_3$}\put(99,-11){$p_4$}\put(129,49){$p_1$}\put(129,-6){$p_2$}
\put(164,49){$p_3$}\put(164,-6){$p_4$} \put(42,20){\line(1,1){10}}\put(40,17){$\bullet$}
\put(42,20){\line(1,-1){10}} \put(60,18){+}
\put(90,31){\line(-1,1){10}}\put(90,31){\line(1,1){10}}
\put(88,29){$\bullet$}\put(90,20){\circle{22}} \put(90,9){\line(-1,-1){10}}
\put(88,6){$\bullet$} \put(90,9){\line(1,-1){10}} \put(115,18){+}
\put(139,20){\line(0,1){12}}\put(139,20){\line(-1,-1){10}} \put(139,32){\line(1,1){18}}
\put(136,17){$\bullet$}\put(150,20){\circle{22}}
\put(162,20){\line(0,1){12}}\put(160,17){$\bullet$}
\put(162,20){\line(1,-1){10}}\put(162,32){\line(-1,1){18}}
\end{picture}\vspace{0.7cm}
\caption{The one-loop vertex diagram\label{vert}}
\end{figure}
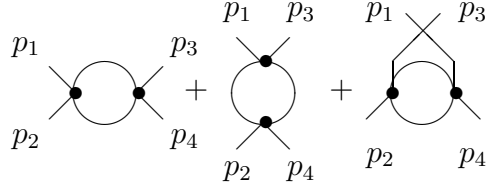

\underline{The vertex:} Here one also has only one diagram but the external momenta can
be adjusted in several ways (see Fig.\ref{vert}).
As a result the total contribution to the vertex function consists of three parts
$$I_1= I_1(s)+I_1(t)+I_1(u),$$
where we introduced the commonly accepted notation for the Mandelstam variables (we
assume here that the momenta  $p_1$ and $p_2$ are incoming and the momenta $p_3$ and
$p_4$ are outgoing)
 $$s=(p_1+p_2)^2=(p_3+p_4)^2,\
t=(p_1-p_3)^2=(p_2-p_4)^2,\ u=(p_1-p_4)^2=(p_2-p_3)^2,$$ and the integral equals
\begin{equation}\label{vert1m}
I_1(s)=\frac{(-i\lambda )^2}{48}\frac{(\mu^2)^\varepsilon }{(2\pi
)^{4-2\varepsilon}}i^2\int \frac{d^{4-2\varepsilon }k}{[k^2-m^2][(p-k)^2-m^2]}
\end{equation} (1/48 is the combinatoric coefficient).
We have already calculated this integral and the answer has the form (\ref{result3}). Now we
perform the calculation in a different and simpler way applicable to the massless
integrals.

Two comments are in order. The first one concerns the evaluation of the combinatoric
coefficient. It comes from the expansion of the S-matrix within the Wick theorem. In the
case when all the particles are different like, for example, in QED, the combinatoric
coefficient is usually 1. For identical particles their permutations are taken into
account already in the Lagrangian (the factors 1/2  and /4! in (\ref{scalar})) and lead
to nontrivial coefficients. There exists a simple method to calculate the combinatoric
coefficient in these cases. The coefficient equals  1/Sym, where Sym is the symmetry
factor of a diagram. Consider the diagram shown in Fig.\ref{vert}. If one does not
distinguish the arrangement of momenta, then the diagram has the following symmetries:
the permutation of external lines entering into the left vertex, the permutation of
external lines entering into the right vertex, the permutation of the vertices, the
permutation of internal lines. Altogether one has: $2\times 2\times 2\times 2=16$. Hence,
the combinatoric coefficient equals $1/16$ but, since we distinguish three different
momentum arrangements, one has $1/48$. The same rule is valid for the multiloop diagrams
and we will use it in the next section.

The second comment is related to the calculation of the massless integrals which are much
simpler, and in some cases one can get the answer without any explicit integration. The
method, which we will describe below, is applicable to a certain type of  massless
integrals and is based on conformal properties of the massless integrals depending on one
external argument and uses the symmetry between the coordinate and momentum
representations.

The key formula is the Fourier-transformation of the propagator of a massless particle
\begin{equation}\label{f}
  \int \frac{d^4p\ e^{ipx}}{p^2}=\frac{i\pi^2}{x^2},
\end{equation}
which can be generalized to an arbitrary dimension and any power of the propagator as
follows:
\begin{equation}\label{ff}
  \int \frac{d^Dp\ e^{ipx}}{(p^2)^\alpha}=i(-\pi)^{D/2}
  \frac{\Gamma(D/2-\alpha)}{\Gamma(\alpha)}\frac{1}{(x^2)^{D/2-\alpha}}.
\end{equation}
Obviously, this formula is also valid for the coordinate integration instead of momentum.
This way the transition from momentum representation to the coordinate one and vice versa
is performed with the help of  (\ref{ff}) and is accompanied by the factor
$\frac{\Gamma(D/2-\alpha)}{\Gamma(\alpha)}$.

Let us go back to the diagram Fig.\ref{vert}. In momentum space it corresponds to the
integral over the  momenta running along the loop. However, in coordinate space it is
just the product of the two propagators and does not contain any integration. Therefore,
the integral in momentum space can be replaced by the Fourier-transform of the square of
the propagator. Since in the massless case all the propagators in both momentum and
coordinate representation are just the powers of $p^2$ or $x^2$, all of them are easily
calculated with the help of relation (\ref{ff}).

In the case of the integral (\ref{vert1m}) for $m=0$ one first has to mentally transform
both the propagators into coordinate space which, according to (\ref{ff}), gives the
factor $(\frac{\Gamma(1-\varepsilon)}{\Gamma(1)})^2$, then multiply the obtained
propagators (this gives  $1/(x^2)^{2-2\varepsilon})$) and transform the obtained result
back into momentum space that gives the factor
$\frac{\Gamma(\varepsilon)}{\Gamma(2-2\varepsilon)}$ and the power of momenta
$1/(p^2)^\varepsilon$ (the same as in the argument of the last $\Gamma$-function).
Besides this, each loop contains the factor $i(-\pi)^{2-\varepsilon}$. Collecting all
together one gets
$$ I_1(s)=\frac{(-i\lambda )^2}{48}\frac{(\mu^2)^\varepsilon i^2}{(2\pi
)^{4-2\varepsilon}}\int\!\! \frac{d^{4-2\varepsilon }k}{k^2(p-k)^2} = \frac{\lambda
^2}{48}\frac{i\pi^{2-\varepsilon}}{(2\pi  )^{4-2\varepsilon}}\left(\frac{\mu
^2}{-s}\right)^\varepsilon\!\! \frac{\Gamma(1-\varepsilon)\Gamma(1-\varepsilon)
\Gamma(\varepsilon)}{\Gamma(1)\Gamma(1) \Gamma(2-2\varepsilon )}$$

$$= \frac{i}{48}\frac{\lambda ^2}{(4\pi)^{2-\varepsilon}}\!\left[\frac{\mu
^2}{-s}\right]^\varepsilon \!\!\frac{1}{\varepsilon(1\!-\!2\varepsilon)}
\frac{\Gamma^2(1\!-\!\varepsilon)\Gamma(1\!+\!\varepsilon)}{\Gamma(1-2\varepsilon)}\!=\!
\frac{i}{48}\frac{\lambda ^2}{16\pi^2}[\frac{1}{\varepsilon}+2-\gamma_E+\log 4\pi+\ln
\frac{\mu^2 }{-s}],$$ which coincides with  (\ref{result3}) at $m=0$.

The described method for calculation of massless integrals is applicable to any integral
depending on one external momentum (propagator type) and allows one to perform the
calculations in any number of loops simply writing down the corresponding factors without
explicit integration. In the case when the integral depends on more than one external
momentum (like for a triangle or a box) and they cannot be put equal to zero the method
is not directly applicable though some modifications are available. We do not consider
them here.

The four-point vertex in the one-loop approximation thus equals   (we take the common
factor  $1/4!\phi^4$ out of the brackets):
\begin{equation}\label{resv}
\Gamma_4\! =\! -i\lambda\left\{1\!-\! \frac{\lambda}{16\pi^2}\left( \frac{3}{2\varepsilon
}+3-\frac 32 \gamma_E\!+\!\frac 32\log 4\pi+\frac{1}{2}\ln\frac{\mu^2 }{-s}+
\frac{1}{2}\ln\frac{\mu^2 }{-t}+\frac{1}{2}\ln\frac{\mu^2 }{-u}\right)\right\} .
\end{equation}
As one can see, the Euler constant and the logarithm of $4\pi$ always accompany the pole
term $1/\varepsilon$ and can be absorbed into the redefinition of $\mu^2$.

\subsection{Quantum electrodynamics}

Consider now the calculation of the diagrams in the gauge theories. We start with quantum
electrodynamics. The QED Lagrangian has the form
\begin{equation}\label{qed}
  {\cal L}_{QED}=-\frac 14 F_{\mu\nu}^2+\bar \psi (i\gamma^\mu \partial_\mu -m)\psi+e\bar
  \psi \gamma^\mu A_\mu \psi-\frac{1}{2\xi}(\partial_\mu A_\mu)^2,
\end{equation}
where the electromagnetic stress tensor is $F_{\mu\nu}=\partial_\mu A_\nu-\partial_\nu
A_\mu$, and the last term in (\ref{qed}) fixes the gauge. In what follows we choose the
Feynman or the diagonal gauge  $(\xi=1)$.

\begin{figure}[ht]\vspace{0.2cm}\hspace*{0.5cm}
\includegraphics[width=0.9\textwidth]{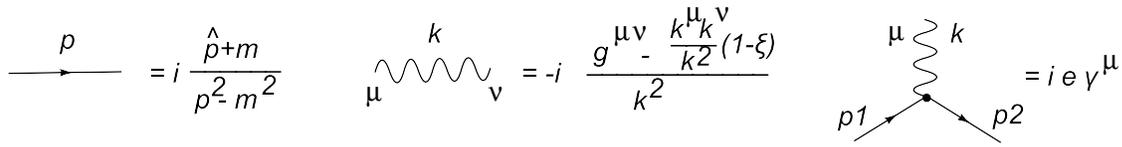}
  \caption{The Feynman rules for QED\label{qedfeyn}}
\end{figure}

The Feynman rules corresponding to the Lagrangian  (\ref{qed}) are shown in
Fig.\ref{qedfeyn}.

\begin{figure}[h]\hspace*{2cm}
 \leavevmode
    \epsfxsize=12cm
 \epsffile{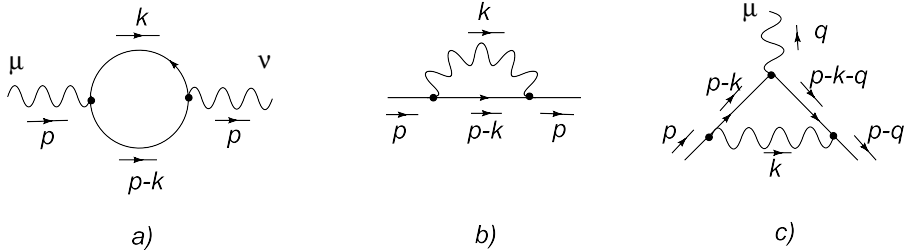}
\caption{The one-loop divergent diagrams in QED} \label{1}
\end{figure}

In quantum electrodynamics the divergences appear only in the photon propagator, the
electron propagator, and the triple vertex. The one-loop divergent diagrams are shown in
Fig.\ref{1}.

We begin with  the vacuum polarization graph. It is given by the diagram shown in Fig.
\ref{1}a). The corresponding expression looks like:
\begin{equation}
\Pi_{\mu\nu}(p)=(-)\frac{e^2}{(2\pi)^4}\int d^4k\frac{Tr [\gamma^\mu(m+\hat k)\gamma^\nu
(m+\hat k-\hat p)]}{[m^2-k^2][m^2-(k-p)^2]},\label{i}
\end{equation}
where the "-" sign comes from the fermion loop and $\hat q \equiv \gamma^\mu q_\mu$. We
first go to dimension $4-2\epsilon$. Then the  integral  (\ref{i}) becomes
\begin{equation}
\Pi_{\mu\nu}^{Dim}(p)=(-)\frac{e^2(\mu^2)^\varepsilon}{(2\pi)^{4-2\varepsilon}} \int
d^{4-2\varepsilon}k\frac{Tr [\gamma^\mu(m+\hat k)\gamma^\nu (m+\hat k-\hat
p)]}{[m^2-k^2][m^2-(k-p)^2]},\label{id}
\end{equation}
Let us put $m=0$ for simplicity. This will allow us to get a simple answer at the end.
First, we calculate the trace of the $\gamma$-matrices: $$Tr\gamma^\mu \hat k \gamma^\nu
(\hat k -\hat p)= Tr(\gamma^\mu \gamma^\rho \gamma^\nu \gamma^\sigma)k^\rho (k-p)^\sigma=
4k^\rho (k-p)^\sigma
[g^{\mu\rho}g^{\nu\sigma}+g^{\mu\sigma}g^{\nu\rho}-g^{\mu\nu}g^{\rho\sigma}].
$$
So the integral now looks like
$$I_{\rho\sigma}^{Dim}(p)=(-)\frac{(\mu^2)^\varepsilon}{(2\pi)^{4-2\varepsilon}}
\int\frac{d^{4-2\varepsilon}k k^\rho (k-p)^\sigma}{k^2(k-p)^2}.$$ Using the Feynman
parametrization  and performing the integration according to the formulae given above one
finds
\begin{equation}
I_{\rho\sigma}^{Dim}(p)=(-)\frac{(\mu^2)^\varepsilon}{(2\pi)^{4-2\varepsilon}}
\int^1_0dx\int \frac{d^{4-2\varepsilon}k k^\rho (k-p)^\sigma}{[k^2-2pkx+p^2x]^2}
\label{in}
\end{equation}
$$=(-)i\frac{(-\mu^2)^\varepsilon \pi^{2-\varepsilon}}{(2\pi)^{4-2\varepsilon}}\left\{
-\Gamma(\varepsilon)\!\int^1_0\!\frac{dx p^\rho p^\sigma
x(1-x)}{[p^2x(1-x)]^\varepsilon}\! +
\!\Gamma(\varepsilon-1)\frac{g^{\rho\sigma}}{2}\int^1_0\!\!\frac{dx
}{[p^2x(1-x)]^{\varepsilon-1}}\right\}.$$

To evaluate the remaining integrals, we use the standard integral for the Euler
beta-function
 $$\int^1_0 dx x^{\alpha -1}(1-x)^{\beta -1}= B(\alpha,\beta )=\frac{\Gamma(\alpha)\Gamma(\beta
)}{\Gamma(\alpha+\beta )},$$ which gives in our case $$ \int^1_0 dx
x^{1-\varepsilon}(1-x)^{1-\varepsilon}= \frac{\Gamma(2-\varepsilon)\Gamma(2-\varepsilon
)}{\Gamma(4-2\varepsilon )}.$$ Thus, the integral (\ref{in}) becomes
\begin{equation}
I_{\rho\sigma}^{Dim}(p)=\frac{i}{16\pi^2}(4\pi)^\varepsilon
\left(-\frac{\mu^2}{p^2}\right)^\varepsilon
\frac{\Gamma^2(2-\varepsilon)\Gamma(\varepsilon)}{\Gamma(4-2\varepsilon)}\left[p^\rho
p^\sigma + \frac{1}{2}\frac{g^{\rho\sigma}p^2}{1-\varepsilon}\right], \label{rr}
\end{equation}
where we have used that $\Gamma(-1+\varepsilon)=-\frac{\displaystyle
\Gamma(\varepsilon)}{\displaystyle 1-\varepsilon}$. Multiplying  eq.(\ref{rr} ) by the
trace $$ [g^{\mu\rho}g^{\nu\sigma}+g^{\mu\sigma}g^{\nu\rho}-g^{\mu\nu}g^{\rho\sigma}]
p^\rho p^\sigma = p^\mu p^\nu+ p^\nu p^\mu-g^{\mu\nu}p^2=2 p^\mu p^\nu-g^{\mu\nu}p^2 ,$$
$$ [g^{\mu\rho}g^{\nu\sigma}+g^{\mu\sigma}g^{\nu\rho}-g^{\mu\nu}g^{\rho\sigma}]
g^{\rho\sigma}p^2 =g^{\mu\nu}p^2+g^{\mu\nu}p^2-g^{\mu\nu}(4-2\varepsilon)
p^2=-(2-2\varepsilon)p^2g^{\mu\nu},$$ we find
\begin{eqnarray}
\Pi_{\mu\nu}^{Dim}(p)&=&i\frac{4e^2}{16\pi^2}(4\pi)^\varepsilon
\left(-\frac{\mu^2}{p^2}\right)^\varepsilon
\frac{\Gamma^2(2-\varepsilon)\Gamma(\varepsilon)}{\Gamma(4-2\varepsilon)}\left[
2p^\mu p^\nu-g^{\mu\nu}p^2-g^{\mu\nu}p^2\right] \nonumber \\
&=&-i\frac{8e^2}{16\pi^2}(4\pi)^\varepsilon \left(-\frac{\mu^2}{p^2}\right)^\varepsilon
(g^{\mu\nu}p^2- p^\mu
p^\nu)\frac{\Gamma^2(2-\varepsilon)\Gamma(\varepsilon)}{\Gamma(4-2\varepsilon)}.
\end{eqnarray}
Expanding now over $\varepsilon$ with the help of
$$\Gamma(\varepsilon)=\frac{1}{\varepsilon}\Gamma(1\!+\!\varepsilon), \
\Gamma(2\!-\!\varepsilon)=(1\!-\!\varepsilon) \Gamma(1\!-\!\varepsilon), \
\Gamma(4\!-\!2\varepsilon)=(3\!-\!2\varepsilon)(2\!-\!2\varepsilon)(1\!-\!2\varepsilon)
\Gamma(1\!-\!2\varepsilon),$$ we finally get
\begin{eqnarray}
\Pi_{\mu\nu}^{Dim}(p)&=&-i\frac{e^2}{16\pi^2}(4\pi)^\varepsilon
\left(-\frac{\mu^2}{p^2}\right)^\varepsilon (g^{\mu\nu}p^2-p^\mu
p^\nu)\frac{4(1+5/3\varepsilon)}{3\varepsilon}e^{-\gamma\varepsilon}  \nonumber \\
&=&-ie^2\frac{g^{\mu\nu}p^2-p^\mu
p^\nu}{16\pi^2}\frac{4}{3}\left[\frac{1}{\varepsilon}-\gamma_E+\log 4\pi
+\log\frac{-\mu^2}{p^2}+\frac{5}{3}\right] , \label{vacuum} \\
&=& i(g^{\mu\nu}p^2-p^\mu p^\nu)\Pi^{Dim}(p^2) ,\nonumber
\end{eqnarray}
where
\begin{equation}\label{pol}
\Pi^{Dim}(p^2)=-\frac{e^2}{16\pi^2}\frac{4}{3}\left[\frac{1}{\varepsilon}-\gamma_E+\log
4\pi +\log\frac{-\mu^2}{p^2}+\frac{5}{3}\right].
\end{equation}

\begin{figure}[htb] \vspace{-0.2cm}\hspace*{1.7cm}
 \leavevmode
    \epsfxsize=12cm
 \epsffile{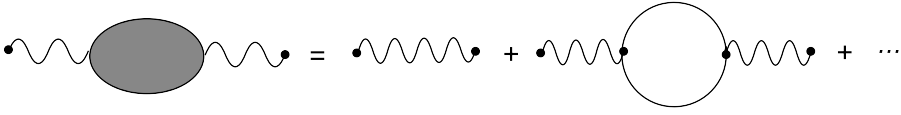}
\caption{The photon propagator in QED\label{propqed}}
\end{figure}

Given the expression for the  vacuum polarization one can construct the photon propagator
as shown in Fig.\ref{propqed}.

\noindent One has
\begin{eqnarray*}
G_{\mu\nu}(p)&=&\frac{-i}{p^2}g^{\mu\nu}+\frac{-i}{p^2}g^{\mu\rho}\Pi_{\rho
\sigma}\frac{-i}{p^2}g^{\sigma\nu}+ \cdots \\
&=&\frac{-i}{p^2}g^{\mu\nu}-\frac{\Pi^{\mu\nu}}{p^4} +\cdots =
\frac{-i}{p^2}g^{\mu\nu}-\frac{i(g^{\mu\nu}-p^\mu p^\nu/p^2)}{p^2}\Pi(p^2)
+ \cdots \\
&=&\frac{-i}{p^2}(g^{\mu\nu}-\frac{p^\mu p^\nu}{p^2})(1+\Pi(p^2)+\cdots)
-\frac{i}{p^2}\frac{p^\mu p^\nu}{p^2},
\end{eqnarray*}
where $\Pi(p^2)$ is given by eq.(\ref{pol}).
 Notice that the radiative corrections  are always proportional to the
transverse tensor $P_{\mu\nu}=g_{\mu\nu}-p_\mu p_\nu/p^2$. This is a consequence of the
gauge invariance and follows from the Ward identities.

Consider now the electron self-energy graph Fig.\ref{1}b). The corresponding integral is
\begin{equation}\label{self}
  \Sigma(\hat p)=-\frac{e^2}{(2\pi)^4}\int d^4k \frac{\gamma^\mu (\hat p-\hat k
  +m)\gamma^\mu}{k^2[(p-k)^2-m^2]}.
\end{equation}
Acting in a usual way we go to dimension  $4-2\varepsilon$, convert the indices of the
$\gamma$-matrices and introduce the Feynman parametrization. The result is
\begin{equation}\label{selfd}
\Sigma^{Dim}(\hat p)=-\frac{e^2(\mu^2)^\varepsilon}{(2\pi)^{4-2\varepsilon}}\int_0^1 dx
\int \frac{d^{4-2\varepsilon}k [-2(1-\varepsilon)(\hat p-\hat k)+(4-2\varepsilon)m]}{[k^2
-2kpx+p^2x-m^2x]^2}.
\end{equation}
The integral over $k$ can now be evaluated according to the standard formulas
\begin{equation}\label{selfr}
\Sigma^{Dim}(\hat
p)=-i\frac{e^2}{16\pi^2}\frac{(-\mu^2)^\varepsilon}{(4\pi)^{-\varepsilon}}\Gamma(\varepsilon)
\int_0^1 dx\frac{-2(1-\varepsilon)\hat p(1-x)+(4-2\varepsilon)m}{[p^2x(1-x)
-m^2x]^\varepsilon}.
\end{equation}
This expression can be expanded in series in $\varepsilon$
\begin{eqnarray}\label{selfrr}
\Sigma^{Dim}(\hat p)&=&-i\frac{e^2}{16\pi^2}\left[-\frac{\hat p-4m}{\varepsilon}+\hat p
-2m -(\hat p -4m)(-\gamma_E+\log(4\pi))\right.\nonumber\\
&+& \left. \int_0^1 dx [2\hat p (1-x)-4m]\log\frac{p^2x(1-x)-m^2x}{-\mu^2}\right].
\end{eqnarray}

Notice that the linear divergence of the integral  manifests itself as a simple pole in
$\varepsilon$, and the coefficient has the dimension equal to 1 and is Lorentz invariant
(this is either  $\hat p$ or $m$).

At last, consider the vertex function Fig.\ref{1}c). The corresponding integral is
\begin{equation}\label{ver}
  \Gamma_1(p,q)=\frac{e^3}{(2\pi)^4}\int d^4k\frac{\gamma^\nu(\hat p-\hat k-\hat q+m)
  \gamma^\mu (\hat p-\hat k+m)\gamma^\nu}{[(p-k-q)^2-m^2][(p-k)^2-m^2]k^2}.
\end{equation}
Transfer to dimension  $4-2\varepsilon$ and introduce the Feynman parametrization. This
gives
\begin{eqnarray}\label{ver2}
  \Gamma^{Dim}_1(p,q)&=&\frac{e^3(\mu^2)^\varepsilon}{(2\pi)^{4-2\varepsilon}}\Gamma(3)
  \int_0^1dx \int_0^x dy \\
  &\times&\int \frac{d^{4-2\varepsilon}k[\gamma^\nu(\hat p-\hat k-\hat q+m)
  \gamma^\mu (\hat p-\hat k+m)\gamma^\nu]}{[((p-k-q)^2-m^2)y+
  ((p-k)^2-m^2)(x-y)+k^2(1-x)]^3}\nonumber .
\end{eqnarray}
The integral over $k$ is straightforward and gives
\begin{eqnarray}\label{ver3}
 && \Gamma^{Dim}_1(p,q)=ie\frac{e^2}{16\pi^2}
 \frac{(-\mu^2)^\varepsilon}{(4\pi)^{-\varepsilon}} \int_0^1dx \int_0^x dy \\
&& \left\{
  \Gamma(1+\varepsilon)
   \frac{[\gamma^\nu(\hat p(1-x)-\hat q(1-y)+m)
  \gamma^\mu (\hat p(1-x)+\hat qy+m)\gamma^\nu]}{
  [(p-q)^2y(1-x)+p^2(1-x)(x-y)+q^2y(x-y)-m^2x]^{1+\varepsilon}
  }\right. \nonumber\\
  &&\left. +\frac{\Gamma(\varepsilon)}{2}
  \frac{\gamma^\nu\gamma^\rho\gamma^\mu\gamma^\rho\gamma^\nu}{
  [(p-q)^2y(1-x)+p^2(1-x)(x-y)+q^2y(x-y)-m^2x]^{\varepsilon}}\right\}.\nonumber
\end{eqnarray}
As one can see, the first integral is finite and the second one is logarithmically
divergent. Expanding in series in $\varepsilon$ we find
\begin{eqnarray}\label{ver4}
 && \Gamma^{Dim}_1(p,q)=ie\frac{e^2}{16\pi^2}\left\{\frac{\gamma^\mu}{\varepsilon}
 -2\gamma^\mu -\gamma^\mu(\gamma_E-\log(4\pi))\right. \\ &-&2\gamma^\mu \int_0^1\!dx
 \int_0^x\! dy\ \log\left[\frac{(p\!-\!q)^2y(1\!-\!x)\!+\!p^2(1\!-\!x)(x\!-\!y)\!+
 \!q^2y(x-y)\!-\!m^2x}{-\mu^2}\right] \nonumber \\
 &+&\left.
 \int_0^1dx \int_0^x dy
   \frac{\gamma^\nu(\hat p(1-x)-\hat q(1-y)+m)
  \gamma^\mu (\hat p(1-x)+\hat qy+m)\gamma^\nu}{
  (p-q)^2y(1-x)+p^2(1-x)(x-y)+q^2y(x-y)-m^2x
  }\right\} \nonumber .
\end{eqnarray}

\subsection{Quantum chromodynamics}

Consider now the non-Abelian gauge theories and, in particular, QCD. The Lagrangian of
QCD has the form
\begin{eqnarray}\label{qcd}
  {\cal L}_{QСD}&=&-\frac 14 (F_{\mu\nu}^a)^2+\bar \psi (i\gamma^\mu \partial_\mu -m)\psi
  +g\bar \psi \gamma^\mu A_\mu^a T^a\psi-\frac{1}{2\xi}(\partial_\mu A_\mu^a)^2\nonumber \\
    &+&\partial_\mu\bar c^a\partial_\mu c^2+gf^{abc}\partial_\mu\bar c^a A_\mu^b c^c,
\end{eqnarray}
where the stress tensor of the gauge field is now $F_{\mu\nu}^a=\partial_\mu
A_\nu^a-\partial_\nu A_\mu^a+gf^{abc}A_\mu^bA_\nu^c$ and the last terms represent the
Faddeev-Popov ghosts.

The Lagrangian (\ref{qcd}) generates the following set of Feynman rules:

\begin{minipage}[h]{5cm}\vspace{-1cm} \hspace*{6.1cm}
%\vbox to 1cm {\special{em:graph s_calc6.bmp}  }
 \leavevmode
    \epsfxsize=3.8cm
 \epsffile{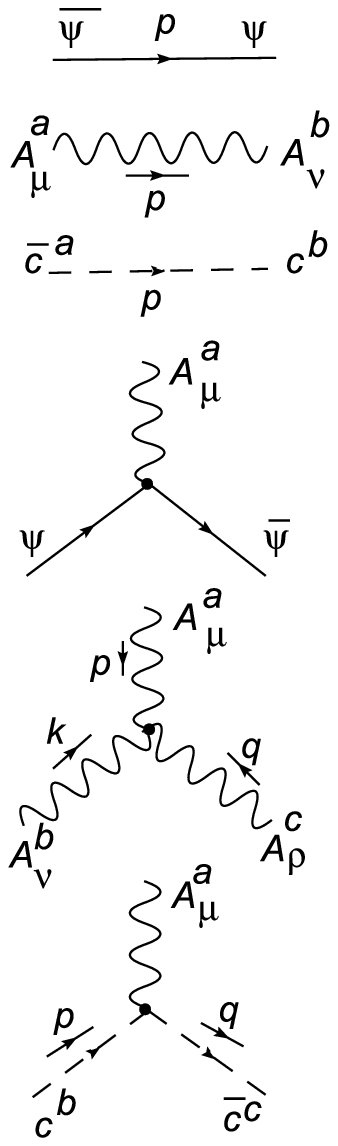}
\end{minipage}\vspace{-13.5cm}
$$\begin{array}{ll} \mbox{the spinor propagator}\hspace{3cm} &
\hspace*{2.5cm}= \frac{\displaystyle i}{\displaystyle \hat p - m }
\\ & \\ \mbox{the vector propagator}&
\hspace*{2.5cm}\frac{\displaystyle -i\delta^{ab}g^{\mu\nu}}{\displaystyle  p^2 }
\\ & \\ \mbox{the ghost
propagator} & \hspace*{2.5cm}= \frac{\displaystyle i\delta^{ab}}{\displaystyle p^2 }
\\ & \\  & \\
\mbox{the spino-gauge vertex} & \hspace*{2.5cm}-ig\gamma^\mu T^a
\\ & \\  & \\  & \\ & \\
 \mbox{the triple gauge vertex} &\hspace*{2.5cm} =-gf^{abc}[(p-q)^\rho
 g^{\mu\nu}\\
 &\hspace*{2.5cm}+ (q-k)^\mu g^{\rho\nu}\\
 &\hspace*{2.5cm}+(k-p)^\nu g^{\mu\rho}]
 \\ & \\  & \\ & \\ & \\
 \mbox{the ghost-gauge vertex} & \hspace*{2.5cm}=-gf^{abc}q^\mu
 \\ &
 \end{array}$$\vspace{1.5cm}

\begin{figure}[htb] \vspace{0.3cm}\hspace*{2cm}
  \leavevmode
     \epsfxsize=12cm
  \epsffile{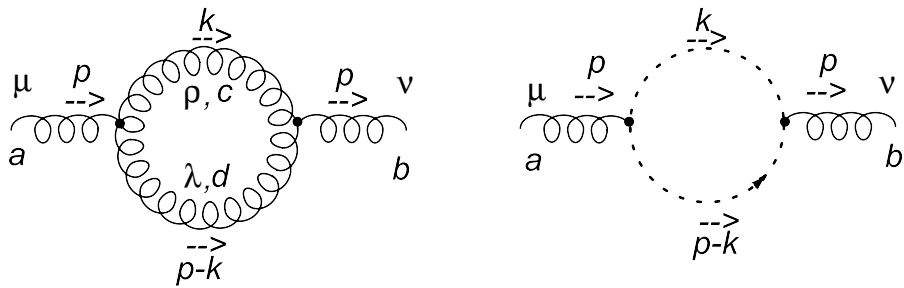}
 %\begin{minipage}[h]{5cm}\hspace*{0cm}
 %\vbox to 5cm {\special{em:graph s_calc3.bmp}  }
 %\end{minipage}
% \vspace{-1cm}
 \caption{The vacuum polarization diagrams in the Yang-Mills theory}\label{3}
 \end{figure}

Consider the one-loop divergent diagrams. We start with the gluon propagator. Besides the
diagram shown in Fig.\ref{1}а), one has additional contributions to the vacuum
polarization from the diagrams shown in Fig.\ref{3}. The first diagram takes into account
the gluon self-interaction and the second one the contribution of the Faddeev-Popov
ghosts. (As has  already been mentioned, the tad-pole diagrams should not be included
since they are automatically zero.) These diagrams depend on the choice of the gauge, and
to evaluate them we have to fix the gauge. In what follows we choose the Feynman gauge
(or the diagonal gauge) for the gluon field.

\noindent Then for the first diagram we have the expression
 \begin{eqnarray}
  \Pi_{\mu\nu}^{ab}(p)&=&\frac{g^2C_A\delta^{ab}}{2(2\pi)^4}\int \frac{d^4k}
 {k^2(k-p)^2} [(2p-k)^\rho g^{\mu\lambda}+(2k-p)^\mu
 g^{\rho\lambda}-(k+p)^\lambda g^{\mu\rho}] \nonumber\\
  &\times&
   [(2p-k)^\rho g^{\lambda\nu}-(k+p)^\lambda g^{\nu\rho }+(2k-p)^\nu
 g^{\rho\lambda} ],  \label{ig}
   \end{eqnarray}
where $1/2$ is a combinatorial factor and $C_2$ is the quadratic Casimir operator which
for the SU(N) group equals  $N$. It  comes  from the contraction of the  gauge group
structure constants $f^{abc}$
    $$f^{abc}f^{dbc}=C_2\delta^{ad}.$$
Contracting the indices and going to $4-2\epsilon $ dimensions, one gets
\begin{eqnarray}
&& \Pi_{\mu\nu}^{Dim\ (ab)}(p) =\delta^{ab}
 \frac{g^2C_A}{2}\frac{(\mu^2)^\varepsilon}{(2\pi)^{4-2\varepsilon}}\int
 \frac{d^{4-2\varepsilon}k}{k^2(k-p)^2}   \{g^{\mu\nu}[4p^2+k^2+(k-p)^2]
 \nonumber \\ &&+(3\!-\!2\varepsilon)(2k\!-\!p)^\mu
 (2k\!-\!p)^\nu\!-\!(2p\!-\!k)^\mu(2p\!-\!k)^\nu\!-\!(k\!+\!p)^\mu(k\!+\!p)^\nu\}\label{igd}
 .
  \end{eqnarray}
To calculate the integrals, one can use the formulas given above. The first step is the
Feynman parametrization, eq.(\ref{ffeyn}),  and then  the momentum integration is
performed according to eqs.(\ref{m}). Applying these rules we get for the integral
(\ref{igd})
\begin{equation}
 \Pi_{\mu\nu}^{Dim \ (ab)}(p)=i\frac{g^2C_A
\delta^{ab}}{(4\pi)^{2-\varepsilon}} \left[\frac{-\mu^2}{p^2}\right]^\varepsilon\!\!
\frac{\Gamma(\varepsilon)\Gamma(1\!-\!\varepsilon)\Gamma(2\!-\!\varepsilon)}{\Gamma(4\!-\!2\varepsilon)}
[g^{\mu\nu}p^2(\frac{19}{2}\!-\!6\varepsilon)-p^\mu p^\nu(11\!-\!7\varepsilon)]
.\label{glu}
\end{equation}
The second diagram corresponds to the integral
\begin{equation}
 \Pi_{\mu\nu}^{Dim \ (ab)}(p)=i\frac{g^2C_A
\delta^{ab}}{(4\pi)^{2-\varepsilon}} \left[\frac{-\mu^2}{p^2}\right]^\varepsilon\!\!
\frac{\Gamma(\varepsilon)\Gamma(1\!-\!\varepsilon)\Gamma(2\!-\!\varepsilon)}{\Gamma(4\!-\!2\varepsilon)}
[g^{\mu\nu}p^2(\frac{19}{2}\!-\!6\varepsilon)-p^\mu p^\nu(11\!-\!7\varepsilon)]
.\label{glu2}
\end{equation}
here the "-" sign comes from the  Fermi statistics of the ghost fields.

Calculation is now straightforward and gives
 \begin{equation}
 \Pi_{\mu\nu}^{Dim \
 (ab)}(p)=i\frac{g^2C_A\delta^{ab}}{(4\pi)^{2-\varepsilon}}
 \left(-\frac{\mu^2}{p^2}\right)^\varepsilon
  \frac{\Gamma(\varepsilon)\Gamma(1-\varepsilon)\Gamma(2-\varepsilon)}{\Gamma(4-
 2\varepsilon)}[
 g^{\mu\nu}p^2/2+p^\mu p^\nu(1-\varepsilon)] .\label{ghost}
 \end{equation}
 Adding up the two contributions together, one finally has
 \begin{equation}
  \Pi_{\mu\nu}^{Dim \
 (ab)}(p)=iC_A\frac{2g^2\delta^{ab}}{16\pi^2}(4\pi)^\varepsilon\!
 \left[\frac{-\mu^2}{p^2}\right]^\varepsilon\!\!
 \frac{\Gamma(\varepsilon)\Gamma(1\!-\!\varepsilon)\Gamma(2\!-\!\varepsilon)}{\Gamma(4-
 2\varepsilon)}
 (5\!-\!3\varepsilon)[g^{\mu\nu}p^2-p^\mu p^\nu] \label{sum}
 \end{equation}
or expanding in $\varepsilon$
 \begin{equation}
 \Pi_{\mu\nu}^{Dim \ (ab)}(p)=iC_A\delta^{ab}g^2
 \frac{g^{\mu\nu}p^2-p^\mu
 p^\nu}{16\pi^2}\frac{5}{3}\left[\frac{1}{\varepsilon}-\gamma_E+\log 4\pi
 +\log\frac{-\mu^2}{p^2}+\frac{31}{15}\right] . \label{res2}
 \end{equation}
Acting the same way as in QED one can calculate the contribution to the gluon propagator.

Notice that the final result for the sum of the two diagrams is again proportional to the
transverse tensor $P_{\mu\nu}=g_{\mu\nu}-p_\mu p_\nu/p^2$. This is not true, however, for
the  diagram with the gauge fields  and is valid only if one takes into account the ghost
contribution. Notice also the opposite sign of the resulting expression compared to that
of eq.(\ref{vacuum}). This is due to a non-Abelian nature of the gauge fields and has
very important consequences to be discussed later.

\begin{figure}[htb]\hspace*{1cm}
\leavevmode
    \epsfxsize=13cm
 \epsffile{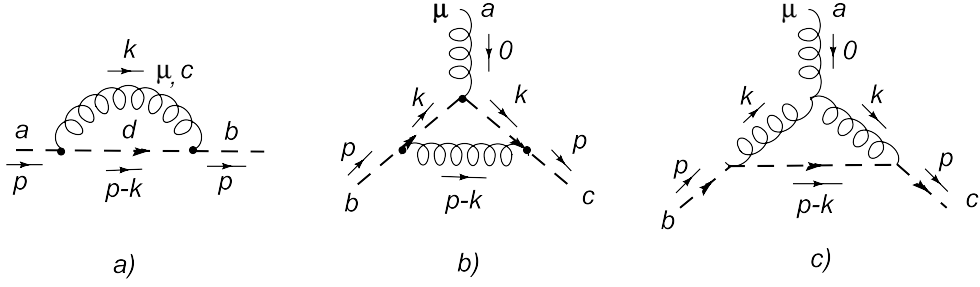}
 \caption{The ghost propagator and the ghost-gluon vertex diagrams in QCD
  \label{ghd}}
  \end{figure}

Consider also the ghost propagator. Here there is only one diagram shown in
Fig.\ref{ghd}a).

It corresponds to the integral
\begin{equation}
\Pi^{Dim \ (ab)}(p) =-C_A\delta^{ab}\frac{g^2(\mu^2)^\varepsilon}
{(2\pi)^{4-2\varepsilon}}\int d^{4-2\varepsilon}k\frac{k^\mu p^\mu}{k^2(k-p)^2}
,\label{gh}
\end{equation}
which equals
\begin{eqnarray}
\Pi^{Dim \ (ab)}(p)
&=&-iC_A\delta^{ab}\frac{g^2}{2(4\pi)^{2-\varepsilon}}\left(\frac{-\mu^2}{p^2}\right)^\varepsilon
p^2\frac{\Gamma(\varepsilon)\Gamma(1-\varepsilon)\Gamma(1-\varepsilon)}{\Gamma(2-
2\varepsilon)} \nonumber \\ &=& - iC_A\delta^{ab}\frac{g^2}{32\pi^2}p^2
\left[\frac{1}{\varepsilon}-\gamma_E+\log 4\pi +\log\frac{-\mu^2}{p^2}+2 \right] .
\label{ghpr}
\end{eqnarray}

Analogously one can calculate the vertex diagrams. We consider in more detail the
calculation of the ghost-gluon vertex as a  simpler one. The corresponding diagrams are
shown in Fig.\ref{ghd}. To simplify the evaluation, we put one of the momenta equal to
zero. Then the first diagram gives the integral
\begin{equation}
V_{1 \rho}^{Dim \ (abc)}(p) =i\frac{C_A}{2}f^{abc}\frac{g^3(\mu^2)^\varepsilon}
 {(2\pi)^{4-2\varepsilon}}\int d^{4-2\varepsilon}k\frac{k^\mu k^\rho
 p^\mu}{(k^2)^2(k-p)^2}
 .\label{v1}
 \end{equation}
Using the equality  $kp=1/2[k^2+p^2-(k-p)^2]$ and substituting it into (\ref{v1}) we find
that the first two terms are reduced to the standard integrals and the last one leads to
the tad-pole structure and is equal to zero. Adding up all together we get
 \begin{eqnarray}
 V_{1 \rho}^{Dim \ (abc)}(p)&=&
-C_A\frac{1}{4}f^{abc}\frac{g^3}{(4\pi)^{2-\varepsilon}}\left(-\frac{\mu^2}{p^2}
 \right)^\varepsilon
 p^\rho\frac{\Gamma(\varepsilon)\Gamma(2-\varepsilon)\Gamma(1-\varepsilon)}{\Gamma(
  3-2\varepsilon)}(1+2\varepsilon) \nonumber \\ &= &
 -C_A\frac{1}{8}f^{abc}\frac{g^3}{16\pi^2}p^\rho
 \left[\frac{1}{\varepsilon}-\gamma_E+\log 4\pi+\log\frac{-\mu^2}{p^2}+4 \right] .
\label{v1r}
\end{eqnarray}
The second diagram gives
\begin{equation} V_{2  \rho}^{Dim \ (abc)}(p)=-i\frac{C_A}{2}f^{abc}\frac{g^3(\mu^2)^\varepsilon}
{(2\pi)^{4-2\varepsilon}}\int d^{4-2\varepsilon}k\frac{(p-k)^\mu p^\nu [ k^\nu
g^{\mu\rho} +k^\mu g^{\nu\rho} -2k^\rho g^{\mu\nu}]}{(k^2)^2(k-p)^2} .\label{v2}
\end{equation}
Contracting the indices in the numerator we have $(p-k)^\rho kp + p^\rho k(p-k)-2k^\rho
p(p-k)$, which after integration leads to
\begin{eqnarray}
V_{2 \rho}^{Dim \ (abc)}(p) &=&
 -C_A\frac{3}{8}f^{abc}\frac{g^3}{(4\pi)^{2-\varepsilon}}\left(-\frac{\mu^2}{p^2}
 \right)^\varepsilon
p^\rho\frac{\Gamma(\varepsilon)\Gamma(1-\varepsilon)\Gamma(1-\varepsilon)}{\Gamma(
2-2\varepsilon)}(1-\frac{2}{3}\varepsilon) \nonumber \\  &=& -C_A\frac{3}{8}f^{abc}
 \frac{g^3}{16\pi^2}p^\rho
 \left[\frac{1}{\varepsilon}-\gamma_E+\log 4\pi+\log\frac{-\mu^2}{p^2}+\frac{4}{3}
 \right] .  \label{v2r}
\end{eqnarray}
Adding up the two contributions together we find
\begin{equation} V^{Dim \
(abc)}_\rho(p)=-C_A\frac{1}{2}f^{abc}\frac{g^3}{16\pi^2}p^\rho
\left[\frac{1}{\varepsilon}-\gamma_E+\log 4\pi+\log\frac{-\mu^2}{p^2} +2\right].
   \label{vr}
\end{equation}
Having in mind that at the tree level the vertex has the form $V^{tree \
(abc)}_\rho(p)=-gf^{abc}p^\rho$ we get the vertex function in the one-loop approximation
as
\begin{equation}\label{vertgh}
V^{(abc)}_\rho(p)=-gf^{abc}p^\rho\left\{1+C_A\frac{1}{2}\frac{g^2}{16\pi^2}
\left[\frac{1}{\varepsilon}-\gamma_E+\log 4\pi+\log\frac{-\mu^2}{p^2} +2\right] \right\}.
\end{equation}
\newpage
\vspace*{1cm}
\section{Lecture IV: Renormalization. General Idea}
\setcounter{equation}{0}

Thus, we have convinced ourselves that the integrals for the radiative corrections are
indeed ultraviolet divergent in accordance with the naive power counting. The question
then is: how to get a sensible result for the cross-sections of the scattering processes,
decay widths, etc? To answer this question let us see what is the reason for divergences
at large values of momenta.  In coordinate space the large values of momenta correspond
to the small distances. Hence, the ultraviolet divergences allow for the singularities at
small distances. Indeed, the simplest divergent loop diagram (Fig.\ref{exm}) in
coordinate space is the product of two propagators. Each propagator is uniquely defined
in momentum as well as in coordinate space, but the square of the propagator has already
an ill-defined Fourier-transform, it is ultraviolet divergent. The reason is that the
square of the propagator is singular as $x^2\to 0$ and behaves like $1/(x^2)^2$. In fact,
the causal Green function (the propagator) is the so-called distribution which is defined
on smooth functions. It has the $\delta$-function like singularities and needs an
additional definition for the product of several such functions at a single point. The
discussed diagram is precisely this product.

The general approach to the elimination of the ultraviolet divergences known as the
${\cal R}$-operation was developed in the 1950s. It consists in the introduction to the
initial Lagrangian of additional local (or quasi-local) terms, called the counter-terms,
which serve the task of the definition of the product of distributions at the coinciding
points. The counter-terms lead to additional diagrams which cancel the ultraviolet
divergences. The  peculiarity of this procedure, being the subject of the
Bogoliubov-Parasiuk theorem, is in  that  the singularities are local in coordinate
space, i.e., are the functions of a single point and can contain only a finite number of
derivatives. In the theories belonging to the renormalizable class, where the number of
divergent structures is finite, the number of types of the counter-terms is also finite,
they repeat the terms of the original Lagrangian. This means that the introduction of the
counter-terms in this case is equivalent to the modification of the coefficients of
various terms., i.e. to the modification of the normalization of these terms. That is why
this procedure was called the {\it renormalization} procedure.

It should be stressed that the parameters of the original Lagrangian like the masses, the
coupling constants and the fields themselves are not, strictly speaking, observable. They
can be infinite. It is important that the renormalized parameters which enter the final
answers are meaningful.

Below we show by several examples of renormalizable theories how one introduces the
counter-terms into the Lagrangian, how they lead to the renormalization of the original
parameters and how the renormalization procedure allows one to get  finite results for
the Green functions.

\subsection{The scalar theory. The one-loop approximation}

We start with the one-loop approximation and consider for simplicity the scalar theory
(\ref{scalar}). It belongs to the renormalizable type and has a finite number of
ultraviolet divergent structures. The one-loop divergent diagrams in this theory were
calculated in the third lecture. Here we are interested in the singular parts, i.e., the
poles in $\varepsilon$. They are given by eqs. (\ref{resp}) and (\ref{resv}.
$$\begin{array}{ll}
\mbox{The propagator}: & Sing\ J_1(p^2)= -im^2(\frac{\lambda}{16\pi^2})(-\frac{1}{2\varepsilon}),\\
&\\ \mbox{The vertex}: & Sing\ \Gamma_4(s,t,u)= -i\lambda (\frac{\lambda}{16\pi^2})
(-\frac{3}{2\varepsilon}). \end{array}$$ Note that the singular parts do not depend on
momenta, i.e. their Fourier-transform has the form of the $\delta$-function in coordinate
space.

\begin{figure}[htb]\hspace*{4cm}
\leavevmode
    \epsfxsize=4cm
 \epsffile{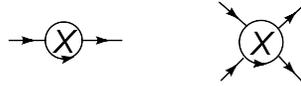}
 \caption{The one-loop counter-terms in the scalar theory
   \label{count}}
  \end{figure}

In order to remove the obtained singularities we add to the Lagrangian (\ref{scalar})
extra terms, the counter-terms equal to the singular parts with the opposite sign (the
factor $i$ belongs to the S-matrix and does not enter into the Lagrangian), namely,
\begin{equation}\label{ct}
  \Delta {\cal L} = \frac{1}{2\varepsilon} \frac{\lambda}{16\pi^2}\
  (-\frac{m^2}{2}\phi^2)
+ \frac{\lambda}{16\pi^2} \frac{3}{2\varepsilon}\ (-\frac{\lambda}{4!}\phi^4).
\end{equation}
These counter-terms correspond to additional vertices shown in Fig.\ref{count},
where the cross denotes the contribution corresponding to (\ref{ct}). With account taken
of the new diagrams the expressions for the propagator  (\ref{resp}) and the vertex
(\ref{resv}) become
\begin{equation}\label{resp2}
J_1(p^2)= \frac{i\lambda}{32\pi^2}m^2\left( 1-\gamma_E+\log(4\pi)-\log(m^2/\mu^2)\right).
\end{equation}
\begin{equation}\label{resv2}
\Delta\Gamma_4 = i\lambda\left\{\frac{\lambda}{16\pi^2}\left( 3-\frac 32 \gamma_E+\frac
32\log(4\pi)+\frac{1}{2}\ln\frac{\mu^2 }{-s}+ \frac{1}{2}\ln\frac{\mu^2
}{-t}+\frac{1}{2}\ln\frac{\mu^2 }{-u}\right)\right\} .
\end{equation}
Notice that the obtained expressions have no infinities but contain the dependence on the
regularization parameter $\mu^2$ which was absent in the initial theory. The appearance
of this dependence on a dimensional parameter is inherent in any regularization and is
called the dimensional transmutation, i.e., an appearance of a new scale in a theory.

What we have done is equivalent to {\it subtraction} of divergences from the diagrams. In
doing this we have subtracted just the singular parts.  This way of subtraction is called
the {\it minimal subtraction scheme} or the $MS$-scheme. One can make the subtraction
differently, for instance,  subtract also the finite parts. It is useful to subtract the
Euler constant and $\log 4\pi$ which accompany the pole terms. This subtraction scheme is
called the {\it modified minimal subtraction scheme} or the $\overline{MS}$-scheme. It is
equivalent to the redefinition of the parameter $\mu^2$. Another popular scheme of
subtraction is the so-called $MOM$-scheme when the subtractions are made for  fixed
values of momenta. For example, in the case of the vertex function one can make the
subtraction at the point $s=t=u=l^2$. This subtraction is called the subtraction at a
symmetric point.

The difference between various subtraction schemes is in the finite parts; in the
one-loop approximation this is just the constant independent of  momentum, however, in
higher loops one already has momentum dependent terms. Therefore, the finite parts of the
Green functions depend on a {\it subtraction scheme}.  Note that this dependence in
general is not reduced to the redefinition of the parameter  $\mu$, since there are
usually a few divergent Green functions and all of them are independent.

Thus, in the three subtraction schemes discussed above we have three different values for
the vertex function
\begin{eqnarray*}
\Gamma_4^{MS} &=& -i\lambda \left\{1\!-\! \frac{\lambda}{16\pi^2}\left[ 3\!-\!\frac 32
\gamma_E\!+\!\frac 32\log 4\pi\!+\!\frac{1}{2}\ln\frac{\mu^2 }{-s}\!+\!
\frac{1}{2}\ln\frac{\mu^2 }{-t}\!+\!\frac{1}{2}\ln\frac{\mu^2 }{-u}\right]\right\} ,\\
\Gamma_4^{\overline{MS}} &=& -i\lambda \left\{1- \frac{\lambda}{16\pi^2}\left[
3+\frac{1}{2}\ln\frac{\mu^2 }{-s}+ \frac{1}{2}\ln\frac{\mu^2
}{-t}+\frac{1}{2}\ln\frac{\mu^2 }{-u}\right]\right\} , \\
\Gamma_4^{MOM} &=& -i\lambda \left\{1- \frac{\lambda}{16\pi^2}\left[
\frac{1}{2}\ln\frac{l^2 }{-s}+ \frac{1}{2}\ln\frac{l^2 }{-t}+\frac{1}{2}\ln\frac{l^2
}{-u}\right]\right\}.
\end{eqnarray*}

The counter-terms are also different. It is useful to write them in the following way
\begin{equation}\label{ct2}
  \Delta {\cal L} = -(Z-1)\frac{m^2}{2}\phi^2-
 (Z_4-1)\frac{\lambda}{4!}\phi^4,
\end{equation}
where for different subtraction schemes one has
\begin{eqnarray}
  Z^{MS}&=&1+\frac{1}{2\varepsilon} \frac{\lambda}{16\pi^2}, \nonumber \\
Z^{\overline{MS}}&=&1+[\frac{1}{2\varepsilon}+1-\gamma_E+\log(4\pi)]
\frac{\lambda}{16\pi^2},
\nonumber \\
Z_4^{MS}&=&1+\frac{3}{2\varepsilon} \frac{\lambda}{16\pi^2},\label{z}  \\
Z_4^{\overline{MS}}&=&1+[\frac{3}{2\varepsilon}-3\gamma_E+3\log(4\pi)]
\frac{\lambda}{16\pi^2},  \nonumber \\
Z_4^{MOM}&=&1+[\frac{3}{2\varepsilon}+3-3\gamma_E+3\log(4\pi)+\frac
32\ln\frac{\mu^2}{\l^2}]\frac{\lambda}{16\pi^2}.\nonumber
\end{eqnarray}

The Lagrangian (\ref{scalar}) together with the counter-terms (\ref{ct2}) can be written
as
\begin{equation}\label{barela}
  {\cal L} + \Delta{\cal L}=Z_2\frac{1}{2}(\partial _\mu \varphi )^2-Z\frac{m^2}{2}\varphi^2
  -Z_4\frac{\lambda
}{4!} \varphi ^4= {\cal L}_{Bare},
\end{equation}
where the renormalization constants $Z$ and $Z_4$ are given by (\ref{z}) and the
renormalization constant $Z_2$ in the one-loop approximation equals 1.

Writing the "bare" Lagrangian in the same form as the initial one but in terms of the
"bare" fields and  couplings
\begin{equation}\label{barel}
  {\cal L}_{Bare} =\frac{1}{2}(\partial _\mu \varphi_B)^2-\frac{m_B^2}{2}\varphi^2_B
  -\frac{\lambda_B
}{4!} \varphi^4_B
\end{equation}
and comparing it with (\ref{barela}), we get the connection between the "bare" and
renormalized quantities
\begin{equation}\label{ren}
  \varphi_B=\sqrt{Z_2}\varphi, \ \ m^2_B=ZZ_2^{-1}m^2, \ \ \lambda_B=Z_4Z_2^{-2}\lambda.
\end{equation}
Equations (\ref{barel}) and (\ref{ren}) imply that the one-loop radiative corrections
calculated from the  Lagrangian (\ref{barel}) with parameters chosen according to
(\ref{ren},\ref{z}) are finite.

\subsection{The scalar theory. The two-loop approximation}

Consider now the two-loop diagrams. For simplicity and in order to complete  all the
integrations  we restrict ourselves to the massless case. Since we are
going to calculate the diagrams off mass shell, no infrared divergences may appear.  \\

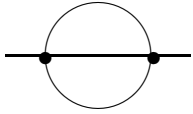
\begin{figure}[ht]\hspace*{6cm}%\vspace*{-1cm}
\begin{picture}(70,30)(0,12)
\put(0,25){\line(1,0){70}}\put(35,25){\circle{40}} \put(12,21){$\bullet$}
\put(53,21){$\bullet$}
\end{picture}
\caption{The two-loop propagator type diagram\label{prop2}}
\end{figure}

 \underline{The propagator:} In this order of PT there is only one diagram shown in
 Fig.\ref{prop2}.

The corresponding integral equals
$$J_2(p^2)=\frac{(-i\lambda )^2}{3!}\frac{i^3(\mu^2)^{2\varepsilon} }{(2\pi
)^{8-4\varepsilon}} \int \frac{d^{4-2\varepsilon }k d^{4-2\varepsilon
}q}{q^2(k-q)^2(p-k)^2},$$ (1/3! is a combinatorial coefficient). Let us use the method of
evaluation of the massless diagrams described above. One has to transform each of the
propagators into coordinate space, multiply them and transform back to momentum space.
This reduces to writing down the corresponding transformation factors. One gets
$$J_2(p^2)=\frac{i\lambda ^2}{6}\frac{(i\pi^2)^{2-\varepsilon} }{(2\pi
)^{8-4\varepsilon}}p^2\left(\frac{\mu^2}{-p^2}\right)^{2\varepsilon }
\frac{\Gamma(1-\varepsilon)\Gamma(1-\varepsilon)\Gamma(1-\varepsilon)
\Gamma(-1+2\varepsilon)}{\Gamma(1)\Gamma(1)\Gamma(1)\Gamma(3-3\varepsilon )}$$
$$=\frac{i}{6}\frac{\lambda ^2}{(16\pi ^2)^2}
\left[\frac{\mu^2}{-p^2}\right]^{2\varepsilon }\!\!\!\frac{p^2}{(2\!-\!3\varepsilon)
(1\!-\!3\varepsilon )(1\!-\!2\varepsilon )2\varepsilon }= \frac{i}{24}\frac{\lambda
^2}{(16\pi ^2)^2}p^2 \left[\frac{1}{\varepsilon }\!+\!\frac{13}{2}\!+\!2\ln
\frac{\mu^2}{-p^2} \right],$$ where the Euler constant and $\log 4\pi$ are omitted.

The appeared ultraviolet divergence, the pole in $\varepsilon$, can be removed via the
introduction of the (quasi)local counter-term
\begin{equation}\label{с2}
  \Delta{\cal L} = \frac 12(Z_2-1)(\partial \phi)^2,
\end{equation}
where the wave function renormalization constant  $Z_2$ in the $\overline{MS}$ scheme is
obtained by taking the singular part of the integral with the opposite sign
\begin{equation}
Z_2=1-\frac{1}{24\varepsilon }\left(\frac{\lambda}{16\pi ^2} \right)^2. \label{zp}
\end{equation}
After that the propagator in the massless case takes the form\vspace{0.3cm}

\begin{picture}(300,30)(-70,20)\put(4,22){$\bullet$}\put(18,22){$\bullet$}
\put(5,25){\line(1,0){15}}\put(32,25){\circle{24}}\put(44,25){\line(1,0){15}}
\put(42,22){$\bullet$} \put(57,22){$\bullet$} \put(88,22){$\bullet$}
\put(70,23){=}\put(90,25){\line(1,0){15}} \put(113,23){+} \put(105,22){$\bullet$}
\put(128,22){$\bullet$}\put(178,22){$\bullet$} \put(141,22){$\bullet$}
\put(165,22){$\bullet$}
\put(130,25){\line(1,0){50}}\put(155,25){\circle{24}}\put(190,23){=}
\put(208,22){$\bullet$} \put(223,22){$\bullet$}
\put(210,25){\line(1,0){15}}\put(230,23){$\Big\{ 1+$}
\put(260,25){\line(1,0){38}}\put(273,25){\circle{24}}\put(258,22){$\bullet$}
\put(310,23){$\Big\}$}\put(282,22){$\bullet$} \put(296,22){$\bullet$} \put(320,23){=}
\put(32,14){\line(0,1){22}} \put(28,15){\line(0,1){20}} \put(24,17){\line(0,1){16}}
\put(40,17){\line(0,1){16}} \put(36,15){\line(0,1){20}}
\end{picture}

\vspace{0.3cm}
\begin{equation}\label{pr2}
  =\frac{i}{p^2}\left\{1-\frac{1}{24}\frac{\lambda ^2}{(16\pi ^2)^2}
\left(\frac{13}{2}+2\ln \frac{\mu^2}{-p^2} \right) \right\}.
\end{equation}\vspace{0.3cm}

\begin{figure}[h]\hspace*{0.7cm}
\begin{picture}(100,40)(0,0)
\put(19,20){\line(-1,1){10}}\put(19,20){\line(-1,-1){10}}
\put(16,17){$\bullet$}\put(30,20){\circle{22}}
\put(54,20){\circle{22}}\put(39,17){$\bullet$}\put(80,18){+ crossed terms}
\put(66,20){\line(1,1){10}}\put(64,17){$\bullet$} \put(66,20){\line(1,-1){10}}
\end{picture}\hspace{4cm}
\begin{picture}(70,40)(-10,-5)
\put(20,20){\oval(30,17)}\put(20,20){\oval(30,40)[b]}
\put(5,20){\line(-1,1){10}}\put(36,20){\line(1,1){10}}
\put(20,0){\line(-1,-1){10}}\put(20,0){\line(1,-1){10}} \put(3,17){$\bullet$}
\put(33,17){$\bullet$} \put(17,-3){$\bullet$}\put(65,13){+ crossed terms}
\end{picture}
\caption{The two-loop vertex diagrams \label{vert2}}
\end{figure}
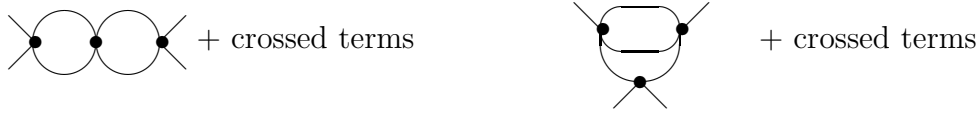

\underline{The vertex:} In the given order there are two diagrams (remind that in the
massless case the tad-poles equal to zero) shown in Fig.\ref{vert2}.

The first diagram by analogy with the one-loop case equals the sum of  $s,t$ and $u$
channels
$$I_{21}= I_{21}(s)+I_{21}(t)+I_{21}(u),$$
where each integral is nothing else but the square of the one-loop integral
\begin{equation}
 I_{21}(s)=\frac{(-i\lambda )^3}{96}\left(\frac{(\mu^2)^\varepsilon }{(2\pi )^{4-2\varepsilon}}i^2\int
\frac{d^{4-2\varepsilon }k}{k^2(p-k)^2}\right)^2 =  -\frac{i}{96}\frac{\lambda
^3}{(16\pi^2)^2} (\frac{1}{\varepsilon}+2+\ln \frac{\mu^2 }{-s})^2 .
\end{equation}
(1/96 is the combinatorial coefficient).

\begin{figure}[htb]\hspace*{4cm}
\leavevmode
    \epsfxsize=6cm
 \epsffile{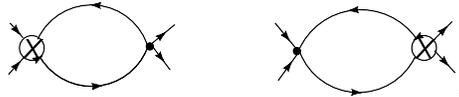}.
 \caption{The diagrams with the counter-terms in the two-loop approximation
   \label{count3}}
  \end{figure}

Opening the bracket we, for the first time here, come across the second order pole term
$1/\varepsilon^2$ and  the single pole $\log(-\mu^2/s)/\varepsilon$ accompanying it. This
latter pole is not harmless since its Fourier-transform is not a local function of
coordinates. This means that it can not be eliminated by a local counter-term. This would
be an unremovable problem if it were not the one-loop counter-terms (\ref{ct}) which
created the new vertices shown in Fig.\ref{count}. In the same order of $\lambda^3$ one
gets additional diagrams presented in Fig.\ref{count3}.

These diagrams lead to the subtraction of divergences in the subgraphs (left and right)
in the first diagram of Fig.\ref{vert2}. The subtraction of divergent subgraphs (the
${\cal R}$-operation without the last subtraction called the ${\cal R'}$-operation) looks
like

\begin{picture}(300,40)(0,18)\put(0,16){${\cal R}'$}
\put(29,20){\line(-1,1){10}}\put(29,20){\line(-1,-1){10}}
\put(40,20){\circle{22}}\put(64,20){\circle{22}}
\put(76,20){\line(1,1){10}}\put(76,20){\line(1,-1){10}}\put(100,16){=}
\put(139,20){\line(-1,1){10}}\put(139,20){\line(-1,-1){10}}
\put(150,20){\circle{22}}\put(174,20){\circle{22}}
\put(186,20){\line(1,1){10}}\put(186,20){\line(1,-1){10}}\put(208,16){-}
\put(277,20){\line(-1,1){10}}\put(277,20){\line(-1,-1){10}}
\put(246,20){\circle{22}}\put(288,20){\circle{22}}
\put(300,20){\line(1,1){10}}\put(300,20){\line(1,-1){10}}\put(315,16){-}
\put(228,0){\dashbox{4}(36,40)} \put(232,17){$\bullet$} \put(255,17){$\bullet$}
\put(339,20){\line(-1,1){10}}\put(339,20){\line(-1,-1){10}}
\put(350,20){\circle{22}}\put(394,20){\circle{22}}
\put(362,20){\line(1,1){10}}\put(362,20){\line(1,-1){10}} \put(376,0){\dashbox{4}(36,40)}
\put(379,17){$\bullet$} \put(403,17){$\bullet$}
\end{picture},
\vspace{1cm}

\noindent where the subgraph surrounded with the dashed line means its singular part, and
the rest of the graph is obtained by shrinking down the singular subgraph to a point. The
result has the form
$${\cal R}'I_{21}(s)= -\frac{i}{4}\frac{\lambda ^3}{(16\pi^2)^2}\left\{
(\frac{1}{\varepsilon}+2+\ln \frac{\mu^2 }{-s})^2 - \frac{2}{\varepsilon}
(\frac{1}{\varepsilon}+2+\ln \frac{\mu^2 }{-s}) \right\}=$$
$$=-\frac{i}{4}\frac{\lambda ^3}{(16\pi^2)^2}\left( -\frac{1}{\varepsilon^2 }
+4+\ln^2 \frac{\mu^2 }{-s}+4 \ln \frac{\mu^2 }{-s} \right).$$ Notice that after the
subtractions of subgraphs the singular part is local, i.e. in momentum space does not
contain $\ln p^2$. The terms with the single pole  $1/\varepsilon $ are absent since the
diagram can be factorized  into  two diagrams of the lower order.

The contribution of a given diagram to the vertex function equals
\begin{eqnarray}\label{ver21}
  \Delta \Gamma_4 &=& -i\lambda \left\{ \frac{1}{4}\frac{\lambda ^2}{(16\pi^2)^2}\left(
-\frac{3}{\varepsilon^2 } +12 \right. \right.\\
&&\left. \left. +\ln^2 \frac{\mu^2 }{-s}+4 \ln \frac{\mu^2 }{-s}+ \ln^2 \frac{\mu^2
}{-t}+4 \ln \frac{\mu^2 }{-t}+ \ln^2 \frac{\mu^2 }{-u}+4 \ln \frac{\mu^2 }{-u}
\right)\right\} \nonumber
\end{eqnarray}
The contribution to the renormalization constant of the four-point vertex in the
$\overline{MS}$ scheme is equal to the singular part with the opposite sign
\begin{equation}
\Delta Z_4=+\frac{3}{4\varepsilon^2 }\left(\frac{\lambda }{16\pi^2} \right)^2.
\label{z4}
\end{equation}

The second diagram with the crossed terms contains 6 different cases. Consider one of
them. Since we are interested  here in the singular parts contributing to the
renormalization constants, we perform some simplification of the original integral. We
use a very important property of the minimal subtraction scheme that the renormalization
constants depend only on dimensionless coupling constants and do not depend on the masses
and the choice of external momenta.  Therefore, we put all the masses equal to zero, and
to avoid artificial infrared divergences, we also put equal to zero one of the external
momenta. Then the diagram becomes  the propagator type one:
\begin{center}
\begin{picture}(70,40)(20,5) \put(-5,17){$p\ \to$} \put(82,17){$\to \ p$}
\put(50,20){\oval(30,25)}\put(65,32){\oval(27,27)[bl]} \put(35,20){\line(-1,0){10}}
\put(50,37){$0$}\put(66,19){\line(1,0){10}}\put(63,16){$\bullet$}\put(33,16){$\bullet$}
\put(49,29){$\bullet$}
\end{picture}
\end{center}
 The corresponding integral is:
$$ I_{22}(p^2)=\frac{(-i\lambda )^3}{48}\frac{(\mu^2)^{2\varepsilon }}{(2\pi
)^{8-4\varepsilon}}i^4\int \frac{d^{4-2\varepsilon }qd^{4-2\varepsilon }k}{q^2
(k-q)^2k^2(p-k)^2},$$ (1/48 is the combinatorial coefficient). Since putting one of the
momenta equal to zero we reduced the diagram to the propagator type, we can again use the
advocated method to calculate the massless integral. One has
$$I_{22}(p^2)=\frac{i\lambda ^3}{48}\frac{(\mu^2)^{2\varepsilon }}{(2\pi)^{8-4\varepsilon}}
i\pi^2 \frac{\Gamma(1-\varepsilon)\Gamma(1-\varepsilon)\Gamma(\varepsilon)
}{\Gamma(1)\Gamma(1)\Gamma(2-2\varepsilon)} \int\frac{d^{4-2\varepsilon
}k}{(k^2)^{1+\varepsilon }(p-k)^2}$$
$$=-\frac{i}{48}\frac{\lambda ^3}{(16\pi ^2)^2}
\left(\frac{\mu^2}{-p^2}\right)^{2\varepsilon }
\frac{\Gamma(1-\varepsilon)\Gamma(1-\varepsilon)\Gamma(\varepsilon)
\Gamma(1-2\varepsilon)\Gamma(1-\varepsilon)\Gamma(2\varepsilon)
}{\Gamma(1)\Gamma(1)\Gamma(2-2\varepsilon)\Gamma(1+\varepsilon)
\Gamma(1)\Gamma(2-3\varepsilon)}$$
$$=-\frac{i}{48}\frac{\lambda ^3}{(16\pi ^2)^2}
\left(\frac{\mu^2}{-p^2}\right)^{2\varepsilon } \frac{1}{2\varepsilon ^2(1-2\varepsilon
)(1-3\varepsilon )}$$
$$ =-\frac{i}{48}\frac{\lambda ^3}{(16\pi ^2)^2}
\left\{\frac{1}{2\varepsilon ^2}+\frac{5}{2\varepsilon }+2+ \frac{\ln (-\mu
^2/p^2)}{\varepsilon }+\ln^2\frac{\mu ^2}{-p^2}+ 5\ln\frac{\mu ^2}{-p^2}\right\}.$$

As one can see, in this case we again have the second order pole in $\varepsilon$ and,
accordingly, the single pole with the logarithm of momentum. The reason of their
appearance is the presence  of the divergent subgraph. Here we again have to look at the
counter-terms of the previous order which eliminate the divergence from the one-loop
subgraph. The subtraction of divergent subgraphs  (the ${\cal R}$-operation without the
last subtraction) looks like

\begin{picture}(200,40)(0,15)
\put(0,17){${\cal R}'$} \put(50,20){\oval(30,25)}\put(65,32){\oval(27,27)[bl]}
\put(35,20){\line(-1,0){10}} \put(66,19){\line(1,0){10}} \put(85,16){=}
\put(130,20){\oval(30,25)}\put(145,32){\oval(27,27)[bl]} \put(115,20){\line(-1,0){10}}
\put(146,19){\line(1,0){10}} \put(165,16){-} \put(200,20){\oval(20,20)}
\put(190,20){\line(-1,0){8}} \put(210,19){\line(1,0){8}} \put(244,26){$\bullet$}
\put(244,8){$\bullet$} \put(247,20){\oval(15,17)} \put(228,0){\dashbox{4}(36,40)}
\end{picture}
\vspace{0.3cm}

\noindent or
$${\cal R}'I_2(s)
=-\frac{i}{2}\frac{\lambda ^3}{(16\pi ^2)^2}\left\{
\left(\frac{\mu^2}{-p^2}\right)^{2\varepsilon } \frac{1}{2\varepsilon ^2(1-2\varepsilon
)(1-3\varepsilon )}- \left(\frac{\mu^2}{-p^2}\right)^{\varepsilon } \frac{1}{\varepsilon
^2(1-2\varepsilon )}\right\}$$
$$=
-\frac{i}{2}\frac{\lambda ^3}{(16\pi^2)^2}\left\{\left( \frac{1}{2\varepsilon
^2}+\frac{5}{2\varepsilon }+2+ \frac{\ln(-\mu^2/p^2)}{\varepsilon }+\ln^2\frac{\mu
^2}{-p^2}+ 5\ln\frac{\mu ^2}{-p^2}\right) \right.$$
$$\left. - \left(
\frac{1}{\varepsilon ^2}+\frac{2}{\varepsilon }+4+ \frac{\ln (-\mu ^2/p^2)}{\varepsilon
}+\frac{1}{2}\ln^2\frac{\mu ^2}{-p^2}+ 2\ln\frac{\mu ^2}{-p^2}\right)\right\}=$$
$$=
-\frac{i}{2}\frac{\lambda ^3}{(16\pi^2)^2}\left\{ -\frac{1}{2\varepsilon
^2}+\frac{1}{2\varepsilon }-2+ \frac{1}{2}\ln^2\frac{\mu ^2}{-p^2}+3\ln\frac{\mu
^2}{-p^2}\right\}.$$
 Once again, after the subtraction of the divergent subgraph the singular
part  is local, i.e. in momentum space does not depend on $\ln p^2$.

The contribution to the vertex function from this diagram is:
\begin{equation}\label{vert22}
  \Delta \Gamma_4 = -i\lambda \left\{
\frac{1}{2}\frac{\lambda ^2}{(16\pi^2)^2}\left( -\frac{3}{\varepsilon
^2}+\frac{3}{\varepsilon }-12+ \frac{1}{2}\ln^2\frac{\mu ^2}{-p^2}+3\ln\frac{\mu
^2}{-p^2} + \dots \right) \right\}
\end{equation}
and, accordingly,
\begin{equation}
\Delta Z_4=(\frac{3}{2\varepsilon^2 }-\frac{3}{2\varepsilon })\left(\frac{\lambda
}{16\pi^2}\right)^2. \label{zzz}
\end{equation}

Thus, due to  (\ref{z}) and (\ref{zzz}) in the two-loop approximation the quartic vertex
renormalization constant in the $\overline{MS}$ scheme looks like:
\begin{equation}\label{4}
Z_4=1+\frac{3}{2\varepsilon }\frac{\lambda }{16\pi^2} +\left(\frac{\lambda
}{16\pi^2}\right)^2(\frac{9}{4\varepsilon ^2}- \frac{3}{2\varepsilon }).
\end{equation}
With taking account of the two-loop renormalization of the propagator (\ref{zp}) one has:
\begin{equation}\label{zl}
Z_\lambda=Z_4Z_2^{-2}=  1+\frac{3}{2\varepsilon }\frac{\lambda }{16\pi^2}
+\left(\frac{\lambda }{16\pi^2}\right)^2(\frac{9}{4\varepsilon ^2}-
\frac{17}{12\varepsilon }).
\end{equation}

The statement is that the counter-terms introduced this way eliminate all the ultraviolet
divergences up to two-loop order and make the Green functions and hence the radiative
corrections finite. In the case of nonzero mass, one should also add the mass
counter-term.

\subsection{The general structure of the  {\cal R}-operation}

We are ready to formulate now the general procedure of getting finite expressions for the
Green functions off mass shell in an arbitrary local quantum field theory. It consists
of:

{\it In any order of perturbation theory in the coupling constant one introduces to the
Lagrangian the (quasi) local counter-terms. They perform the subtraction of divergences
in the diagrams of a given order. The subtraction of divergences in the subgraphs is
provided by the counter-terms of the lower order. After the subtraction of divergences in
the subgraphs the rest of the divergences are always local. The Green functions  of the
given order calculated on the basis of the initial Lagrangian with account of the
counter-terms are ultraviolet finite.}

The structure of the counter-terms as functions of the field operators depends on the
type of a theory. According to the classification discussed in the first lecture, the
theories are divided into three classes: {\it superrenormalizable} (a finite number of
divergent diagrams), {\it renormalizable} (a finite number of types of divergent
diagrams) and {\it non-renormalizable} (a infinite number of types of divergent
diagrams). Accordingly, in the first case one has a finite number of counter-terms; in
the second case, a infinite number of  counter-terms but they repeat the structure of the
initial Lagrangian, and in the last case,  one has an infinite number of structures with
an increasing number of the fields and derivatives.

In the case of renormalizable and superrenormalizable theories, since the counter-terms
repeat the structure of the initial Lagrangian, the result of the introduction of
counter-terms can be represented as
\begin{equation}\label{lagr}
  {\cal L}+\Delta {\cal L} = {\cal L}_{Bare}={\cal L}(\phi_B,\{g_B\},\{m_B\}),
\end{equation}
i.e., ${\cal L}_{Bare}$ is the same Lagrangian ${\cal L}$ but with the fields, masses and
coupling constants being the "bare" ones related to the renormalized quantities by the
multiplicative equalities
\begin{equation}\label{con}
  \phi_i^{Bare}=Z_i^{1/2}(\{g\},1/\varepsilon)\phi,  \ \
  g_i^{Bare}=Z_g^i(\{g\},1/\varepsilon)g_i, \ \ m_i^{Bare}=Z_m^i(\{g\},1/\varepsilon)m_i,
\end{equation}
where the renormalization constants $Z_i$ depend on the renormalized parameters and the
parameter of regularization (for definiteness we have chosen  $1/\varepsilon$). In some
cases the renormalization can be nondiagonal and the renormalization constants become
matrices.

The renormalization constants are not unique and depend on the renormalization scheme.
This arbitrariness, however, does not influence the observables expressed through the
renormalized quantities. We will come  back to this  problem later when discussing the
group of renormalization. In the gauge theories   $Z_i$ may depend on the choice of the
gauge though in the minimal subtraction scheme the renormalizations of the masses and the
couplings are gauge invariant.

In the minimal schemes the renormalization constants do not depend on dimensional
parameters like masses and do not depend on the arrangement of external momenta in the
diagrams. This property allows one to simplify the calculation of the counter-terms
putting the masses and some external momenta to zero, as it was  exemplified above by
calculation of the two-loop diagrams.  In making this trick, however, one has to be
careful not to create artificial infrared divergences. Since in dimensional
regularization they also have the form of poles in $\varepsilon$, this may lead to the
wrong answers.

In renormalizable theory the finite Green function is obtained from the "bare" one, i.e.,
is calculated from the "bare" Lagrangian by multiplication on the corresponding
renormalization constant
\begin{equation}\label{green}
  \Gamma(\{p^2\},\mu ^2,g_\mu )=Z_\Gamma (1/\varepsilon ,g_\mu )
\Gamma_{Bare}(\{p^2\},1/\varepsilon ,g_{Bare}),
\end{equation}
where in the n-th order of perturbation theory the "bare" parameters in the r.h.s. have
to be expressed in terms of the renormalized ones with the help of relations (\ref{con})
taken in the (n-1)-th order. The remaining constant $Z_\Gamma$ creates the counter-term
of the n-th order of the form $\Delta {\cal L}=(Z_\Gamma-1)O_\Gamma$, where the operator
$O_\Gamma$ reflects the corresponding Green function. If the Green function is finite by
itself (for instance, has many legs), then one has to remove the divergences only in the
subgraphs and the corresponding renormalization constant $Z_\Gamma=1$.

Note that since the propagator is inverse to the operator quadratic in fields in the
Lagrangian, the renormalization of the propagator is also inverse to the renormalization
of the 1-particle irreducible two-point Green function
\begin{equation}\label{propren}
  D(p^2,\mu ^2,g_\mu )=Z_2^{-1}(1/\varepsilon ,g_\mu )
D_{Bare}(p^2,1/\varepsilon ,g_{Bare}).
\end{equation}
The propagator renormalization constant is also the renormalization constant of the
corresponding field, but the fields themselves, contrary to the masses and couplings, do
not enter into the expressions for observables.

We would like to stress once more that the ${\cal R}$-operation works independently on
the fact renormalizable or non-renormalizable the theory is. In local theory the
counter-terms are local anyway. But only in renormalizable theory the counter-terms are
reduced to the multiplicative renormalization of the finite number of fields and
parameters.

One can perform the ${\cal R}$-operation for each diagram separately. For this purpose
one has first of all to subtract the divergences in the subgraphs and then subtract the
divergence in the diagram itself which has to be local. This serves as a good test that
the divergences in the subgraphs are subtracted correctly. In this case the ${\cal
R}$-operation can be symbolically written in a factorized form
\begin{equation}\label{fact}
  {\cal R} G = \prod_{div.subgraphs}(1-M_\gamma)G,
\end{equation}
where $G$ is the initial diagram, $M$ is the subtraction operator (for instance,
subtraction of the singular part of the regularized diagram) and the product goes over
all divergent subgraphs including the diagram itself. By a subgraph we mean here the
1-particle irreducible diagram consisting of the vertices and lines of the diagram which
is UV divergent. The 1-particle irreducible is called the diagram which can not be made
disconnected by deleting of one line.

\begin{figure}[htb]
\begin{center} \leavevmode
    \epsfxsize=9cm
 \epsffile{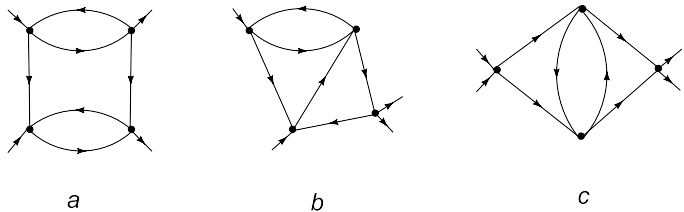}
 \caption{The multiloop diagrams in the $\phi^4$ theory
   \label{rop}}
   \end{center}
  \end{figure}

We have demonstrated above the application of the ${\cal R}$-operation to the two--loop
diagrams in a scalar theory. Consider some other examples of diagrams with larger number
of loops shown in Fig.\ref{rop}. They appear in the $\phi^4$ theory in the three-loop
approximation.\vspace{0.5cm}

\begin{figure}[htb]\hspace*{0.5cm}
\leavevmode
    \epsfxsize=15cm
 \epsffile{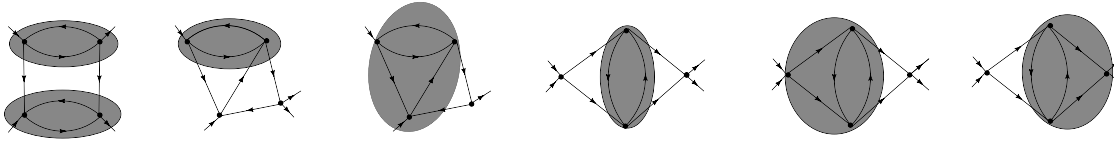}
 \caption{The divergent subgraphs in the diagrams of Fig.\ref{rop}
   \label{sub}}
  \end{figure}

In order to perform the ${\cal R}$-operation for these diagrams one first has to find out
the divergent subgraphs. They are shown in Fig.\ref{sub}.

Let us use the factorized representation of the ${\cal R}$-operation in the form of
(\ref{fact}). For the three chosen diagrams one has, respectively,
\begin{eqnarray*}
  R G_a&=&(1-M_G)(1-M_{\gamma_1})(1-M_{\gamma'_1})G_a, \\
  R G_b&=&(1-M_G)(1-M_{\gamma_2})(1-M_{\gamma_1})G_б, \\
  R G_c&=&(1-M_G)(1-M_{\gamma_2})(1-M_{\gamma'_2})(1-M_{\gamma_1})G_в,
\end{eqnarray*}
where $\gamma_1$ and $\gamma_2$ are the one- and two-loop divergent subgraphs shown in
Fig.\ref{sub}.

The result of the application of the  ${\cal R}$-operation without the last subtraction (
${\cal R'}$-operation) for the diagrams of interest graphically is as
follows:\vspace{0.3cm}
\begin{figure}[htb]
\leavevmode
    \epsfxsize=16cm
 \epsffile{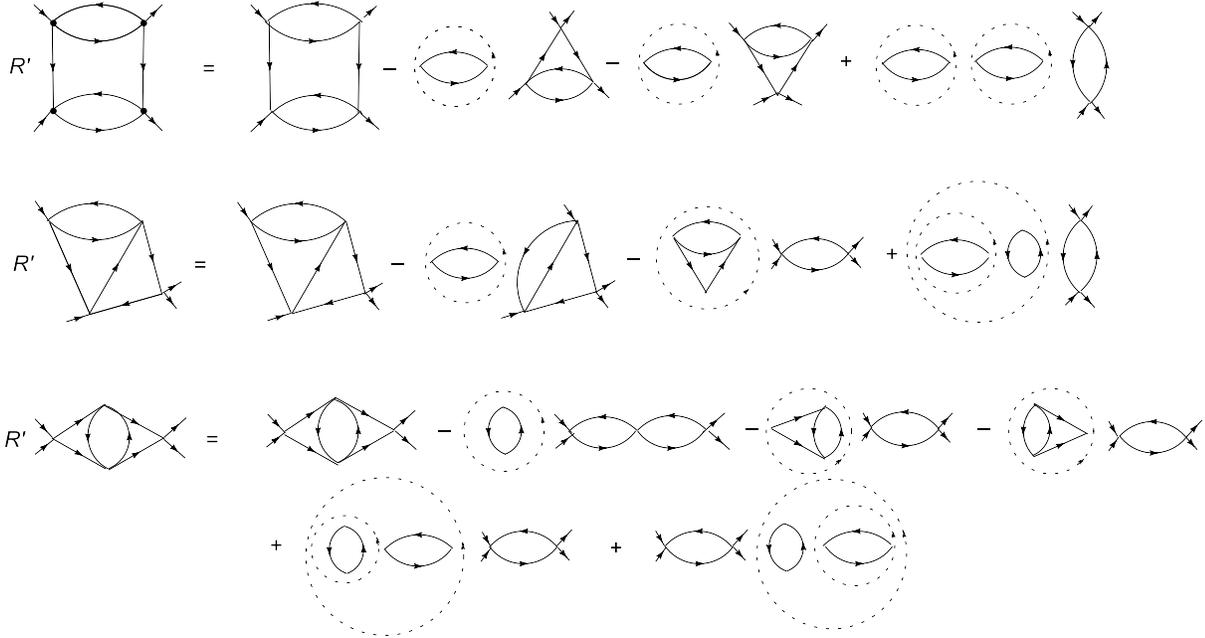}\vspace{0.3cm}
 \caption{The ${\cal R'}$-operation for the multiloop diagrams
   \label{rpr}}
  \end{figure}

Here, as before, the graph surrounded with the dashed circle means its singular part and
the remaining graph is obtained by shrinking the singular subgraph to a point.

Let us demonstrate how the  ${\cal R'}$-operation works for the diagram Fig.\ref{rop}a).
Since the result of the  ${\cal R'}$-operation does not depend on external momenta, we
put two momenta on the diagonal to be equal to zero so that the integral takes the
propagator form. Then we can use the method based on Fourier-transform, as it was
explained above. One has\vspace{0.3cm}
\newpage

 \epsfxsize=12cm
 \epsffile{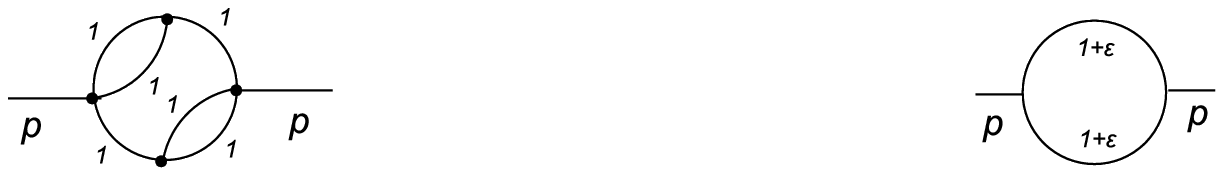}\vspace{-1.4cm}

 \hspace*{4.0cm}$=\left(\Gamma(1-\varepsilon)\frac{\Gamma^2(1-\varepsilon)
 \Gamma(\varepsilon)}{\Gamma(2-2\varepsilon)}\right)^2$\vspace{1cm}

 \hspace*{0.5cm}$=\left(\Gamma(1\!-\!\varepsilon)\frac{\Gamma^2(1-\varepsilon)
 \Gamma(\varepsilon)}{\Gamma(2-2\varepsilon)}\right)^2
 \left(\Gamma(1\!-\!\varepsilon)\frac{\Gamma^2(1-2\varepsilon)
 \Gamma(3\varepsilon)}{\Gamma^2(1+\varepsilon)\Gamma(2-4\varepsilon)}\right)
 (\frac{\mu^2}{p^2})^{3\varepsilon} \cong\frac{1}{\varepsilon^3(1-2\varepsilon)^2(1-4\varepsilon)}
 (\frac{\mu^2}{p^2})^{3\varepsilon}.$\vspace{0.5cm}

We use here the angular integration measure in the $4-2\varepsilon$ dimensional space
accepted above, which results in the multiplication of the standard expression by
$\Gamma(1-\varepsilon)$ in order to avoid the unwanted transcendental functions.
Following the scheme shown in Fig.\ref{rpr} we get\vspace{0.3cm}

 \epsfxsize=13cm
 \epsffile{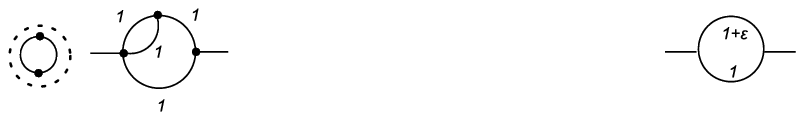}\vspace{-1.5cm}

 \hspace*{5.0cm}$=\frac{1}{\varepsilon}\Gamma(1-\varepsilon)\frac{\Gamma^2(1-\varepsilon)
 \Gamma(\varepsilon)}{\Gamma(2-2\varepsilon)}$\vspace{1cm}

 \hspace*{1.0cm}$=\frac{1}{\varepsilon}\Gamma(1-\varepsilon)\frac{\Gamma^2(1-\varepsilon)
 \Gamma(\varepsilon)}{\Gamma(2-2\varepsilon)}\Gamma(1-\varepsilon)
 \frac{\Gamma(1-\varepsilon)
 \Gamma(1-2\varepsilon)\Gamma(2\varepsilon)}{\Gamma(1+\varepsilon)
 \Gamma(2-3\varepsilon)}(\frac{\mu^2}{p^2})^{2\varepsilon}
\cong\frac{1}{\varepsilon^3(1-2\varepsilon)(1-3\varepsilon)}
 (\frac{\mu^2}{p^2})^{2\varepsilon}.
 $
 \vspace{1cm}

\epsfxsize=5cm
 \epsffile{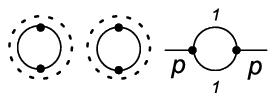}\vspace{-1.4cm}

\hspace*{5.5cm}$=\frac{1}{\varepsilon^2}\Gamma(1-\varepsilon)\frac{\Gamma^2(1-\varepsilon)
 \Gamma(\varepsilon)}{\Gamma(2-2\varepsilon)}(\frac{\mu^2}{p^2})^{\varepsilon}
 \cong\frac{1}{\varepsilon^3(1-2\varepsilon)}
 (\frac{\mu^2}{p^2})^{\varepsilon}.$\vspace{1cm}

Combining all together one finds\vspace{0.9cm}

\epsfxsize=3.5cm
 \epsffile{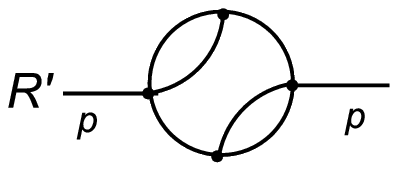}\vspace{-1.1cm}

 \hspace*{3.7cm}$\cong\frac{1}{\varepsilon^3(1-2\varepsilon)^2(1-4\varepsilon)}
 (\frac{\mu^2}{p^2})^{3\varepsilon}-2\frac{1}{\varepsilon^3(1-2\varepsilon)}
 (\frac{\mu^2}{p^2})^{\varepsilon}+\frac{1}{\varepsilon^3(1-2\varepsilon)}
 (\frac{\mu^2}{p^2})^{\varepsilon}$

 $$=\frac{1-\varepsilon-\varepsilon^2}{\varepsilon^3}.$$
\noindent Note the cancellation of all nonlocal contributions. The singular part after
the $R'$-operation is always local.

The realization of the ${\cal R'}$-operation for each diagram $G$ allows one to find the
contribution of a given diagram to the corresponding counter-term and, in the case of a
renormalizable theory, to find the renormalization constant equal to
\begin{equation}\label{dz}
  Z = 1-{\cal K\ R'}G,
\end{equation}
where ${\cal K}$ means the extraction of the singular part. Adding the contribution of
various diagrams we get the resulting counter-term of a given order and, accordingly, the
renormalization constant.
\newpage
\vspace*{1cm}
\section{Lecture V: Renormalization. Gauge Theories and the Standard Model}
\setcounter{equation}{0}

Consider now the gauge theories. The difference from the scalar case is in the relations
between various renormalization constants which follow from the gauge invariance. If the
regularization and the renormalization scheme do not break the symmetry these relations
hold automatically. In the opposite case, this is an additional requirement imposed on
the counter-terms.

\subsection{Quantum electrodynamics}

Quantum electrodynamics (\ref{qed}) is a renormalizable theory; hence, the counter-terms
repeat the structure of the Lagrangian. They can be written as
\begin{equation}\label{qedc}
 \Delta {\cal L}_{QED}=-\frac{Z_3-1}{4} F_{\mu\nu}^2+(Z_2-1)i\bar \psi\hat{\partial}\psi
  -m(Z-1)\bar \psi\psi+e(Z_1-1)\bar \psi\hat A \psi.
\end{equation}
The term that fixes the gauge is not renormalized. In the leading order of perturbation
theory we calculated the corresponding diagrams with the help of dimensional
regularization (see (\ref{pol}),(\ref{selfrr}),(\ref{ver4})). Their singular parts with
the opposite sign give the proper renormalization constants. They are, respectively,
\begin{eqnarray}\label{const}
  Z_1&=&1-\frac{e^2}{16\pi^2}\frac{1}{\varepsilon}, \nonumber \\
  Z_2&=&1-\frac{e^2}{16\pi^2}\frac{1}{\varepsilon}, \nonumber\\
  Z_3&=&1-\frac{e^2}{16\pi^2}\frac{4}{3\varepsilon},\label{zqed}\\
  Z&=&1-\frac{e^2}{16\pi^2}\frac{4}{\varepsilon}. \nonumber
\end{eqnarray}
Adding (\ref{qedc}) with (\ref{qed}) we get
\begin{eqnarray}\label{qedb}
&&{\cal L}_{QED}+\Delta {\cal L}_{QED}=-\frac{Z_3}{4} F_{\mu\nu}^2+Z_2i\bar
\psi\hat{\partial}\psi -mZ\bar \psi\psi+eZ_1\bar \psi\hat A
\psi-\frac{1}{2\xi}(\partial_\mu A_\mu)^2 \nonumber \\
&&=-\frac{1}{4} F_{\mu\nu B}^2+i\bar \psi_B\hat{\partial}\psi_B -mZZ_2^{-1}\bar
\psi_B\psi_B+eZ_1Z_2^{-1}Z_3^{-1/2}\bar \psi_B\hat A_B \psi_B \nonumber\\ &&\ \ \
-\frac{Z_3^{-1}}{2\xi}(\partial_\mu A_{\mu B})^2,
\end{eqnarray}
that gives
\begin{equation}\label{qedbb}
  \psi_B=Z_2^{1/2}\psi, \ \ A_B=Z_3^{1/2}A, \ \ m_B=ZZ_2^{-1}m,\ \
  e_B=Z_1Z_2^{-1}Z_3^{-1/2}e, \ \ \xi_B=Z_3\xi.
\end{equation}
The gauge invariance here manifests itself in two places. First, the transversality of
the radiative correction to the photon propagator means that the gauge fixing term is not
renormalized and, hence, the gauge parameter $\xi$ is renormalized as a gauge field.
Second, the gauge invariance connects the vertex Green function and the fermion
propagator (the Ward identity), which leads to the identity $Z_1=Z_2$. Since the
dimensional regularization which we use throughout the calculations does not break the
gauge invariance, this identity is satisfied automatically  (see (\ref{const})). This
means that the renormalization of the coupling (\ref{qedbb}) is defined by the photon
propagator only. Note, however, that this is not true in general in a non-Abelian theory.

\subsection{Quantum chromodynamics}

The complications which appear in non-Abelian theories are caused by the presence of many
vertices with the same coupling as it follows from the gauge invariance. Hence, they have
to renormalize the same way, i.e there appear new identities, called the Slavnov-Taylor
identities. The full set of the counter-terms in QCD looks like
\begin{eqnarray}\label{qcdc}
 && \Delta{\cal L}_{QСD}=-\frac{Z_3-1}{4} (\partial_\mu A_\nu^a-\partial_\nu A_\mu^a)^2
  -g(Z_1-1)f^{abc}A_\mu^a A_\nu^b \partial_\mu A_\nu^c \nonumber \\
  &&-\!(Z_4\!-\!1)\frac{g^2}{4}f^{abc}f^{ade}
  A_\mu^b A_\nu^c A_\mu^d A_\nu^e  +
    (\tilde{Z}_3\!-\!1)\partial_\mu\bar c^a\partial_\mu c^a+g(\tilde{Z}_1\!-\!1)
    f^{abc}\partial_\mu\bar c^a A_\mu^b c^c\nonumber \\
     &&+ i(Z_2-1)\bar \psi \hat \partial\psi
  -m(Z-1)\bar \psi \psi +g(Z_{1\psi}-1)
  \bar \psi \hat A^a T^a\psi,
\end{eqnarray}
that being added to the initial Lagrangian gives
\begin{eqnarray}\label{qcdcс}
&&\hspace*{-1cm}{\cal L}_{QСD}+ \Delta{\cal L}_{QСD}=-\frac{Z_3}{4} (\partial_\mu
A_\nu^a-\partial_\nu A_\mu^a)^2
  -gZ_1f^{abc}A_\mu^a A_\nu^b \partial_\mu A_\nu^c \nonumber \\
 && -Z_4\frac{g^2}{4}f^{abc}f^{ade}
  A_\mu^b A_\nu^c A_\mu^d A_\nu^e -\tilde{Z}_3\partial_\mu\bar c^a\partial c^a-g\tilde{Z}_1
    f^{abc}\partial_\mu\bar c^a A_\mu^b c^c \nonumber\\
  &&   +iZ_2\bar \psi \hat \partial\psi
  -mZ\bar \psi \psi +gZ_{1\psi}
  \bar \psi \hat A^a T^a\psi-\frac{1}{2\xi}(\partial_\mu A_\mu^a)^2\nonumber\\
 &&\hspace*{-0.5cm}=-\frac{1}{4} (\partial_\mu
A_{\nu B}^a-\partial_\nu A_{\mu B}^a)^2
  -gZ_1Z_3^{-3/2}f^{abc}A_{\mu B}^a A_{\nu B}^b \partial_\mu A_{\nu B}^c \nonumber \\
 && \hspace*{-0.5cm}-Z_4Z_3^{-2}\frac{g^2}{4}f^{abc}f^{ade}
  A_{\mu B}^b A_{\nu B}^c A_{\mu B}^d A_{\nu B}^e
    +\partial_\mu\bar c^a_B\partial_\mu
    c^a_B+g\tilde{Z}_1\tilde{Z}_3^{-1}Z_3^{-1/2}
    f^{abc}\partial_\mu\bar c^a_B A_{\mu B}^b c^c_B \nonumber \\
    &&\hspace*{-0.5cm} +\frac{Z_3^{-1}}{2\xi}(\partial_\mu A_{\mu B}^a)^2
 +i\bar \psi_B \hat \partial\psi_B
  -mZZ_2^{-1}\bar \psi_B \psi_B +gZ_{1\psi}Z_2^{-1}Z_3^{-1/2}
  \bar \psi_B \hat A_{B}^a T^a\psi_B.
\end{eqnarray}
This results in  the relations between the renormalized and the "bare" fields and
couplings
\begin{eqnarray}\label{qcdbb}
 && \psi_B=Z_2^{1/2}\psi, \ \ A_B=Z_3^{1/2}A, \ \ c_B=\tilde{Z}_3^{1/2}c, \nonumber\\
 && m_B=ZZ_2^{-1}m,\ \ g_B=Z_1Z_3^{-3/2}g, \ \ \xi_B=Z_3\xi, \\
 && Z_1Z_3^{-1}=\tilde{Z}_1\tilde{Z}_3^{-1}, \
 \ Z_4=Z_1^2Z_3^{-1}, \ \ Z_{1\psi}Z_2^{-1}=Z_1Z_3^{-1}. \nonumber
\end{eqnarray}
The last line of equalities follows from the requirement of identical renormalization of
the coupling in various vertices and represents the Slavnov-Taylor identities for the
singular parts.

The explicit form of the renormalization constants in the lowest approximation follows
from the one-loop diagrams  calculated earlier (see (\ref{vacuum}), (\ref{selfrr}),
(\ref{ver4}), (\ref{res2}), (\ref{ghpr}), (\ref{vertgh}). Aa usual, one has to take the
singular part with the opposite sign.  One has in the $\overline{MS}$ scheme
\begin{eqnarray}\label{constqcd}
  Z_2&=&1-\frac{g^2}{16\pi^2}\frac{C_F}{\varepsilon}, \nonumber\\
Z_3&=&1+\frac{g^2}{16\pi^2}(\frac{5}{3\varepsilon }C_A-\frac{4}{3\varepsilon
}T_fn_f) ,\nonumber\\
  Z&=&1-\frac{g^2}{16\pi^2}\frac{4C_F}{\varepsilon}, \nonumber\\
  \tilde{Z}_1&=& 1-\frac{g^2}{16\pi^2}\frac{C_A}{2\varepsilon }, \\
\tilde{Z}_2&=& 1+\frac{g^2}{16\pi^2}\frac{C_A}{2\varepsilon }, \nonumber\\
Z_g&=&\tilde{Z}_1\tilde{Z}_2^{-1}Z_3^{-1/2}=1-\frac{g^2}{16\pi^2}(\frac{11}{6\varepsilon
}C_A-\frac{4}{3\varepsilon }T_fn_f),\nonumber
\end{eqnarray}
where the following notation for the Casimir operators of the gauge group is used
$$f^{abc}f^{dbc}=C_A\delta^{ad}, \ \ (T^aT^a)_{ij}=C_F\delta_{ij}, \ \
Tr(T^aT^b)=T_F\delta^{ab}.$$ For the $SU(N)$ group and the fundamental representation of
the fermion fields they are equal to
$$ C_A=N, \ \ C_F=\frac{N^2-1}{2N}, \ \ T_F=\frac 12.$$

\subsection{The Standard Model of fundamental interactions}

In the Standard Model of fundamental interactions besides the gauge interactions and the
quartic interaction of the Higgs fields there are also Yukawa type interactions of the
fermion fields with the Higgs field. These interactions are also renormalizable and is
characterized by the Yukawa coupling constants, one for each fermion field. The
peculiarity of the SM is that the masses of the fields appear as a result of spontaneous
symmetry breaking when the Higgs field develops a vacuum expectation value.  As a result
the masses are not independent but are expressed via the coupling constant multiplied by
the vacuum expectation value. Here there are two possibilities: to treat the Yukawa
couplings as independent quantities and to renormalize them in a usual way and then
express the renormalized masses via the renormalized couplings or to start with the
masses of particles and to treat the Yukawa couplings as  secondary quantities. The first
approach is usually used within the minimal subtraction scheme where the renormalizations
do not depend on masses. On the contrary, in the МОМ scheme when the subtraction is
carried out on mass shell (the so-called "on-shell" scheme), one usually takes masses of
particles as the basis. Under this way of subtraction the pole of the propagator is not
shifted and the renormalized mass coincides with the mass of a physical particle. Below
we consider the renormalizations in the SM in the $\overline{MS}$ scheme and concentrate
on the renormalization of the fields and the couplings.

Another property of the Standard Model is that it has the gauge group  $SU_c(3)\times
SU_L(2)\times U_Y(1)$ which is spontaneously broken to $SU_c(3)\times U_{EM}(1)$. In the
theories with spontaneously broken symmetry, according to the Goldstone theorem there are
massless particles, the goldstone bosons. These particles indeed are present in the SM
but they are not the physical degrees of freedom and due to the Higgs effect are absorbed
by vector bosons turning into  longitudinal degrees of freedom of massive vector
particles.

Thus, there are two possibilities to formulate the SM as a theory with spontaneous
symmetry breaking: the {\it unitary} formulation in which nonphysical degrees of freedom
are absent and vector bosons have three degrees of freedom, and the so-called {\it
renormalizable} formulation in which goldstone bosons are present in the spectrum and
vector fields have two degrees of freedom. These two formulations correspond to two
different choices of the gauge in spontaneously broken theory.

In unitary gauge we have only physical degrees of freedom, i.e., the theory is
automatically unitary, hence the name of this gauge. However, the propagator of the
massive vector fields in this case has the form
$$G_{\mu\nu}(k)=-i\frac{g^{\mu\nu}-\frac{k^\mu k^\nu}{M^2}}{k^2-M^2},$$
i.e., does nor decrease when momentum goes to infinity. This leads to the increase in the
power of divergences and the theory happens to be formally nonrenormalizable despite
 the coupling constant being dimensionless.  We have mentioned this fact in the first
lecture.

 On the other hand, in renormalizable gauge, where the vector fields have two
degrees of freedom, the propagator behaves as
$$G_{\mu\nu}(k)=-i\frac{g^{\mu\nu}-\frac{k^\mu k^\nu}{k^2}}{k^2-M^2},$$
which obviously leads to a renormalizable theory which explains the name of this gauge.
However, the presence of the goldstone bosons calls into question the unitarity  of the
theory since transitions between the physical and unphysical states become possible.

Since all the gauges are equivalent, one can work in any of them but in the unitary gauge
one has to prove the renormalizability while in the renormalizable gauge one has to prove
unitarity. The gauge invariance of observables preserved in a spontaneously broken theory
should guarantee the fulfilment of both the requirements simultaneously. Note that in
spontaneous symmetry breaking the symmetry of the Lagrangian is preserved, it is the
boundary condition that breaks the symmetry.

The rigorous proof of that the theory is simultaneously renormalizable and unitary is not
so obvious and eventually was awarded the Nobel prize, but can be seen by using  some
intermediate gauge called the $R_\xi$-gauge. The gauge fixing term in this case is chosen
in the form
$$-\frac{1}{2\xi}(\partial_\mu A^a_\mu-\xi g F^a_i\chi_i)^2, \ \ \
gF^a_i=\frac v2\left(
\begin{array}{ccc} g&0&0\\0&g&0\\0&0&g\\0&0&g'\end{array}\right),$$
where $v$ is the vacuum expectation value of the Higs field, and $\chi_i$ are the
goldstone bosons. In this gauge the vector propagator has the form
$$G_{\mu\nu}(k)=-i\frac{g^{\mu\nu}-\frac{k^\mu k^\nu}{k^2-\xi M^2}(1-\xi)}{k^2-M^2},$$
and at $\xi=0$ corresponds to the renormalizable gauge while as $\xi\to\infty$ it
corresponds to the unitary one. Since all the observables do not depend on $\xi$, we can
choose $\xi=0$ when investigating the renormalizability properties and choose $\xi =
\infty$ in examining the unitarity. Since we are interested here in the renormalizability
of the SM, in what follows we will work in a renormalizable gauge.

The Lagrangian of the Standard Model consists of the following three parts:
 \begin{equation}
{\cal L} ={\cal L}_{gauge} + {\cal L}_{Yukawa} + {\cal L}_{Higgs}, \label{SM}
 \end{equation}
The gauge part is totally fixed by the requirement of the gauge invariance leaving only
the values of the couplings as  free parameters
 \begin{eqnarray}
{\cal L}_{gauge} & = & -\frac{1}{4} G_{\mu\nu}^aG_{\mu\nu}^a - \frac{1}{4}
W_{\mu\nu}^iW_{\mu\nu}^i -\frac{1}{4} B_{\mu\nu}B_{\mu\nu}\\ & &  +
 i\overline{L}_{\alpha}\gamma^{\mu}D_{\mu}L_{\alpha} +
i\overline{Q}_{\alpha}\gamma^{\mu}D_{\mu}Q_{\alpha} +
i\overline{E}_{\alpha} \gamma^{\mu}D_{\mu}E_{\alpha} \nonumber \\
 & &  + i\overline{U}_{\alpha}\gamma^{\mu}D_{\mu}U_{
\alpha} + i\overline{D}_{\alpha}\gamma^{\mu}D_{\mu}D_{\alpha} +
(D_{\mu}H)^{\dagger}(D_{\mu}H),  \nonumber
 \end{eqnarray}
where the following notation for the covariant derivatives is used
 \begin{eqnarray*}
G_{\mu\nu}^a & = &
\partial_{\mu}G_{\nu}^a-\partial_{\nu}G_{\mu}^a+g_{s}
f^{abc}G_{\mu}^bG_{\nu}^c,\\ W_{\mu\nu}^i & = &
\partial_{\mu}W_{\nu}^i-\partial_{\nu}W_{\mu}^i+g\epsilon
^{ijk}W_{\mu}^jW_{\nu}^k,\\ B_{\mu\nu} & = &
\partial_{\mu}B_{\nu}-\partial_{\nu}B_{\mu},\\ D_{\mu}L_{\alpha} &
= &
(\partial_{\mu}-i\frac{g}{2}\tau^iW_{\mu}^i+i\frac{g'}{2}B_{\mu})L_{\alpha},\\
D_{\mu}E_{\alpha} & = & (\partial_{\mu}+ig'B_{\mu})E_{\alpha},\\
D_{\mu}Q_{\alpha} & = & (\partial_{\mu}-i\frac{g}{2}\tau^iW_{\mu}^i-i\frac{g'}{6}B_{\mu}-
i\frac{g_s}{2}\lambda^aG_{\mu}^a)Q_{\alpha},\\ D_{\mu}U_{\alpha} & = &
(\partial_{\mu}-i\frac{2}{3}g'B_{\mu}-i\frac{g_s}{2}\lambda^aG_{\mu}^a) U_{\alpha},\\
D_{\mu}D_{\alpha} & = & (\partial_{\mu}+i\frac{1}{3}g'
B_{\mu}-i\frac{g_s}{2}\lambda^aG_{\mu}^a)D_{\alpha}.
 \end{eqnarray*}

The Yukawa part of the Lagrangian which is needed for the generation of the quark and
lepton masses is also chosen in the gauge invariant form and contains  arbitrary Yukawa
couplings (we ignore the neutrino masses, for simplicity)
 \begin{equation}
{\cal L}_{Yukawa} = y_{\alpha\beta}^L\overline{L}_{\alpha}E_{\beta}H + y_{\alpha
\beta}^D\overline{Q}_{\alpha}D_{\beta}H + y_{\alpha\beta}^U\overline{Q}_{\alpha}
U_{\beta}\tilde{H} + h.c., \label{yukawa}
 \end{equation}
where $\tilde{H}=i\tau_2H^{\dagger}$.

At last the Higgs part of the Lagrangian contains the Higgs potential which is chosen in
such a way that the Higgs field acquires the vacuum expectation value and the potential
itself is stable
 \begin{equation}
{\cal L}_{Higgs} = - V = m^2H^{\dagger}H - \frac{\lambda}{2}(H^{\dagger}H)^2.\label{hig}
 \end{equation}
Here there are two arbitrary parameters: $m^2$  и $\lambda$.  The ghost fields and the
gauge fixing terms are omitted.

The Lagrangian of the SM contains the following set of free parameters:
 \begin{itemize}
     \item 3 gauge couplings $g_s, g, g' $;
  \item 3 Yukawa matrices $y_{\alpha\beta}^L, y_{\alpha\beta}^D, y_{\alpha\beta}^U
$;
     \item Higgs coupling  constant $\lambda$;
     \item Higgs mass parameter $m^2$;
     \item the number of the matter fields (generations).
 \end{itemize}

All particles obtain their masses due to spontaneous breaking of the $SU_{left}(2)$
symmetry group via a nonzero vacuum expectation value (v.e.v.) of the Higgs field
 \begin{equation}
<H> = \left(\begin{array}{c}v\\ 0\end{array}\right),\ \ \ \ v=m/\sqrt{\lambda}.
\label{vac}
 \end{equation}
As a result,  the gauge group of the SM is spontaneously  broken down to $$SU_c(3)\otimes
SU_L(2)\otimes U_Y(1) \Rightarrow SU_c(3)\otimes U_{EM}(1).$$ The physical weak
intermediate bosons are  linear combinations of the gauge ones
 \begin{equation} W_{\mu}^{\pm} =  \frac{W_{\mu}^1\mp iW_{\mu}^2}{\sqrt{2}},\
\ \ \ Z_{\mu} = -\sin{\theta_W}B_{\mu} + \cos{\theta_W}W_{\mu}^3
 \end{equation}
with masses
 \begin{equation}
m_W=\frac{1}{\sqrt{2}}gv,\ \ \ \  m_Z=m_W/\cos{\theta_W},\ \ \ \ \tan{\theta_W}=g'/g,
\label{Z}
 \end{equation}
while the photon field
 \begin{equation}
\gamma_{\mu} = \cos{\theta_W}B_{\mu} + \sin{\theta_W}W_{\mu}^3
 \end{equation}
remains massless.

The matter fields acquire masses proportional to the corresponding Yukawa couplings:
 \begin{equation}
M_{\alpha\beta}^u = y_{\alpha\beta}^uv,\ M_{\alpha\beta}^d = y_{\alpha\beta}^dv,\
M_{\alpha\beta}^l = y_{\alpha\beta}^lv, \ m_H = \sqrt{2}m. \label{mass}
 \end{equation}
The mass matrices have to be diagonalized to get the quark and lepton masses.

The explicit mass terms in the Lagrangian are forbidden because they are not
$SU_{left}(2)$ symmetric. They would destroy the gauge invariance and, hence, the
renormalizability of the Standard Model. To preserve the gauge invariance we use the
mechanism of spontaneous symmetry breaking  which, as was explained above, allows one to
get the renormalizable theory with massive fields.

The Feynman rules in the SM include the ones for QED and QCD with additional new vertices
corresponding to the $SU(2)$ group and the Yukawa interaction, as well as the vertices
with goldstone particles if one works in the renormalizable gauge. We will not write them
down due to their complexity, though the general form is obvious.

\begin{figure}[htb]
\leavevmode
    \epsfxsize=16cm
 \epsffile{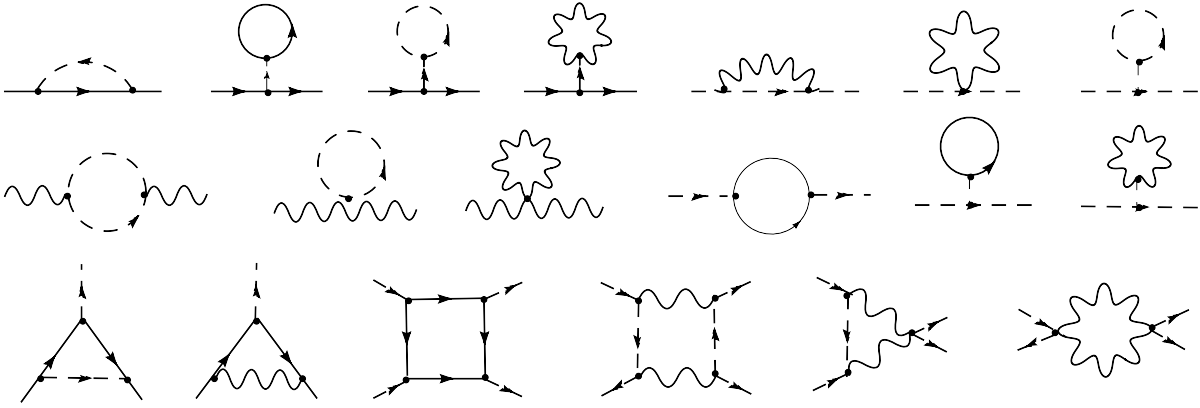}
 \caption{Some divergent one-loop diagrams in the SM. The dotted line denotes the Higgs field,
 the solid line - the quark and lepton fields, and the wavy line - the gauge fields \label{smd}}
  \end{figure}

Consider the one-loop divergent diagrams in the SM. Besides the familiar diagrams in QED
and QCD discussed above one has the diagrams presented in Fig.\ref{smd}. The diagrams
containing the goldstone bosons are omitted.
The calculation of these diagrams is similar to what we have done above. Therefore, we
show only the results for the renormalization constants of the fields and the coupling
constants. They have the form (for the gauge fields we use the Feynman gauge)
\begin{eqnarray}\label{constsm}
  Z_{2Q_L}&=&1-\frac{1}{\varepsilon}\frac{1}{16\pi^2}[\frac {1}{36} g'^2+ \frac 34 g^2
  +\frac 43 g_s^2+\frac 12 y_U^2+\frac 12 y_D^2], \nonumber \\
  Z_{2u_R}&=&1-\frac{1}{\varepsilon}\frac{1}{16\pi^2}[\frac 49 g'^2+ \frac 43 g_s^2
  + y_U^2],  \nonumber \\
  Z_{2d_R}&=&1-\frac{1}{\varepsilon}\frac{1}{16\pi^2}[\frac 19 g'^2+ \frac 43 g_s^2 + y_D^2],
  \nonumber \\
  Z_{2L_L}&=&1-\frac{1}{\varepsilon}\frac{1}{16\pi^2}[\frac 14 g'^2+ \frac 34 g^2 +
  \frac 12 y_L^2],  \nonumber \\
  Z_{2e_R}&=&1-\frac{1}{\varepsilon}\frac{1}{16\pi^2}[g'^2 + y_L^2],
\nonumber \\
  Z_{2H}&=&1+\frac{1}{\varepsilon}\frac{1}{16\pi^2}[\frac 12 g'^2+ \frac 32 g^2 - 3 y_U^2
   -3 y_D^2 - y_L^2],  \nonumber \\
  Z_{3B}&=&  1 - \frac{1}{\varepsilon}\frac{1}{16\pi^2}[ \frac{20}{9} N_F +
  \frac{1}{6} N_H] g'^2  \qquad \mathrm{U(1)_Y~boson}\nonumber \\
  Z_{3A}&=&  1 + \frac{1}{\varepsilon}\frac{1}{16\pi^2}[ 3 - \frac{32}{9} N_F ] e^2
  \qquad \mathrm{photon}\nonumber \\
  Z_{3W}&=&1+\frac{1}{\varepsilon}\frac{1}{16\pi^2}[\frac{10}{3} - \frac 13 (N_F + 3 N_F)
  - \frac 16 N_H ] g^2,  \nonumber \\
  Z_{3G}&=&1+\frac{1}{\varepsilon}\frac{1}{16\pi^2}[ 5 - \frac 43 N_F] g_s^2,  \nonumber \\
  Z_{g_3^2}&=&1+\frac{1}{\varepsilon}\frac{1}{16\pi^2}[-11+\frac 43 N_F]g_s^2, \nonumber\\
  Z_{g_2^2}&=&1+\frac{1}{\varepsilon}\frac{1}{16\pi^2}[-\frac{22}{3}+\frac 43 N_F+\frac
16N_H]g^2, \nonumber\\
  Z_{g'^2}&=&1+\frac{1}{\varepsilon}\frac{1}{16\pi^2}[\frac{20}{9}N_F+\frac{1}{6}N_H]g'^2,
\nonumber\\
  Z_{y_U^2}&=&1+\frac{1}{\varepsilon}\frac{1}{16\pi^2}[-\frac{17}{12}g'^2-\frac 94 g^2
  -8 g_s^2 +\frac 92 y_U^2 + \frac 32 y_D^2 + y_L^2 ], \nonumber\\
  Z_{y_D^2}&=&1+\frac{1}{\varepsilon}\frac{1}{16\pi^2}[-\frac{5}{12}g'^2-\frac 94 g^2
  -8 g_s^2 +\frac 32 y_U^2 + \frac 92 y_D^2 + y_L^2], \nonumber\\
  Z_{y_L^2}&=&1+\frac{1}{\varepsilon}\frac{1}{16\pi^2}[-\frac{15}{4} g'^2- \frac 94 g^2
   + \frac 94 y_L^2 + 3 y_U^2 + 3 y_D^2],
\nonumber\\
  Z_{\lambda}&=&1+\frac{1}{\varepsilon}\frac{1}{16\pi^2}[-\frac{3}{2}g'^2- \frac{9}{2}g^2
  + 2 ( 3 y_U^2 + 3 y_D^2 +  y_L^2)+6\lambda\nonumber \\
&& -2  ( 3 y_U^4 + 3 y_D^4 + y_L^4)/\lambda
   +(\frac{3}{8}g'^4+\frac 98 g^4+
   \frac{3}{4}g^2g'^2)/\lambda], \nonumber
\end{eqnarray}
where, for simplicity, we ignored the mixing between the generations and assumed the
Yukawa matrices to be diagonal.

The difference from the expressions considered above is that the renormalization constant
of the scalar coupling contains the terms of the type $g^4/\lambda$ and $y^4/\lambda$.
This is because writing the counter-term for the quartic vertex we factorized $\lambda$.
The counter-terms themselves are proportional to $g^4$ and $y^4$ and are not equal to
zero. Thus, the quantum corrections generate a new interaction even if it is absent
initially. Since the gauge and Yukawa interactions belong to the renormalizable type, the
number of types of the counter-terms is finite and the only new interaction which is
generated this way, if it was absent, is the quartic scalar one. With allowance for this
interaction the model is renormalizable.

Since the masses of all the particles are equal to the product of the gauge or Yukawa
couplings and the vacuum expectation value of the Higgs field, in the minimal subtraction
scheme the mass ratios are renormalized the same way as the ratio of couplings. To find
the renormalization of the mass itself, one should know how the v.e.v. is renormalized or
find explicitly the mass counter-term from Feynman diagrams. In this case, one has also
to take into account the tad-pole diagrams shown in Fig.\ref{smd}, including the diagrams
with goldstone bosons.

For illustration we present the renormalization constant of the $b$-quark mass in the SM
\begin{eqnarray}
Z_{m_b}&=&1+\frac{1}{\varepsilon}\frac{1}{16\pi^2}[\sum_l\frac{y_l^4}{\lambda}+
3\sum_q\frac{y_q^4}{\lambda}-\frac 32\lambda+\frac 34(y_b^2-y_t^2)\nonumber
\\&-&\frac{3}{16} \frac{(g^2+g'^2)^2}{\lambda}-\frac 38
\frac{g^4}{\lambda}-3Q_b(Q_b-T^3_b)g'^2-4g_s^2].\label{mq}
\end{eqnarray}
The result for the $t$-quark can be obtained by replacing $b$ by $t$. For the light
quarks the Yukawa constants are very small and can be ignored in eq.(\ref{mq}).

Note that here we again have the Higgs self-interaction coupling $\lambda$ in the
denominator. It appears from the tad-pole diagrams but, contrary to the previous case,
the renormalization constant $Z_{m_q}$ is not multiplied by $\lambda$ and the denominator
is not cancelled. This does not lead to any problems in perturbation theory since by
order of magnitude $\lambda \sim g^2 \sim y^2$ and the loop expansion is still valid.
\newpage
\vspace*{1cm}
\section{Lecture VI: Renormalization Group}
\setcounter{equation}{0}

The procedure formulated above allows one to eliminate the ultraviolet divergences and
get the finite expression for any Green function  in any local quantum field theory. In
renormalizable theories this procedure is reduced to the multiplicative renormalization
of parameters (masses and couplings) and multiplication of the Green function by its own
renormalization constant. This is true for any regularization and subtraction scheme.
Thus, for example, in the cutoff regularization and dimensional regularization the
relation between the "bare" and renormalized  Green functions looks like
\begin{equation}\label{cut}
\Gamma(\{p^2\},\mu ^2,\{g_\mu\} )=Z_\Gamma (\Lambda^2/\mu ^2 ,\{g_\mu\} )
\Gamma_{Bare}(\{p^2\},\Lambda ,\{g_{Bare}\})
\end{equation}
\begin{equation}\label{dim}
\Gamma(\{p^2\},\mu ^2,\{g_\mu\} )=Z_\Gamma (1/\varepsilon ,\{g_\mu\} )
\Gamma_{Bare}(\{p^2\},1/\varepsilon ,\{g_{Bare}\}), \label{fg}
\end{equation}
where $\{p^2\}$ is the set of external momenta, $\{g\}$ is the set of masses and
couplings, and
$$ g_{Bare}=Z_g(( \Lambda^2/\mu ^2,\{g_\mu\})g\ \  \ \mbox{or} \ \ \
g_{Bare}=Z_g((1/\varepsilon ,\{g_\mu\})g.$$

It is obvious that the  operation of multiplication by the constant $Z$ obeys the group
property. Indeed, after the elimination of divergences one can multiply the couplings,
masses and the Green functions by finite constants and this will be equivalent to the
choice of another renormalization scheme. Since these finite constants can be changed
continuously, we have a continuous Lie group which got the name of renormalization group.
The group transformations of multiplication of the couplings and the Green functions are
called the Dyson transformations.

\subsection{The group equations and solutions via the method of characteristics}

In what follows we stick to dimensional regularization and rewrite relation (\ref{dim})
in the form
\begin{equation}\label{inv}
   \Gamma_{Bare}(\{p^2\},1/\varepsilon, \{g_{Bare}\})=
Z_{\Gamma}^{-1}(1/\varepsilon ,\{g_\mu\}) \Gamma(\{p^2\}, \mu^2, \{g_\mu\}).
\end{equation}
It is obvious that the l.h.s. of this equation does not depend on the parameter of
dimensional transmutation $\mu$ and, hence, the r.h.s. should not also depend on it. This
allows us to write the functional equation for the renormalized Green function.
Differentiating it with respect to the continuous parameter $\mu$ one can get the
differential equation which has a practical value: solving this equation one can get the
improved expression for the Green function which corresponds to summation of an infinite
series of Feynman diagrams.

Consider an arbitrary Green function $\Gamma$ obeying equation (\ref{dim}) with the
normalization  condition
$$\Gamma(\{p^2\},\mu ^2,0 )=1. $$
Differentiating (\ref{dim}) with respect to $\mu^2$ one gets:
$$\mu^2 \frac{d}{d\mu ^2}\Gamma=\left(\mu^2 \frac{\partial }{\partial
\mu ^2} + \mu^2 \frac{\partial g}{\partial \mu ^2}\frac{\partial }{\partial  g}\right)
\Gamma = \mu^2 \frac{d \ln Z_\Gamma }{d \mu ^2}Z_\Gamma \Gamma_{Bare},$$ or
\begin{equation}
\left(\mu^2 \frac{\partial }{\partial \mu ^2} + \beta (g)\frac{\partial }{\partial  g}
+\gamma_\Gamma \right) \Gamma(\{p^2\},\mu ^2,g_\mu ) =0, \label{rg}
\end{equation}
where we have introduced the so-called beta function $\beta (g)$ and the anomaly
dimension of the Green function $\gamma_\Gamma(g)$ defined as
\begin{eqnarray}
\beta &=& \mu^2  \frac{d g}{d \mu ^2}\vert_{g_{bare}}, \label{db}\\ \gamma_\Gamma &=&
-\mu^2 \frac{d \ln Z_\Gamma }{d \mu ^2} \vert_{g_{bare}}. \label{dg}
\end{eqnarray}
Equation (\ref{rg}) is called the {\it renormalization group equation} in partial
derivatives (in Ovsyannikov form). In the western literature it is also called the
Callan-Simanzik equation.

The solution of the renormalization group equation can be written in terms of
characteristics:
\begin{equation}
\Gamma\left(e^{\displaystyle t}\frac{\{p^2\}}{\mu^2},g\right)=
\Gamma\left(\frac{\{p^2\}}{\mu ^2}, \bar{g}(t,g)\right)\mbox{\Large e}^{\displaystyle \
\int\limits_{0}^{t}\gamma_\Gamma(\bar{g}(t,g))dt},\label{solut}
\end{equation}
where the characteristic equation is (for definiteness we restrict ourselves to a single
coupling)
\begin{equation}
\frac{d}{dt}\bar{g}(t,g)=\beta (\bar g) , \ \ \bar{g}(0,g)=g. \label{ch}
\end{equation}
The quantity $\bar{g}(t,g)$ is called the {\it effective charge} or {\it effective
coupling}.

We will consider the useful properties of this solution (\ref{solut}) later and we first
derive several other similar equations. Since the vertex function usually comes with the
coupling, one can consider the product
\begin{equation}
g\Gamma\left(\frac{\{p^2\}}{\mu^2},g\right). \label{pr}
\end{equation}
If $\Gamma $ is the n-point function, then the renormalization of the coupling $g$ is
given by
$$g_{Bare}=Z_\Gamma Z_2^{-n/2}g,$$
and the product (\ref{pr}) is renormalized as
$$ g\Gamma=Z_2^{n/2} g_{Bare}\Gamma_{Bare}.$$
Hence, one has the same equation as (\ref{fg}) with solution (\ref{solut}) but with
$Z_\Gamma=Z_2^{n/2}$ and $\gamma_\Gamma=-n/2\gamma_2$. (Recall that the anomalous
dimension $\gamma _2$ is defined with respect to the renormalization constant
$Z_2^{-1}$.)

Furthermore, one can construct the so-called {\it invariant charge} by multiplying the
product (\ref{pr}) by the corresponding propagators
\begin{equation}
\xi=g\Gamma\left(\frac{\{p^2\}}{\mu^2},g\right)
\prod_{i}^{n}D^{1/2}\left(\frac{p^2_i}{\mu^2},g\right).  \label{ic}
\end{equation}
The invariant charge $\xi$, being RG-invariant,  obeys the RG equation without the
anomalous dimension and plays an important role in the formulation of the renormalization
group together with the effective charge. In some cases, for instance in the MOM
subtraction scheme, the effective and invariant charges coincide.

The usefulness of  solution (\ref{solut}) is that it allows one to sum up an infinite
series of logs coming from the Feynman diagrams in the infrared ($t \to -\infty$) or
ultraviolet ($t \to \infty$) regime and  improve the usual perturbation theory
expansions. This in its turn extends the applicability of perturbation theory and allows
one to study the infrared or the ultraviolet asymptotics of the Green functions.

To demonstrate the power of the RG, let us consider the invariant charge in a theory with
a single coupling and restrict ourselves to the massless case. Let the perturbative
expansion be
\begin{equation}
\xi(\frac{p^2}{\mu^2},g) = g(1+bg\ln \frac{p^2}{\mu^2}+ ...). \label{pt}
\end{equation}
The $\beta $ function in the one-loop approximation is given by
\begin{equation}
\beta (g)=bg^2. \label{beta}
\end{equation}
Notice that the coefficient $b$  of the logarithm in eq.(\ref{pt}) coincides with that of
the $\beta $ function. Alternatively the $\beta $ function can be defined as the
derivative of the invariant charge with respect to logarithm of momentum
\begin{equation}
\beta (g)=p^2\frac{d}{dp^2}\xi(\frac{p^2}{\mu^2 },g)\vert_{p^2=\mu^2}. \label{def}
\end{equation}

This definition is useful in the MOM scheme where the mass is not considered as a
coupling but as a parameter and the renormalization constants depend on it. We will come
back to the discussion of this question below when considering different definitions of
the mass.

According to eq.(\ref{solut}) (with vanishing anomalous dimension) the RG-improved
expression for the invariant charge corresponding to the perturbative expression
(\ref{pt}) is:
\begin{equation}
\xi_{RG}(\frac{p^2}{\mu^2},g)=\xi_{PT}(1,\bar{g}(\frac{p^2}{\mu^2},g))=
\bar{g}(\frac{p^2}{\mu^2},g),
\end{equation}
where we have put in eq.(\ref{solut})  $p^2=\mu ^2$ and then replaced $t$ by $t=\ln
p^2/\mu ^2$. The effective coupling is a solution of the characteristic equation
\begin{equation}
\frac{d}{dt}\bar{g}(t,g)=b\bar{g}^2 , \ \ \bar{g}(0,g)=g, \ \ t\equiv \ln \frac{p^2}{\mu
^2}.
\end{equation}
The solution of this equation is
\begin{equation}
\bar{g}(t,g)=\frac{g}{1-bgt} . \label{gp}
\end{equation}
Being expanded over $t$, the geometrical progression (\ref{gp}) reproduces the expansion
(\ref{pt}); however, it sums the infinite series of terms of the form $g^nt^n$. This is
called the leading log approximation (LLA) in QFT. To get the correction to the LLA, one
has to consider the next term in the expansion of the $\beta $ function. Then one can sum
up the next series of terms of the form $g^nt^{n-1}$ which is called the next to leading
log approximation (NLLA), etc. This procedure allows one to describe the leading
asymptotics of the Green functions for $t \to \pm \infty$.

Consider now the Green function with non-zero anomalous dimension. Let its perturbative
expansion be
\begin{equation}
\Gamma(\frac{p^2}{\mu^2},g)=1+cg\ln \frac{p^2}{\mu^2}+...
\end{equation}
Then in the one-loop approximation the anomalous dimension is
\begin{equation}
\gamma (g) = cg. \label{gamma}
\end{equation}
Again the coefficient  of the logarithm coincides with that of the anomalous dimension.
In analogy with eq.(\ref{def}) the anomalous dimension can be defined as a derivative
with respect to the logarithm of momentum
\begin{equation}
\gamma (g)=p^2\frac{d}{dp^2}\ln \Gamma(\frac{p^2}{\mu^2 },g)\vert_{p^2=\mu^2}.
\label{def2}
\end{equation}

Substituting  (\ref{gamma}) into eq.(\ref{solut}), one has in the exponent
$$\int\limits_{0}^{t}\gamma(\bar{g}(t,g)dt = \int\limits_{g}^{\bar
g} \frac{\gamma (g)}{\beta (g)}dg=\int\limits_{g}^{\bar g}\frac{cg}{bg^2}dg=
\frac{c}{b}\ln \frac{\bar g}{g}.$$ This gives for the Green function the improved
expression
\begin{equation}
\Gamma_{RG}=\left(\frac{\bar g}{g}\right)^{-c/b}= \left(\frac{1}{1-bgt}\right)^{c/b}
\approx 1+ct+... \label{gf2}
\end{equation}
Thus, one again reproduces the perturbative expansion, but expression (\ref{gf2}) again
contains the whole infinite sum of the leading logs. To get the NLLA, one has to take
into account the next term in eq.(\ref{gamma}) together with the next term of expansion
of the $\beta $ function.

All the formulas can be easily generalized to the case of multiple couplings and masses.

\subsection{The effective coupling}

\begin{figure}[htb]\hspace*{4cm}
\begin{center}
 \leavevmode
    \epsfxsize=14cm
 \epsffile{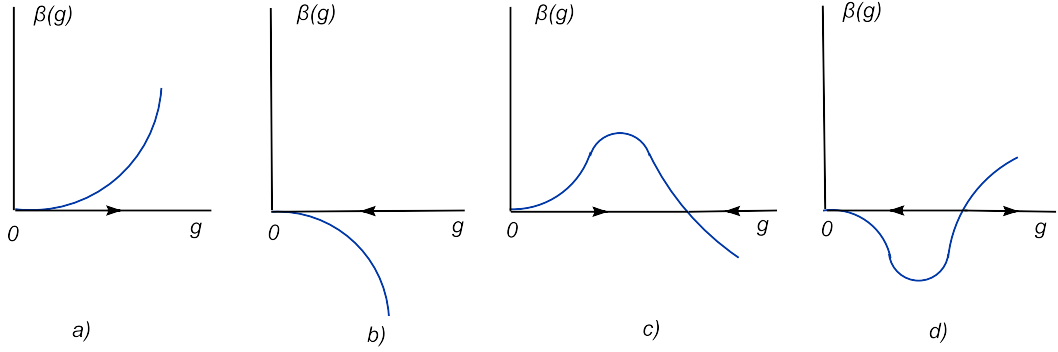}
\caption{The possible form of the $\beta $-function. The arrows show the behaviour of the
effective coupling in the ultraviolet regime ($t \to \infty$)\label{ec}}
\end{center}
\end{figure}

By virtue of the  central role played  by the effective coupling in RG formulas, consider
it in more detail.  The behaviour of the effective coupling is determined by the $\beta$
function. Qualitatively, the $\beta $ function can exhibit the behaviour shown in
Fig.\ref{ec}. We restrict ourselves to the region of small couplings.

In the first case, the $\beta$-function is positive. Hence, with increasing momentum the
effective coupling unboundedly increases. This situation is typical of most of the models
of QFT in the one-loop approximation when $\beta(g)=bg^2$  and $b>0$. The solution of the
RG equation for the effective coupling in this case has the form of a geometric
progression (\ref{gp}). It is characterized by the presence of a pole at high energies,
called the Landau pole. We will consider this pole in detail later.

In the second case, the $\beta$-function is negative and, hence, the effective coupling
decreases with increasing momentum. This situation appears in the one-loop approximation
when $b<0$, which takes place in the gauge theories. Here we also have a pole but in the
infrared region.

In the third case, the $\beta$-function has zero: at first, it is positive and then is
negative. This means that for small initial values the effective coupling increases; and
for large ones, decreases. In both the cases, with increasing momentum it tends to the
fixed value defined by the zero of the $\beta$-function. This is the so-called
ultraviolet stable {\it fixed point}. It appears in some models in higher orders of
perturbation theory.

Eventually, in the last case one also has the fixed point but now for the small initial
coupling it decreases and for the large one it increases, i.e., with increasing momentum
the effective coupling moves away from the fixed point, it is ultraviolet unstable. On
the contrary, with decreasing momentum it tends to the fixed point, i.e., it is infrared
stable. It appears in some models in lower dimensions, for instance, in statistical
physics.

\subsection{Dimensional regularization and the $\overline{MS}$ scheme}

Consider now the calculation of the $\beta $ function and the anomalous dimensions in
some particular models within the dimensional regularization and the minimal subtraction
scheme. Note that in transition from dimension  $4$ to $4-2\varepsilon $ the dimension of
the coupling is changed and the "bare" coupling acquires the dimension
$[g_B]=2\varepsilon$. That is why the relation between the "bare" and renormalized
coupling contains the factor $(\mu^2)^\varepsilon$
\begin{equation}\label{charged}
  g_B=(\mu ^2)^\varepsilon Z_g g.
\end{equation}
Hence, even before the renormalization when $Z_g=1$, in order to compensate this factor
the dimensionless coupling $g$ should depend on $\mu$. Differentiating (\ref{charged})
with respect to $\mu^2$ one gets
$$ 0=\varepsilon Z_g g + \frac{d\log Z_g}{d\log \mu^2}Z_g g+Z_g\frac{dg}{d\log \mu^2},$$
i.e.,
\begin{equation}\label{betad}
  \beta_{4-2\varepsilon}(g)\equiv\frac{dg}{d\log \mu^2}=-\varepsilon g + g\frac{d\log Z_g}{d\log \mu^2}
=-\varepsilon g+\beta_4(g).
\end{equation}

In the $\overline{MS}$ scheme the renormalization constants are given by the pole terms
in $1/\varepsilon $ expansion and so does the bare coupling. They can be written as
\begin{equation}
Z_\Gamma= 1+ \sum_{n=1}^{\infty}\frac{c_n(g)}{\varepsilon ^n}= 1+
\sum_{n=1}^{\infty}\sum_{m=n}^{\infty}\frac{c_{nm}g^m}{\varepsilon ^n}. \label{zz}
\end{equation}
And similarly
\begin{equation}
g_{Bare}= (\mu ^2)^\varepsilon \left[g+\sum_{n=1}^{\infty}\frac{a_n(g)}{\varepsilon
^n}\right]= (\mu ^2)^\varepsilon
\left[g+\sum_{n=1}^{\infty}\sum_{m=n}^{\infty}\frac{a_{nm}g^{m+1}}{\varepsilon
^n}\right]. \label{gg}
\end{equation}

Differentiating eq.(\ref{zz}) with respect to $\ln \mu^2 $ and having in mind the
definitions (\ref{db}) and (\ref{dg}), one has:
$$-[1+ \sum_{n=1}^{\infty}\frac{c_n(g)}{\varepsilon ^n}]
\gamma_\Gamma(g)=\left[-\varepsilon g+\beta(g)\right]\frac{d}{dg}
\sum_{n=1}^{\infty}\frac{c_n(g)}{\varepsilon ^n}.$$ Equalizing the  coefficients of equal
powers of $\varepsilon $, one finds
\begin{eqnarray}
\gamma_\Gamma (g)&=&  g\frac{d}{dg}c_1(g), \label{polgam}\\
g\frac{d}{dg}c_n(g)&=& [\gamma_\Gamma(g) + \beta (g)\frac{d}{dg}]c_{n-1}(g), \ \ n \geq
2.
\end{eqnarray}

One sees that the coefficients of higher poles  $c_n, \ n\geq 2$ are completely defined
by that of the lowest pole $c_1$ and the $\beta $ function. In its turn the
$\beta$-function is also defined by the lowest pole. To see this, consider eq.(\ref{gg}).
Differentiating it with respect to $\ln \mu ^2$ one has
\begin{equation}
\varepsilon \left[g+\sum_{n=1}^{\infty}\frac{a_n(g)}{\varepsilon ^n}\right]+
\left[-\varepsilon g+\beta(g)\right]
\left[1+\frac{d}{dg}\sum_{n=1}^{\infty}\frac{a_n(g)}{\varepsilon ^n}\right] =0 .
\end{equation}
Equalizing the coefficients of equal powers of $\varepsilon $, one finds
\begin{eqnarray}
\beta (g)&=&(g\frac{d}{dg}-1)a_1(g), \label{polbet}\\
(g\frac{d}{dg}-1)a_n(g)&=&\beta (g)\frac{d}{dg}a_{n-1}(g), \ \ n\geq 2.\label{poln}
\end{eqnarray}

Thus, knowing the coefficients of the lower poles one can reproduce all the higher order
divergences. This means that they are not independent, all the information about them is
connected in the lowest pole. In particular, substituting in (\ref{poln}) the
perturbative expansion (\ref{gg}) one can solve the recurrent equation and find for the
highest pole term
\begin{equation}
a_{nn}= a_{11}^n, \label{pole}
\end{equation}
i.e. in the leading order one has the geometric progression
\begin{equation}\label{bare}
  g_B=\mu^{2\varepsilon} \frac{g}{1-ga_{11}/\varepsilon},
\end{equation}
which reflects the fact that the effective coupling in the LLA is also given by a
geometric progression (\ref{gp}).

The pole equations are easily generalized for the multiple couplings case, the higher
poles are also expressed through the lower ones though the solutions of the RG equations
are more complicated.

Consider now some particular models and calculate the corresponding $\beta$-functions and
the anomalous dimensions.\vspace{0.3cm}

\underline{The $\phi^4$ theory}

The renormalization constants in the $\overline{MS}$ scheme up to two loops are given by
eqs. (\ref{zp},\ref{z4},\ref{zl}). ($g\equiv \lambda /16\pi ^2$)
\begin{eqnarray}
Z_4&=&1+\frac{3}{2\varepsilon }g+g^2(\frac{9}{4\varepsilon
^2}-\frac{3}{2\varepsilon }), \\
Z_2^{-1}&=&1+\frac{g^2}{24\varepsilon },\\
Z_g&=&1+\frac{3}{2\varepsilon }g+g^2(\frac{9}{4\varepsilon ^2}-\frac{17}{12\varepsilon}).
\end{eqnarray}
Notice that the higher pole coefficient $a_{22}=9/4$ in the last expression is the square
of the lowest pole one $a_{11}=3/2$ in accordance with eq.(\ref{pole}).

Applying now eqs.(\ref{polgam}) and (\ref{polbet}) one gets
\begin{eqnarray}
\gamma_4(g)&=&\frac{3}{2}g-3g^2, \\
\gamma_2(g)&=&\frac{1}{12}g^2, \\
\beta (g)&=&g(\gamma_4+2\gamma_2)=\frac{3}{2}g^2-\frac{17}{6}g^2. \label{res}
\end{eqnarray}
One can see from eq.(\ref{res}) that the first coefficient of the $\beta$-function is
$3/2$, i.e., the $\phi^4$ theory belongs to the type of theories shown in Fig.\ref{ec}a).
In the leading log approximation (LLA) one has a Landau pole behaviour.  In the two-loop
approximation (NLLA) the $\beta$-function gets a non-trivial zero and the effective
coupling possesses an UV fixed point like the one shown in Fig.\ref{ec}в). However, this
fixed point is unstable with respect to higher orders and is not reliable. Here we
encounter the problem of divergence of perturbation series in quantum field theory, they
are the so-called asymptotic series which have a zero radius of convergence. \\

\underline{QED}

In QED in the one-loop approximation the renormalization constants in the Feynman gauge
are given by eq.(\ref{zqed}).  Due to the Ward identities the renormalization of the
coupling is defined by the photon wave function renormalization constant $Z_3$ and is
gauge invariant. Equation (\ref{zqed}) allows one to determine the anomalous dimensions
and the $\beta$-function
\begin{eqnarray}
\gamma_1(\alpha)&=&-\alpha, \\
\gamma_2(\alpha)&=&\alpha, \\
\gamma_3(\alpha)&=&\frac 43 \alpha,\\
\gamma_m(\alpha)&=&-4\alpha,\\
 \beta_\alpha(\alpha)&=&\frac 43 \alpha^2,
\label{resqed}
\end{eqnarray}
where we use the notation $\alpha \equiv e^2/16\pi ^2$.

Thus, in QED in the one-loop approximation the effective coupling behaves the same way a
in the  $\phi^4$ theory and has a Landau pole in the LLA. In this theory, the next term
of expansion of the $\beta$-function is also calculated. It has the same sign.\\

\underline{QCD}

In QCD the calculation of the $\beta $ function can be based on various vertices. The
result should be the same due to the gauge invariance. To simplify the calculations, we
choose the ghost-ghost-vector vertex. The renormalization constants in the one-loop
approximation in the Feynman gauge are given by (\ref{constqcd}) and lead to the
following anomalous dimensions and the $\beta$-function:
\begin{eqnarray}
\tilde{\gamma}_1(\alpha)&=&-\frac{C_2}{2}\alpha , \\
\tilde{\gamma}_2(\alpha)&=&-\frac{C_2}{2}\alpha , \\
\gamma_3(\alpha)&=&-(\frac{5}{3}C_2-\frac{2}{3}n_f)\alpha , \\
\beta_\alpha(\alpha) &=&\alpha (2\tilde{\gamma}_1+2\tilde{\gamma}_2+\gamma_3)=
-(\frac{11}{3}C_А-\frac{2}{3}n_f)\alpha ^2,\label{ress2}
\end{eqnarray}
where like in QED we take $\alpha\equiv g^2/16\pi^2$, the Casimir operator $C_А$ in the
case of SU(3) groups is equal to  3, and  $n_f$ is the number of quark flavours.

One can see from eq.(\ref{ress2}) that if the number of flavours is less than
$\frac{11}{2}C_2 = \frac{33}{2}$, the $\beta$-function is negative and the effective
coupling decreases and tends to zero with increasing momentum. This type of behaviour of
the effective coupling is called the {\it asymptotic freedom}. It takes place only in
gauge theories.

\subsection{$\Lambda_{QCD}$}

\begin{figure}[htb]
\begin{center}
 \leavevmode
    \epsfxsize=7cm
 \epsffile{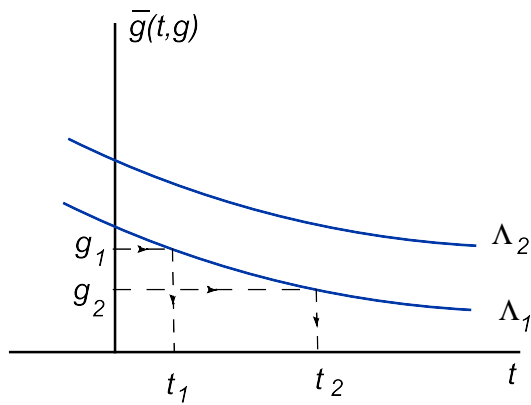}
\caption{Different parametrizations of the effective coupling. Each curve is
characterized by a single parameter $\Lambda$ \label{lam}}
\end{center}
\end{figure}

The solution of the characteristic equation for the effective coupling, which is a
differential equation of the first order, depends on initial conditions. Therefore, the
solution  (\ref{gp}) depends on the choice of the initial point and the value of the
coupling at this point. However, this choice is not unique and one can choose another
initial point and another value of the coupling and still get the same solution, as it is
shown in Fig.\ref{lam}.

In fact, every curve is not characterized  by two numbers (the initial point and the
coupling), but by one number and the transition from one curve to another is defined by
the change of this number. To see this, consider the one-loop expression for the
effective coupling in a gauge theory and rewrite it in equivalent form
\begin{equation}
\bar g(\frac{Q^2}{\mu^2},g_\mu)= \frac{g_\mu}{1-\beta_0g_\mu\ln \frac{Q^2}{\mu^2}}=
\frac{1}{\frac{1}{g_\mu}-\beta_0\ln \frac{Q^2}{\mu^2}} \equiv -\frac{1}{\beta_0\ln
\frac{Q^2}{\Lambda^2}}= \bar g(\frac{Q^2}{\Lambda^2}),
\end{equation}
where we have introduced the notation
\begin{equation}\label{l}
\Lambda^2 = \mu^2 e^{\displaystyle - \frac{1}{\beta_0 \alpha_\mu}}.
\end{equation}
This quantity is called $\Lambda_{QCD}$ in quantum chromodynamics and can be introduced
in any model. The numerical value of $\Lambda$ is defined from experiment.

Equation (\ref{l}) can be generalized to any number of loops. For this purpose, let us
rewrite the RG equation for the effective coupling in the Gell-Mann -- Low form. One has
\begin{equation}
\ln\frac{Q^2}{\mu^2}=
 \int^{g_Q}_{g_\mu}\frac{dg}{\beta_g(g)}.
\end{equation}
Combining the lower limit with $\ln \mu^2$ one gets
\begin{equation}
\ln\frac{Q^2}{\Lambda^2}= \int^{g_Q}_{}\frac{dg}{\beta_g(g)},
\end{equation}
where
\begin{equation}\label{ll}
\Lambda^2 = \mu^2 exp\left( \int^{g_\mu}
 \frac{dg}{\beta_g(g)}\right),
\end{equation}
which is the generalization of eq.(\ref{l}) for an arbitrary number of loops.

The quantity $\Lambda$, introduced this way, is $\mu$-independent but depends on the
renormalization scheme  due to the scheme dependence of the $\beta$-function. However,
the scheme dependence of $\Lambda$ is given exactly (!) in one-loop order. Indeed, since
$\Lambda$ does not depend on $\mu$, let us choose $\mu$ in such a way that $g_\mu \to 0$.
Then for the $\beta$-function one can use the perturbative expansion
$$\beta_\alpha(\alpha)= \beta_0 \alpha^2 +\beta_1 \alpha^3 + ...$$
or
$$\int \frac{d\alpha}{\beta(\alpha)} = -\frac{1}{\beta_0\alpha} + \ln \alpha
+ O(\alpha).$$ In this limit the ratio of two  parameters $\Lambda$ belonging to two
different schemes is
\begin{equation}\label{s}
\ln \frac{\Lambda_1^2}{\Lambda_2^2} = - \frac{1}{\beta_0}\left[ \frac{1}{\alpha_1} -
\frac{1}{\alpha_2}\right] = - \frac{1}{\beta_0}\left[ c_1-c_2 \right],\end{equation}
where the coefficients $c_1$ and $c_2$ are calculated in the one-loop order. They can be
found from perturbative expansion of any physical quantity in two different schemes
\begin{eqnarray*}
 R&=& g_1(1+c_1g_1 + ...)\\
 &=& g_2(1+c_2g_2 + ...).
\end{eqnarray*}
Since $\Lambda$ does not depend on $g$, one can take any value of $g$, and eq.(\ref{s})
is always valid. The difference $c_1-c_2$ does not depend on a particular choice of $R$
(though each of them depends) and is universal.

It should be noted that the quantities like the invariant or effective coupling, the
$\beta$-function, etc. are not directly observable. Therefore, their dependence on the
subtraction scheme does not contradict the independence of predictions of the method of
calculations. We perform the perturbative expansion over the coupling which is scheme
dependent, but the coefficients are also scheme dependent. As a result, within the given
accuracy defined by the order of perturbation theory the answer is universal.

In the minimal subtraction schemes when the renormalizations depend only on dimensionless
couplings, the one-loop renormalization constants and hence the anomalous dimensions and
the $\beta$-function are the same in all schemes; the difference starts from two loops.
The exception is the  $\beta$-function in a theory with a single coupling like QED, QCD
or the $\phi^4$ theory, where the difference starts from three loops. Indeed, if one has
two subtraction schemes $M_1$ and $M_2$ so that the couplings in two schemes are related
by
$$g_2=q(g_1)=g_1+cg_1^2+O(g_1^3),$$
then the $\beta$-functions $\beta_1(g_1)$ and $\beta_2(g_2)$ are connected by the
relation
$$\beta_2(g_2)=\frac{dq(g_1)}{dg_1}\beta_1(g_1)$$
and their perturbative expansions are
\begin{eqnarray*}
 \beta_1(g_1)&=& \beta_0g_1+\beta_1g_1^2 + \beta_2g_1^3+...,\\
\beta_2(g_2)&=& \beta_0g_2+\beta_1g_2^2 + \beta_2'g_2^3+....
 \end{eqnarray*}
so that the first two terms of the $\beta$-function are universal.

As for the further terms of expansion, they depend on the renormalization scheme and one
can use this dependence as discretion, for instance, one can put all of them equal to
zero. Then we would have an exact $\beta$-function. However, one should have in mind that
it is not valuable by itself but rather in the aggregate with the PT expansion for the
Green functions for which we construct the solution of the RG equation. This expansion in
our "exact" scheme is unknown.

\subsection{The running masses}

In the minimal subtraction scheme the renormalization of the mass is performed the same
way as the renormalization of the couplings, i.e., the mass is treated as an additional
coupling  and is renormalized multiplicatively, namely,
 $$m_{Bare} = Z_m m,$$
where the mass renormalization constant $Z_m$ is independent of the mass parameters and
depends only on dimensionless couplings. Then, in full analogy with the effective
coupling one can introduce the {\em effective} or the "running" mass
\begin{equation}
\frac{d}{dt}\bar{m}(t,g)=\bar m\gamma_m (\bar g) , \ \ \bar{m}(0,g)=m_0. \label{ma}
\end{equation}
Solving this equation together with the equation for the effective coupling (\ref{ch})
one has
\begin{equation}
\bar m(t,g)= m_0\mbox{\Large e}^{\displaystyle \
\int\limits_{0}^{t}\gamma_m(\bar{g}(t,g))dt}=m_0\mbox{\Large e}^{\displaystyle \
\int\limits_{g}^{\bar g} \frac{\gamma (g)}{\beta (g)}dg}.
\end{equation}

In the one-loop order
$$ \beta(\alpha) = b\alpha^2, \ \ \ \gamma_m(\alpha) = c\alpha$$
and the solution is
$$m(t) =m_0 \left(\frac{\alpha(t)}{\alpha_0}\right)^{c/b}.$$ This is the
running mass!

The natural question arises: what is the physical mass measured in experiment and how is
it related to the running mass and at what scale?

To answer this question, consider why the mass is running. This is due to the radiative
corrections. If one considers the value of momentum which is bigger than the mass, i.e.
$p^2 > m^2$, then the particles are created, they are running inside the loops and give
the  contribution to the running. On the contrary, if $p^2 <m^2$, particles are not
created, they "decouple" and do not contribute to the running. In the MOM scheme this
takes place automatically because for the momentum smaller than the mass the diagram
simply disappears. In the minimal scheme, on the contrary, this does not happen. Hence,
it is quite natural in this case to stop the running at the value of $p^2=m^2$ and to
identify the physical mass with the running mass at the scale of the mass, i.e
 $$m^2 = \bar m^2 (m^2).$$

However, this is true only up to finite corrections. Let us come back to the definition
of the mass term in the Lagrangian. It is chosen in such a way that the propagator of a
particle, which is the inverse to the quadratic form, has the pole at $p^2=m^2$.
Therefore, a more appropriate definition of the physical mass is the position of the pole
of the propagator with allowance for the radiative corrections, .i.e.,

\begin{center}\large physical mass $\equiv$ pole mass \end{center}\vspace{0.1cm}

This definition of a mass does not depend on a scale and it is also scheme independent
and may have physical meaning. The pole mass can be expressed through the running mass at
the scale of a mass with finite and calculable corrections.

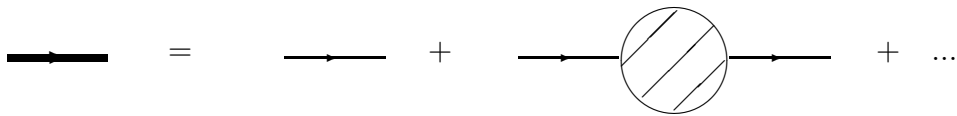
\begin{figure}[h]\vspace{-0.5cm}
\begin{picture}(100,40)(-20,0)
   \put(0,0){\vector(1,0){20}}\put(20,0){\line(1,0){18}}
    \put(0,1){\vector(1,0){20}}\put(20,1){\line(1,0){18}}
   \put(0,2){\vector(1,0){20}}\put(20,2){\line(1,0){18}}
   \hspace*{2cm} =
   \put(35,1){\vector(1,0){20}}\put(55,1){\line(1,0){18}}
   \hspace{3cm} +  \put(25,1){\vector(1,0){20}}\put(45,1){\line(1,0){18}}
   \put(84,0){\circle{50}}\put(84,-19){\line(1,1){19}}
   \put(72,-14){\line(1,1){27}}\put(64,-2){\line(1,1){21}}
   \put(105,1){\vector(1,0){20}}\put(125,1){\line(1,0){18}}
   \hspace{5.5cm} + \ \ ...
\end{picture}\vspace{0.8cm}

\caption{The quark propagator}\label{11}
\end{figure}

Consider as an example the quark mass in QCD. The quark propagator is graphically
presented in Fig.\ref{11}.
%\vspace{1.3cm}

\noindent The corresponding  expression is
\begin{eqnarray*}
&&G(\hat p,m)= \frac{i}{\hat p - m}+ \frac{i}{\hat p - m}(iA\hat p +iB m)
\frac{i}{\hat p - m} + ... \\
&& =\frac{i}{\hat p - m}\left[1 -\frac{A\hat p +B m}{\hat p -m}+ ... \right] =
\frac{i}{\hat p - m}\ \frac{1}{1+\frac{A\hat p+Bm}{\hat p -m}} = \frac{i}{\hat p -m
+A\hat p +Bm}.
\end{eqnarray*}
The pole mass is now defined as a root of the equation
\begin{equation}\label{polm}
  \hat p (1+A(p^2))-m(1-B(p^2))=0,
\end{equation}
which gives in the lowest order
$$m_{pole} = m\frac{1-B(m^2)}{1+A(m^2)} = m[1-A(m^2)-B(m^2)].$$

\begin{figure}[htb]\hspace*{4cm}
\begin{center}
 \leavevmode
    \epsfxsize=4.5cm
 \epsffile{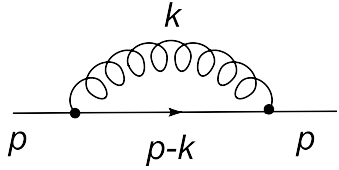}
 \caption{The quark propagator in one loop in QCD}\label{12}
      \end{center}
   \end{figure}

To calculate the functions $A$ and $B$, consider the one-loop diagram shown in
Fig.\ref{12}.

The corresponding expression is
\begin{equation}
   \Sigma =-\frac{g_s^2}{(2\pi)^4}C_F\int \frac{dk \ \gamma^\mu(\hat p-\hat
   k +m)\gamma^\nu}{[(p-k)^2-m^2]} \frac{g^{\mu\nu}}{k^2}
\end{equation}
and was calculated earlier. The result has the form (\ref{selfrr})
\begin{eqnarray}
A(p^2,m^2)&=& \frac{g_s^2}{16\pi^2}C_F\left[\frac 1\varepsilon\!
-\!1\!-\!2\int_0^1\!dx(1\!-\!x)\log\frac{p^2x(1-x)\!-\!m^2}{-\mu^2}\right], \\
B(p^2,m^2)&=& \frac{g_s^2}{16\pi^2}C_F\left[-\frac 4\varepsilon
+2+4\int_0^1dx\log\frac{p^2x(1-x)-m^2}{-\mu^2}\right].
\end{eqnarray}
After subtraction of divergences in the $\overline{MS}$-scheme one has
\begin{eqnarray}
A^{\overline{MS}}(p^2,m^2)&=&- \frac{g_s^2}{16\pi^2}C_F\left[
1\!+\!2\int_0^1\!\! dx(1\!-\!x)\log\frac{p^2x(1\!-\!x)\!-\!m^2}{-\mu^2}\right], \\
B^{\overline{MS}}(p^2,m^2)&=& \frac{g_s^2}{16\pi^2}C_F\left[
2+4\int_0^1dx\log\frac{p^2x(1-x)-m^2}{-\mu^2}\right].
\end{eqnarray}
Substituting $p^2=m^2$, one finds
\begin{equation}\label{AB}
  A^{\overline{MS}}(m^2,m^2) = 2 +\ln \frac{\mu^2}{m^2}, \ \ \
B^{\overline{MS}}(m^2,m^2) = -6-4\ln \frac{\mu^2}{m^2}.
\end{equation}
Thus, for the radiative correction to the pole mass we have
\begin{equation}
   m_{pole} = m(\mu)\left[1+\frac{\alpha_sC_F}{4\pi}( 4 +
   3 \ln \frac{\mu^2}{m^2})\right].
\end{equation}
Substituting $C_F=4/3$ and $\mu^2=m^2$ one obtains the desired relation between the pole
mass and the running mass at the mass scale
\begin{equation}
   m_{pole} = m(m)\left[1+ \frac{4}{3}\frac{\alpha_s}{\pi}
   \right].
\end{equation}
\newpage
\vspace*{1cm}
\section{Lecture VII: Zero Charge and Asymptotic Freedom}
\setcounter{equation}{0}

Since the behaviour of the effective coupling has so essential consequences we consider
two typical examples which are realized in quantum field theory in the one-loop
approximation and presumably take place in a full theory. Usually, one speaks about the
zero charge behaviour or the asymptotic freedom. We explain below what  it means.

\subsection{The zero charge}

\begin{figure}[htb]\hspace*{4cm}
\begin{center}
 \leavevmode
    \epsfxsize=15cm
 \epsffile{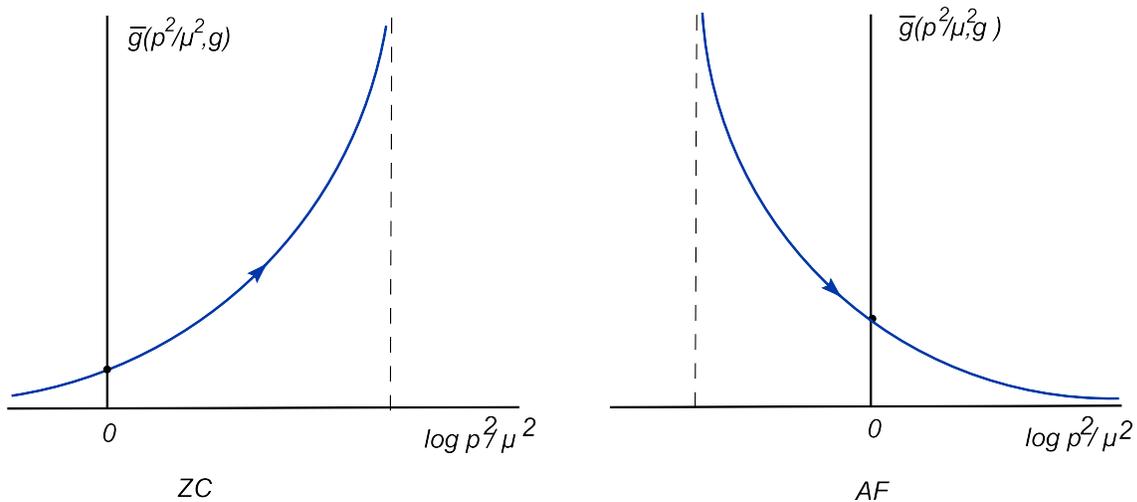}
\caption{The behaviour of the effective coupling: the zero charge (left) and the
asymptotic freedom (right) \label{charge}}
\end{center}
\end{figure}

The notion of the zero charge appeared in QED in the leading log approximation. This is
what takes place within the renormalization group method in the one-loop approximation.
If one writes down the expression for the renormalized  coupling as a function of the
"bare" coupling, i.e. inverts  eq.(\ref{bare}), one gets
\begin{equation}\label{nz}
  g=\frac{g_B}{1+\beta_0 g_B/\varepsilon}=\frac{g_B}{1+\beta_0 g_B\log \Lambda^2},
\end{equation}
where the first coefficient of the $\beta$-function $\beta_0>0$. Then, removing the
regularization, i.e., for $\varepsilon \to 0$ or $\Lambda\to\infty$, the renormalized
coupling tends to zero independently of the value of the "bare" coupling. This is what is
called  the {\it zero charge}. For the effective coupling considered above the zero
charge corresponds to the behaviour shown on the left panel of Fig.\ref{charge} which is
characterized by the Landau pole at high energies.

The zero charge behaviour is typical of QED, the $\phi^4$ theory for  positive quartic
coupling and also  the Yukawa type interactions, i.e., in those theories where the
$\beta$-function is positive.

It is obvious that in the vicinity of the pole the perturbation theory does not work and,
hence, the one-loop formula is not applicable.  However, for small momenta transfer the
one-loop approximation is reliable. For instance, in QED the effective expansion
parameter is $e^2/16\pi^2=\alpha/4\pi\approx 1/137/4\pi\approx 5.8\cdot 10^{-4}$ and the
next loop corrections (which have the same sign) do not play any essential role. The
behaviour of the effective coupling in QED in the region up to 100 GeV has got the
experimental confirmation in measuring the fine structure constant at the LEP
accelerator. At the scale equal to the mass of the Z-boson $M_Z$ the fine structure
constant is not $1/137$ but $\alpha(M_Z)\approx1/128$, which is in a good agreement with
the one-loop formula.

The large momenta transfer in this case are limited by the pole provided the pole does
not disappear in a full theory. It is still unclear how higher orders of perturbation
theory influence this behaviour since the perturbation series is divergent and it is
impossible to make definite conclusions without additional nonperturbative information.

The presence of the Landau pole indicates the presence of unphysical ghost states. To see
this, consider the photon propagator in QED which due to the Ward identities coincides
with the invariant charge and in the leading log approximation has the form of a
geometric progression
\begin{equation}\label{phpr}
  G(p^2)=-i\frac{g^{\mu\nu}-p^\mu p^\nu/p^2}{p^2}
  \frac{1}{1-\frac{4}{3}\sum Q^2\frac{\alpha_0}{4\pi}\log(-p^2/m^2)},
\end{equation}
where $Q$ is the electric charge of a particle (in the units of electron charge) running
round the loop.

This expression has a pole in the Euclidean region at
$p^2=-m^2exp(\frac{3\pi}{\alpha_0Q^2n_f})$. Substituting $m=m_e=0.5$ MeV, $\alpha_0\simeq
1/137$ and $\sum Q^2=[(4/9+1/9)3+1)3]=8$, one gets $p^2\simeq -(5\cdot 10^{31})^2$
GeV$^2$. That is the pole is very far off, even beyond the Planck scale, and at low
energies one can ignore it. However, the presence of the pole indicates the presence of a
new asymptotic state and the residue at the pole defines the norm of this state. In the
case of the Landau pole the residue is negative, i.e., the new state is a ghost, it has
the wrong sign of the kinetic term in the Lagrangian. This fact, in its turn, leads to
negative probabilities, which indicates internal inconsistency of the theory.

Usually, it is assumed that there are two ways out of this trouble: either the higher
order corrections  improve the behaviour of the theory at high momenta so that the Landau
pole disappears, or that the zero charge theory is contradictory by itself, but at high
energies it is part of a more general theory where the behaviour of the coupling is
improved. The example of such a behaviour is given by the Grand Unified Theories where
QED is one of the branches of a non-Abelian gauge theory with the asymptotically free
behaviour. In both the cases the theory at high energies is modified.  At the same time,
the zero charge theory is infrared free, i.e. for small momenta transfer the coupling
goes to zero.

\subsection{The asymptotic freedom}

The name {\it asymptotic freedom} originates from  the non-Abelian gauge theories where
it was found that  the sign of the first coefficient of the $\beta$-function is negative.
The effective coupling in this case behaves as is shown in the right panel of
Fig.\ref{charge} and tends to zero at high momenta transfer. This means that  quarks in
QCD  are quasi-free particles, i.e., practically do not interact. This way one explains
the success of the so-called parton model of the strong interactions at high energies,
according to which the proton behaves as a set of free partons, and at high energies the
interaction takes place with the individual partons and their interaction does not play
any role.

\begin{figure}[ht]\vspace{0.4cm}
 \begin{center}
 \leavevmode
  \epsfxsize=9cm
 \epsffile{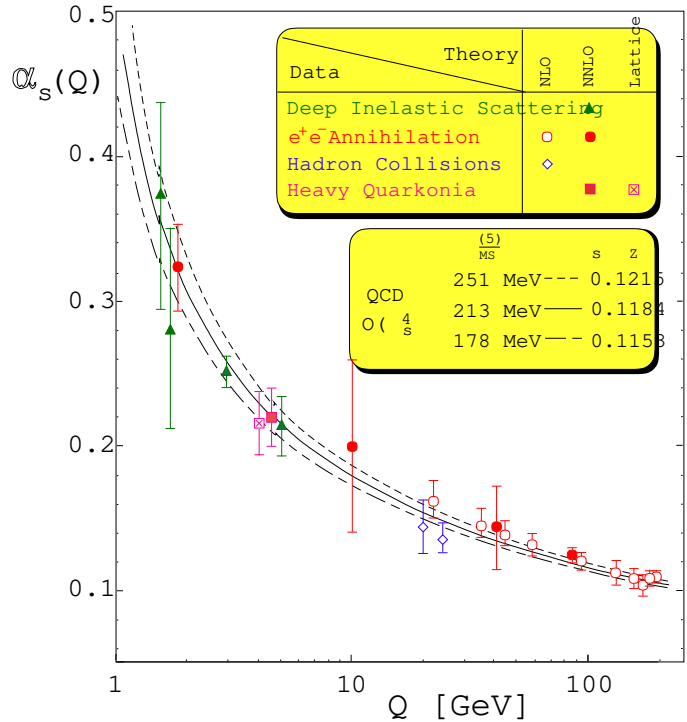}
 \end{center}
 \caption{The variation of the effective coupling of the strong interactions
  $\alpha_s$ with energy}
 \label{alpha}
 \end{figure}

The behaviour of the effective coupling in QCD at high energies was tested at various
accelerators and in various experiments and the validity of the renormalization group
formula was confirmed. The accuracy of modern measurements assumes the inclusion of the
next terms of perturbative expansion. In QCD in the  $\overline{MS}$ scheme the four
terms of the $\beta$-function are known. Below we present the two-loop expression
\begin{equation}\label{betaqcd}
  \beta_\alpha(\alpha_s)=-\frac{1}{4\pi}[11-\frac 23n_f]\alpha_s^2-\frac{1}{(4\pi)^2}[
  102-\frac{38}{3} n_f]\alpha_s^3+O(\alpha_s^4).
\end{equation}
As one can see, if the number of quarks in not too big, both the coefficients of the
$\beta$-function are negative. All the experimental data fit a single curve for the
effective coupling with the parameter $\Lambda_{QCD}\simeq 200$ MeV (see Fig.\ref{alpha})

\begin{figure}[ht]
 \begin{center}
 \leavevmode
  \epsfxsize=5cm
 \epsffile{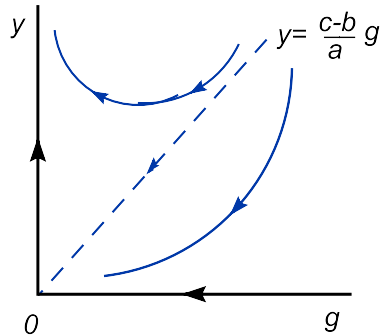}
 \end{center}\vspace{-0.6cm}
 \caption{The behaviour of the Yukawa and gauge couplings for various initial conditions
  \label{ph}}
 \end{figure}

In four-dimensional space the asymptotic freedom occurs only in non-Abelian gauge
theories. But in the case when one has several interactions, like in the Standard Model,
the non-Abelian coupling may draw other couplings into the asymptotically free region.
Consider, for instance, the behaviour of the Yukawa couplings in the SM. For simplicity,
let us take a single Yukawa coupling for the t-quark and a single gauge coupling. Then in
the one-loop approximation the equations for the effective couplings look like
\begin{eqnarray}\label{eq}
  \frac{dg}{dt}&=&-bg^2, \ \ \  g\equiv \frac{g^2_s}{16\pi^2}, \\
\frac{dy}{dt}&=&y(ay-cg),\ \ \  y\equiv \frac{y^2_t}{16\pi^2}, \ \
t\equiv\log\frac{q^2}{q_0^2}, \nonumber
\end{eqnarray}
where the coefficients $b,a$ and $c$ are always positive and for the SM are equal to
$7,9/2$ and $8$, respectively. The solutions to these equations are
\begin{eqnarray}\label{sol}
  g&=&\frac{g_0}{1+bg_0t}, \ \ y=\frac{y_0E}{1-ay_0F}, \\
   E(t)&=&(g/g_0)^{c/b},  \ \ F(t)=\int_0^tE(t')dt'.\nonumber
\end{eqnarray}
In the case of a single Yukawa coupling it can be written in an explicit form
\begin{equation}\label{yuk}
  y=\frac{y_0(\frac{g}{g_0})^{c/b}}{1+\frac{y_0}{g_0}\frac{a}{c-b}
  [(\frac{g}{g_0})^{c/b-1}-1]}.
\end{equation}
Graphically, it can be presented in a phase diagram shown in Fig.\ref{ph}.
For the initial condition such that $y_0>(c-b)/a\ g_0$ the Yukawa coupling increases with
 momenta and has the Landau pole, while for $y_0\leq(c-b)/a\ g_0$ it
demonstrates the asymptotically free behaviour. In a similar way in the Grand Unified
Theories one can reach the asymptotic freedom for all the couplings.

The back side of the asymptotic freedom at high energies is the presence of a pole at low
energies or the infrared pole. In this region, we also go beyond the validity of
perturbation theory since the coupling increases. To find the true behaviour of the
coupling one has to attract independent nonperturbative information. However, in QCD the
region near the infrared pole $p\sim \Lambda_{QCD}$ is in the phase of hadronization,
i.e., in this region the quark-gluon description is no more adequate. Therefore, the
behaviour of the effective coupling in this region  is not described by perturbative QCD.

\subsection{The screening and anti-screening of the charge}

The variation of the coupling with momenta transfer or with the scale, which is the
characteristic feature of quantum field theory, has its analog in a classical theory.
This analogy allows one to understand the qualitative reason for the variation of the
coupling.

\begin{figure}[ht]
 \begin{center}
 \leavevmode
  \epsfxsize=15cm
 \epsffile{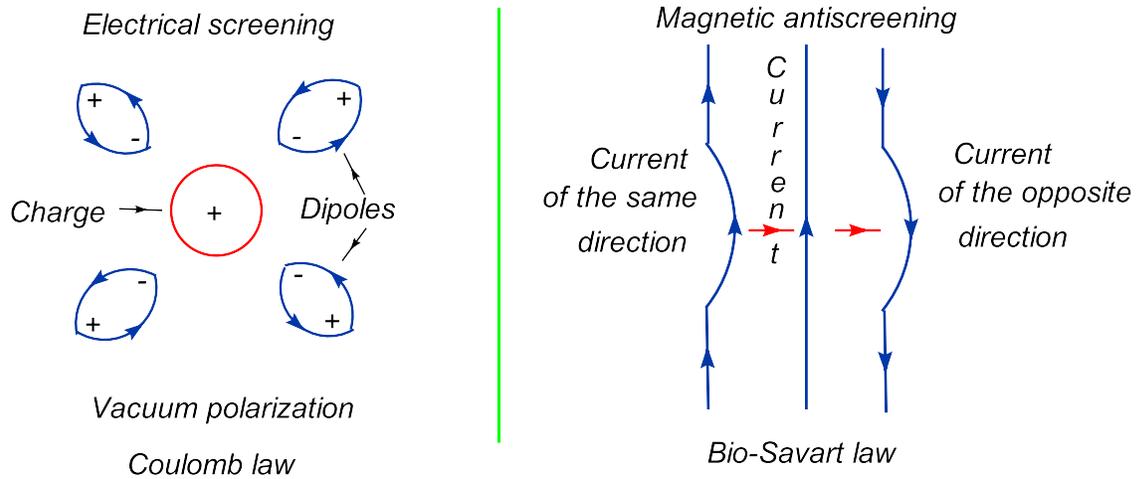}
 \end{center}
 \caption{The electric screening and magnetic anti-screening}
 \label{fig:run}
 \end{figure}

Indeed, let us consider the electromagnetic phenomena. Consider the dielectric medium and
put the test electric charge in it. The medium will be polarized. The electric dipoles
present in the medium will be rearranged in such a way as to screen the charge (see
Fig.\ref{fig:run}). This is a consequence of the Coulomb law: the opposite charges are
attracted and the same charges are repulsed. This is the essence of the electric
screening phenomena.

The opposite situation occurs in magnetic medium. According to the Bio-Savart law, the
electric currents of the same direction are attracted and the opposite direction are
repulsed (see Fig.\ref{fig:run}). This leads to the anti-screening in magnetic medium.

In quantum field theory the role of the medium is played by the vacuum. The vacuum is
polarized in the presence of created virtual pairs. The matter particles as well as
transversely polarized quanta of the gauge fields act like the electric dipoles in the
dielectric and cause the screening of the charge. At the same time, the longitudinal
quanta of the gauge fields behave like currents and cause the anti-screening. These two
effects are in competition (see eq.(\ref{res2}) above) and, for instance, in QCD with a
small number of quarks the effect of anti-screening prevails.

Thus, the couplings become the functions of the distance or momentum transfer described
by the renormalization group equations.

\newpage
\vspace*{1cm}
\section{Lecture VIII: Anomalies}
\setcounter{equation}{0}

The gauge invariance leads to numerous relations between various operators and their
vacuum averages, i.e., the Green functions. We have already come across such relations
called the Ward or the Slavnov-Taylor identities. They are the consequences of the gauge
symmetry of the classical theory. In case when one has divergences in a theory and is
bound to use some regularization, the validity of these identities depends on invariance
of the regularization. However, one can always perform the subtraction of divergences in
such a way that the finite parts obey these relations.

The exception from this rule is the so-called {\it anomalies}. By anomalies one usually
means the violation in quantum theory of some relation, for instance, the conservation of
the current or the Ward identity following from the symmetry properties of a classical
theory. The well-known examples of quantum anomalies is the anomaly of the trace of the
energy-momentum tensor or the axial anomaly. The characteristic feature of the anomaly is
the impossibility of its removing  by the redefinition of any quantities or parameters.

\subsection{The axial anomaly}

Consider quantum electrodynamics. Let us define the vector and the axial vector currents
\begin{equation}\label{cur}
  j_\mu=\bar \psi \gamma^\mu \psi, \ \ \ j^5_\mu=\bar \psi \gamma^\mu \gamma^5\psi.
\end{equation}
In classical theory the equations of motion lead to the conservation or partial
conservation of the current
\begin{equation}\label{cons}
  \partial_\mu  j_\mu=0, \ \ \  \partial_\mu  j_\mu^5=2imj^5,
\end{equation}
where $j^5=\bar \psi \gamma^5\psi$.

On the other hand, as a  consequence of the gauge invariance, the vector and the axial
vertices obey the Ward identities
\begin{eqnarray}
(p-p')^\mu\Gamma_\mu(p,p')&=&S^{-1}(p)-S^{-1}(p'), \label{iden}\\
(p-p')^\mu\Gamma_\mu^5(p,p')&=&S^{-1}(p)\gamma^5+\gamma^5S^{-1}(p')+2m\Gamma^5(p,p')
 \label{iden2},
\end{eqnarray}
where  $\Gamma_\mu,\Gamma_\mu^5$ and $\Gamma^5$ are the vector, axial and pseudoscalar
vertices, respectively, and  $S$ is the fermion propagator.

\begin{figure}[ht]
 \begin{center}
 \leavevmode
  \epsfxsize=9cm
 \epsffile{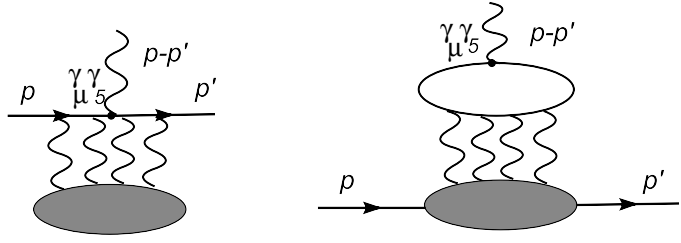}
 \end{center}
 \caption{The diagrams with the axial current in external and internal fermion lines}
 \label{anom}
 \end{figure}

\begin{figure}[ht]
 \begin{center}
 \leavevmode
  \epsfxsize=3.5cm
 \epsffile{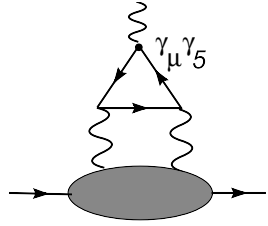}
 \end{center}
 \caption{The anomalous triangle diagram for the axial current}
 \label{tr}
 \end{figure}

If one looks how the identities (\ref{iden},\ref{iden2}) are fulfilled in perturbation
theory, one first of all has to introduce some regularization due to the presence of the
ultraviolet divergences. If the regularization is gauge invariant, then the vector Ward
identity is satisfied in any order of PT. For the axial identity there are two types of
diagrams: in the first one the axial current is in the outgoing fermion line, and in the
second one the axial current is in the internal loop  (see Fig.\ref{anom}).
For the first type of a diagram the identity (\ref{iden2}) is satisfied, and for the
second type there exists \underline{one} famous triangle diagram (see Fig.\ref{tr}) where
it is violated due to the ultraviolet divergence of the  integral.
\\

Indeed, the corresponding integral in momentum space looks like \\
\hspace*{0.7cm}\epsffile{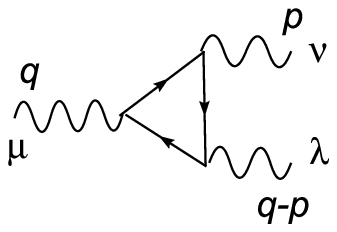}\vspace{-2.5cm}
\begin{equation}\label{int}
\hspace{3cm} = (-)(-ie)^2\int \frac{d^4k}{(2\pi)^4}Tr\left[\gamma^\mu\gamma^5 \frac{i\hat
k}{k^2}\gamma^\nu\frac{i(\hat k+\hat p)}{(p+k)^2}\gamma^\lambda\frac{i(\hat k+\hat
q)}{(q+k)^2}\right]
\end{equation}
\vspace{0.5cm}

\noindent and is formally divergent requiring the regularization.

To preserve the conservation of the gauge invariance, it is useful to introduce the
dimensional regularization; however, here we for the first time face a problem since the
$\gamma^5$ matrix has no natural and consistent continuation to non-integer dimension.
Two properties of the $\gamma^5$ matrix, namely, the anticommutation with all
 $\gamma^\mu$, $\mu=0,1,2,3$ and the property of the trace
$Tr(\gamma^5\gamma^\mu\gamma^\nu\gamma^\rho\gamma^\sigma)=-4i\epsilon^{\mu\nu\rho\sigma}$
are in contradiction if the dimension is noninteger. To calculate the axial anomaly, we
use the following trick: we use the formula for the trace but reject the property of
anticommutativity of  $\gamma^5$. This allows one to perform al the calculations in a
consistent and unambiguous way.

The divergence of the axial current can be obtained by multiplication of (\ref{int}) by
$iq^\mu$ which gives
\begin{equation}\label{int2}
e^2\int \frac{d^4k}{(2\pi)^4}\frac{Tr\left[\hat q\gamma^5 \hat k\gamma^\nu(\hat k+\hat
p)\gamma^\lambda(\hat k+\hat q)\right]}{k^2(k+p)^2(k+q)^2}
\end{equation}
Using the cyclic property of the trace we move  $\hat q$ to the right and write it as
$\hat q=(\hat q+\hat k)-\hat k$. Then the first term multiplied by $\hat k +\hat q$ gives
$(k+q)^2$ and cancels with the denominator.  As a result, one gets the integral
$$ \int\frac{d^4k}{(2\pi)^4}\frac{Tr\left[\hat q\gamma^5 \hat k\gamma^\nu(\hat k+\hat
p)\gamma^\lambda \right]}{k^2(k+p)^2},$$ which depends only on $p$ and after the
integration turns to zero due to the antisymmetry of the trace with the $\gamma^5$
matrix.

In the second term we will drag  $\hat k$ to the left until it is multiplied by $\hat k$
giving $k^2$. As a result, at each step we always get the trace of four $\gamma$-matrices
with $\gamma^5$ for which we have the formula with the $\epsilon$-tensor. We obtain in
the numerator
$$-4i\epsilon^{\alpha\nu\beta\lambda}k^\alpha(k+p)^\beta [(k+q)^2-q^2]
+8i\epsilon^{\alpha\nu\beta\rho}k^\alpha(k+p)^\beta q^\rho k^\lambda
-4i\epsilon^{\alpha\nu\lambda\rho} k^\alpha q^\rho [(k+p)^2-p^2]$$
$$ -4i\epsilon^{\nu\alpha\lambda\rho} p^\alpha q^\rho k^2
+8i\epsilon^{\alpha\beta\lambda\rho}k^\alpha p^\beta q^\rho k^\nu.$$ Despite the fact
that the integral is formally divergent, using a dimensional regularization and
collecting all terms together we finally get the finite answer equal to
\begin{equation}\label{int3}
-\frac{e^2}{4\pi^2}\epsilon^{\mu\nu\rho\lambda}p^\mu
q^\rho=-\frac{e^2}{4\pi^2}\epsilon^{\mu\nu\rho\lambda}p^\mu (q-p)^\rho,
\end{equation}

One has to add to this expression the same diagram but with the replacement $p
\leftrightarrow q-p, \nu\leftrightarrow\lambda$ and take  the sum, but  the answer is
already invariant with respect to this replacement.  Multiplying (\ref{int3}) by
$A_\nu(p)A_\lambda((q-p)$ and transforming to the coordinate representation, one gets
\begin{equation}\label{div}
  \partial_\mu j^5_\mu=\frac{e^2}{4\pi^2}\epsilon^{\mu\nu\rho\lambda}\partial_\mu A_\nu
  \partial_\rho A_\lambda=\frac{e^2}{16\pi^2}\epsilon^{\mu\nu\rho\lambda}F_{\mu\nu}
  F_{\rho\lambda}.
\end{equation}

As a result one has the following modification of  equations for the divergence of the
axial current and the axial vertex
\begin{equation}\label{anomcur}
  \partial_\mu
  j_\mu^5=2imj^5+\frac{\alpha}{4\pi}F_{\mu\nu}F_{\rho\sigma}\epsilon^{\mu\nu\rho\sigma},
\end{equation}
\begin{equation}\label{iden3}
(p-p')^\mu\Gamma_\mu^5(p,p')=S^{-1}(p)\gamma^5+\gamma^5S^{-1}(p')+2m\Gamma^5(p,p')-i
\frac{\alpha}{4\pi}F(p,p'),
\end{equation}
where $F(p,p')$ is the vertex with insertion of the operator $F\tilde F$. The appearance
of the r.h.s in these equations is called {\it anomaly} known as the Adler-Bell-Jackiw or
triangle anomaly.

The most essential here is not the violation of the Ward identity but the fact that
subtracting the anomaly and restoring the "normal" Ward identity for the axial vertex we
violate the conservation of the vector current. In other words, it is impossible to
satisfy the conservation of axial and vector currents {\it simultaneously}.

Notice that the violation of the conservation of the axial current preserving the
conservation of the vector current (\ref{anomcur}) can be obtained by accurately
calculating the matrix element for the divergence of the axial current in x-space
splitting the arguments of the field operators. Consider the vacuum average of the
divergence of the axial current, and to avoid the singularity for the product of two
operators at coinciding points, split the arguments. Then to preserve the gauge
invariance, we have to insert between the operators the exponent of the Wilson line. The
axial current then takes the form
\begin{equation}\label{ac}
  j_\mu^5(x)=\lim_{\varepsilon\to 0}\{\bar \psi(x+\varepsilon/2)\gamma^\mu\gamma^5
  \exp[-ie\int\limits_{x-\varepsilon/2}^{x+\varepsilon/2}dz^\nu A_\nu(z)]\psi(x-\varepsilon/2)\},
\end{equation}
and for the divergence we get
\begin{eqnarray}\label{acd}
  \partial_\mu j_\mu^5(x)&=&\lim_{\varepsilon\to 0}\{\partial_\mu\bar \psi(x+\varepsilon/2)
  \gamma^\mu\gamma^5 \exp[-ie\int_{x-\varepsilon/2}^{x+\varepsilon/2}dz^\nu A_\nu(z)]
  \psi(x-\varepsilon/2)\nonumber\\&+&\bar \psi(x+\varepsilon/2) \gamma^\mu\gamma^5
  \exp[-ie\int_{x-\varepsilon/2}^{x+\varepsilon/2}dz^\nu A_\nu(z)]
  \partial_\mu\psi(x-\varepsilon/2)\\ &+&\bar \psi(x+\varepsilon/2)
  \gamma^\mu\gamma^5 [-ie\varepsilon^\nu \partial_\mu A_\nu(x)]
  \exp[-ie\int_{x-\varepsilon/2}^{x+\varepsilon/2}dz^\nu A_\nu(z)]
  \psi(x-\varepsilon/2)\}.\nonumber
\end{eqnarray}
Using the equations of motion
$$\gamma^\mu \partial_\mu \psi = -ie\hat A\psi, \ \ \
 \partial_\mu \bar \psi \gamma^\mu= ie\bar \psi \hat A$$
and keeping the terms of the order of $\varepsilon$ we find
\begin{eqnarray}\label{acdd}
  \partial_\mu j_\mu^5(x)&=&\lim_{\varepsilon\to 0}\{\partial_\mu\bar \psi(x+\varepsilon/2)
   [-ie\hat A(x+\varepsilon/2)-ie\hat A(x-\varepsilon/2)\nonumber\\
   && \hspace{3cm} -ie\varepsilon^\nu \gamma^\mu \partial_\mu A_\nu(x)]
  \gamma^5\psi(x-\varepsilon/2)\}\nonumber\\
&=&\lim_{\varepsilon\to 0}\{\bar \psi(x+\varepsilon/2)[-ie\varepsilon^\nu \gamma^\mu
(\partial_\mu A_\nu-\partial_\nu A_\mu)]\gamma^5\psi(x-\varepsilon/2)\}
\end{eqnarray}
Now we have to calculate the vacuum average over the fermion vacuum (the photon field is
assumed to be external) which means that we have to permute the fermion operators. The
permutation function of the fermion operators is singular and this is the reason for
appearance of a nonzero term similarly to the appearance of triangle anomaly due to
divergency of the integral. Indeed, calculating the propagator of the fermion in external
field and keeping the terms linear in the photon field, we get
\begin{equation}\label{fprop}
  S(y-z)=\int \frac{d^4k}{(2\pi)^4} e^{ik(y-z)}\frac{i\hat k}{k^2}\\
+\int\frac{d^4k}{(2\pi)^4} \frac{d^4p}{(2\pi)^4}e^{i(k+p)y}e^{-ikz}\frac{i(\hat k+\hat
p)}{(k+p)^2}(-ie\hat A(p)\frac{i\hat k}{k^2}+...
\end{equation}
The propagator (\ref{fprop}) is singular as $y\to z$; however, the first term does not
give a contribution to the divergence, while the second one leads to
\begin{eqnarray}\label{aver}
&&\langle \bar \psi(x+\varepsilon/2)\gamma^\mu \gamma^5 \psi(x-\varepsilon/2) \rangle =
\nonumber \\&&= \int\frac{d^4k}{(2\pi)^4}
\frac{d^4p}{(2\pi)^4}e^{ipx}e^{-ik\varepsilon}Tr[\frac{i(\hat k+\hat p)}{(k+p)^2}(-ie\hat
A(p))\frac{i\hat k}{k^2}\gamma^\mu\gamma^5]\nonumber \\
&&=\int\frac{d^4k}{(2\pi)^4} \frac{d^4p}{(2\pi)^4}e^{ipx}e^{-ik\varepsilon}
\frac{4e\epsilon^{\mu\nu\rho\sigma}(k+p)_{\nu}A_{\rho}(p)k_\sigma}{(k+p)^2k^2}.
\end{eqnarray}
To find the limit as $\varepsilon\to 0$, one can expand the integrand for large $k$,
which gives
\begin{eqnarray}\label{average}
\langle \bar \psi(x+\varepsilon/2)\gamma^\mu \gamma^5 \psi(x-\varepsilon/2) \rangle &=&
4e\epsilon^{\mu\nu\rho\sigma}\int \frac{d^4p}{(2\pi)^4}e^{ipx}p_{\nu}A_{\rho}(p)\int
\frac{d^4k}{(2\pi)^4}e^{-ik\varepsilon}\frac{k_\sigma}{k^4}\nonumber\\
&&\hspace{-4cm}=-4e\epsilon^{\mu\nu\rho\sigma}i\partial_\nu
A_\rho(x)\frac{2\varepsilon_\sigma}{16\pi^2\varepsilon^2}=-e\epsilon^{\mu\nu\rho\sigma}
iF_{\nu\rho}(x)\frac{\varepsilon_\sigma}{4\pi^2\varepsilon^2},
\end{eqnarray}
Substituting this expression into (\ref{acdd}) we find
\begin{equation}\label{re}
  \partial_\mu j^5_\mu=\lim_{\varepsilon\to 0}\{-e\epsilon^{\mu\nu\rho\sigma}
iF_{\nu\rho}(x)\frac{\varepsilon_\sigma}{4\pi^2\varepsilon^2}(-ie\varepsilon^\tau
F_{\mu\tau})\} =\frac{e^2}{16\pi^2}\epsilon^{\mu\nu\rho\sigma}F_{\nu\rho}F_{\sigma\mu},
\end{equation}
that coincides with (\ref{anomcur}).

\begin{figure}[ht]
 \begin{center}
 \leavevmode
  \epsfxsize=15cm
 \epsffile{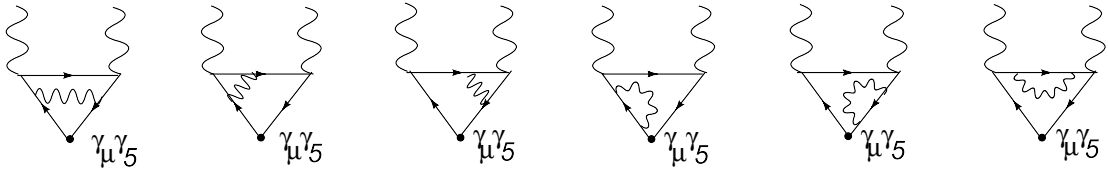}
 \end{center}\vspace{-0.3cm}
 \caption{Cancellation of radiative corrections to the axial anomaly}
 \label{anocan}
 \end{figure}

The axial anomaly has one very important property: the obtained formulas (\ref{anomcur})
and (\ref{iden3}) are {\it exact} in all orders of perturbation theory, i.e., have no
radiative corrections. More rigorous statement is: there exists such a renormalization
scheme (and it was constructed explicitly) that the radiative corrections to the axial
anomaly are absent. This statement is the subject of the Adler-Bardeen theorem.
Graphically, this means the cancellation of the contributions of the diagrams shown in
Fig.\ref{anocan},
which was checked by explicit calculation.

The Adler-Bardeen theorem is valid also in non-Abelian theories. It has important
consequences: if the anomaly is compensated in the lowest order, it will not appear
further.

\subsection{Consequences of the axial anomaly}

Let us ask the question what are the consequences of the axial anomaly? Here one has to
distinguish two cases: when the operator of the axial current is an external operator
with respect to the Lagrangian and when it is present in the interaction Lagrangian.

In the first case, the presence of anomaly does not lead to any troubles and even may be
useful. Thus, for instance, in the current algebra which describes the low energy hadron
interactions, the axial anomaly is responsible for the neutral pion decay $\pi^0\to
2\gamma$ and is in agreement with the experiment.

\begin{figure}[ht]
 \begin{center}
 \leavevmode
  \epsfxsize=3.5cm
 \epsffile{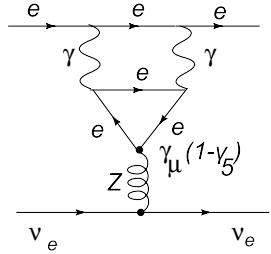}
 \end{center}\vspace{-0.3cm}
 \caption{The anomaly in the process of lepton scattering in the Standard Model}
 \label{anosc}
 \end{figure}

In the second case, the triangle anomaly leads to  that the ultraviolet renormalizations
of the vector vertex do not remove all divergences from the axial vertex. This has
destructive consequences for the renormalizability of the whole theory. To see this,
compare the two processes of the elastic scattering of leptons: $\nu_e+e\to\nu_e+e$ and
$\nu_\mu+e\to\nu_e+\mu$ in the Standard Model. Graphically, in the lowest order they
differ by one diagram containing the triangle anomaly (See Fig.\ref{anosc}).

As a result, after the renormalization the amplitude of  $\nu_\mu e$-scattering has
finite radiative corrections, while that of $\nu_е e$-scattering is divergent. This led
to nonrenormalizability of the theory and was a serious problem for the left-right
nonsymmetric model with $SU_L(2)\times U(1)$ symmetry before the introduction of the
$с$-quark. Remarkably, the $с$-quark introduced by Glashow, Iliopoulos and Maiani for
suppression of the neutral current changing strangeness leads to the compensation of the
contributions of quarks and leptons to triangle anomaly and restores the
renormalizability of the theory.

In the Standard Model due to its left-right asymmetry the presence of the axial currents
for quarks and leptons leads to several kinds of triangle anomalies where all three gauge
fields may be in the vertices of the triangle. However, not all of them lead to
anomalies. In general, the anomaly is proportional to the trace
$$ Tr\ T^a\{T^b,T^c\},$$
where the matrix $T^a$ is the generator of the corresponding gauge group in the
representation corresponding to the fields that run inside the triangle. The necessary
condition of the existence of anomaly is the presence of the complex representations and
the nontrivial anticommutator of the generators of the group. Among the simple Lie groups
which satisfy this requirement,  only the groups $SU(n),\ SO(4n+2)$ and $E_6$ have
complex representations and out of them only the  $SU(n), n>2$ and $SO(6)$ groups have a
symmetric invariant needed for the construction of the anomaly. The gauge theories built
on other groups are free from anomalies.

\begin{figure}[ht]
 \begin{center}
 \leavevmode
  \epsfxsize=10cm
 \epsffile{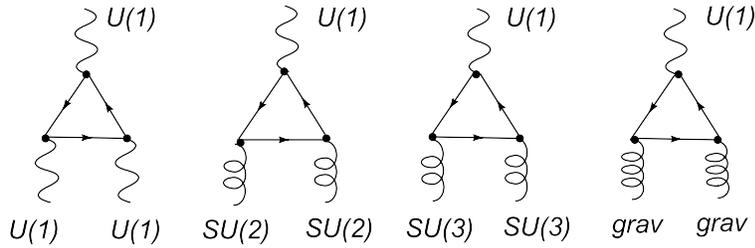}
 \end{center}\vspace{-0.3cm}
 \caption{The triangle anomaly in the Standard Model}
 \label{smanom}
 \end{figure}

The non-vanishing anomalies corresponding to the symmetry group of the Standard Model
$SU_c(3)\times SU_L(2)\times U_Y(1)$ are presented in Fig.\ref{smanom} where the gauge
fields adjusted to the groups $U(1)$ and $SU(2)$ are shown prior to mixing. The particles
that run over the triangle can be either left or right quarks and leptons. Particles of
different helicity give the opposite sign contribution to the axial anomaly.

In the first case, the anomaly is proportional to the trace of the cube of hypercharge
$Tr Y^3=TrY_L^3-TrY_R^3$ and its absence is achieved by the cancellation of the
contributions of quarks and leptons in each generation
\begin{eqnarray}\label{tracey}
  Tr Y^3=&=& 3\left[(\frac{1}{3})^3+(\frac{1}{3})^3-(\frac{4}{3})^3-
  (-\frac{2}{3})^3\right]+(-1)^3+(-1)^3-(-2)^3=0 .\nonumber\\
  && \uparrow \ \ \  \uparrow \ \ \ \ \ \ \ \ \uparrow \ \ \ \ \ \ \ \ \uparrow
   \ \ \ \ \ \ \ \ \ \
  \uparrow \ \ \ \ \ \ \ \ \ \ \ \uparrow  \ \ \ \ \ \ \ \ \ \ \uparrow
  \ \ \ \ \ \ \ \ \ \uparrow \\
  && \hspace{-0.4cm}colour \ \  u_L \ \ \ \ \ d_L \ \ \ \ \ \ u_R \ \ \ \ \ \ \ \  d_R\ \ \
  \ \ \ \ \ \  \nu_L \ \ \ \ \ \ \ \
   \ \  e_L \ \ \ \ \ \ \ \ e_R . \nonumber
\end{eqnarray}
In further diagrams the anomaly is proportional to, respectively,
\begin{eqnarray}\label{traсeanom}
  Tr Y_L&=& 3\left(\frac{1}{3}+\frac{1}{3}\right)-1-1=0 , \nonumber\\
  Tr Y_q&=& 3\left(\frac{1}{3}+\frac{1}{3}-\frac{4}{3}-(-\frac{2}{3})\right)=0,\\
  Tr Y&=& 3\left(\frac{1}{3}+\frac{1}{3}-\frac{4}{3}-(-\frac{2}{3})\right)-1-1-(-2)=0.
  \nonumber
\end{eqnarray}
This way the anomaly is miraculously canceled in all the cases and does not break the
renormalizability of the SM.

\subsection{The conformal anomaly}

Another example of quantum anomaly is the conformal anomaly or the anomaly of the trace
of the energy-momentum tensor. The requirement of conformal (scale) invariance means the
invariance of the action with respect to the transformation
\begin{equation}\label{conf}
  x_\mu \to x_\mu e^{-\sigma}, \ \ \phi(xe^{-\sigma}) \to e^{\Delta \sigma}\phi(x),
\end{equation}
where $\Delta$ is the dimension of a field. This condition is fulfilled in the classical
Lagrangian if it has no dimensional parameters. In this case, according to the Noether
theorem, there exists a conserved current called the dilatation current
$D^\mu=\Theta^{\mu\nu}x_\nu$, so that
$$\partial_\mu D^\mu=\Theta^\mu_\mu,$$
where $\Theta^\mu_\nu$ is the symmetric energy-momentum tensor.

The easiest way to see it is to define the energy-momentum tensor as a variation of the
action of the matter fields with respect to the space-time metric in the external
gravitational filed
\begin{equation}\label{ten}
  \Theta^{\mu\nu}=2\frac{\delta}{\delta g_{\mu\nu}}\int d^4x\ {\cal L}(x).
\end{equation}
The scale transformation can be realized as a variation of the metric
\begin{equation}\label{met}
  \ \ \
  g_{\mu\nu}(x) \to e^{2\sigma}g_{\mu\nu}(x).
\end{equation}
This means that the variation of the Lagrangian under this transformation is the trace of
$\Theta^{\mu\nu}$. The deviation of the trace of the energy-momentum tensor from zero
indicates the violation of the scale (and hence conformal) invariance.

In the quantum case, due the presence of the ultraviolet divergences the new scale
appears. This is the same phenomenon of dimensional transmutation discussed above.
Therefore, the scale invariance of the action is violated.

Since the coupling constant becomes scale dependent, its variation with the scale
 (\ref{conf}) takes the form
\begin{equation}\label{var}
  \delta g =\sigma \mu \frac{dg}{d\mu}=\sigma \beta(g).
\end{equation}
Hence, for the variation of the Lagrangian we get
\begin{equation}\label{varlag}
  \delta \ {\cal L}= \sigma \frac{\delta {\cal L}}{\delta g_i}\beta_i(\{g\}),
\end{equation}
i.e.,
\begin{equation}\label{trace}
  \partial_\mu D^\mu=\Theta^\mu_\mu=\frac{\delta {\cal L}}{\delta g_i}\beta_i(\{g\}).
\end{equation}
This relation is known as {\it the trace anomaly} of the energy-momentum tensor.

Similarly to the axial anomaly,  relation (\ref{trace}) can be checked by perturbation
theory. However, in this case the result is defined by the full  $\beta$-function
calculated in all orders of PT.
\newpage
\vspace*{1cm}
\section{Lecture IX: Infrared Divergences}
\setcounter{equation}{0}

One more problem that we encounter on the way of calculating the finite expressions for
the probabilities of physical processes is the presence of the so-called {\it infrared
divergences}. They  appear when calculating the matrix elements of the scattering matrix
on shell, i.e., when the squares of external momenta are equal to the corresponding
masses squared and the theory contains massless particles like photons or gluons. The
infrared divergences can be of two types: the divergences for small values of momenta
(the genuine infrared divergences) and the divergences at parallel momenta (the collinear
divergences). Contrary to the ultraviolet divergences, the infrared divergences have a
clear physical meaning: a massless particle with  a very small momentum can not be
registered and with momentum parallel to another particle cannot be distinguished. For
this reason in the theories with massless particles one has to define the physical
process  to be evaluated in a proper way.

\subsection{The double logarithmic asymptotics}

\begin{figure}[ht]
 \begin{center}
 \leavevmode
  \epsfxsize=16cm
 \epsffile{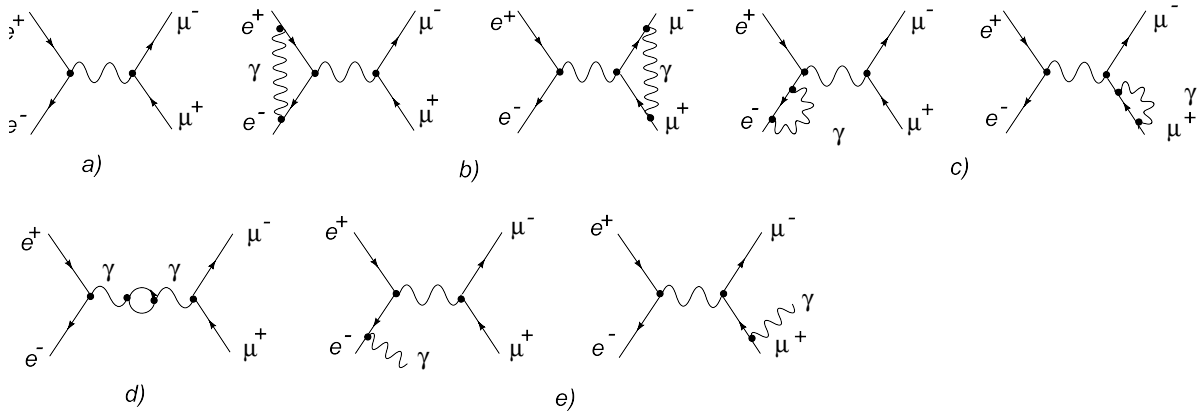}
 \end{center}\vspace{-0.3cm}
 \caption{The diagrams contributing to the process $e^+e^-\to \mu^+\mu^-$ in QED: a)
 the leading order, b)- d) the virtual corrections of the order of $\alpha$, e) the real
 corrections of the order of $\alpha$.}
 \label{mu}
 \end{figure}

For illustration consider the process of creation of a muon pair in the  $e^+e^-$
annihilation. The leading diagrams for this process are shown in Fig.\ref{mu}.

The first diagram is the tree amplitude, it gives the contribution in the leading order.
The radiative corrections due to emission of virtual photons (Fig.\ref{mu} b)) are the
corrections to the vertex function considered above (see (\ref{ver4})). It is easy to see
that if one puts in this formula all fermion momenta on mass shell, i.e.
$p^2=(p-q)^2=m^2$, then in the second integral in the denominator one gets
$[-m^2x^2+q^2y(x-y)]$. Performing the change of variables  $y\to yx$ so that all the
integrations are performed within the limits [0,1], we get $[-m^2x^2+q^2x^2y(1-y)]$, and
the integral (with account of the Jacobian  = x) is logarithmically divergent as $x\to
0$.

The appeared divergence has the infrared nature. Like the ultraviolet one it can be
regularized, for instance, by introducing the nonzero photon mass or cutting the integral
over momenta at the lower limit, or with the help of dimensional regularization but it
cannot be removed by any renormalization.

Let us calculate this diagram on mass shell introducing the nonzero photon mass $m_{ph}$
into the virtual photon line. This will not  break the gauge invariance since, as it will
be clear later, after the cancellation of the IR divergences one can put the mass of a
photon equal to zero.

Let us go back to eq.(\ref{ver4}), remove the UV divergence by the minimal subtraction
and go to the mass shell for the fermion fields taking into account that the external
fermion operators obey the Dirac equation $(\hat p -m)u(p)=0$ and $ \bar{u}(p-q)(\hat p -
\hat q - m)=0$. Then after some exercise we obtain for the vertex function the following
expression:
\begin{equation}\label{form}
\Gamma^{R}_1(p,q)=ie\left[F_1(q^2)\gamma^\mu
+iF_2(q^2)\frac{\sigma^{\mu\nu}q^\nu}{2m}\right], \ \ \ \ \ \sigma^{\mu\nu}\equiv
i\frac{\gamma^\mu\gamma^\nu-\gamma^\nu\gamma^\mu}{2},
\end{equation}
where the form-factors $F_i(q^2)$ have the form
\begin{eqnarray}\label{for}
 F_1(q^2) &=&\frac{e^2}{16\pi^2}\left[-2 -2\int_0^1dx\!\! \int_0^1 \!\!dy\ x\
 \log\left(\frac{-m^2x^2+q^2x^2y(1-y)}{-\mu^2}\right) \right.
 \nonumber\\
 &+&\left.
 \int_0^1\!\!dx \int_0^1\!\! dy\ x\
   \frac{2m^2(2\!-\!2x\!-\!x^2)\!-\!2q^2(1\!-\!xy)(1\!-\!x\!+\!xy)}{-m^2x^2+q^2x^2y(1-y)-m_{ph}^2(1-x)}\right],\\
F_2(q^2)&=&\frac{e^2}{16\pi^2}\left[\int_0^1\!\!dx \int_0^1\!\! dy\ x\
   \frac{-4m^2x(1-x)}{-\!m^2x^2\!+\!q^2x^2y(1\!-\!y)\!-\!m_{ph}^2(1\!-\!x)}\right].
\end{eqnarray}

The form factor $F_2$ is IR convergent and does not need any regularization. Substituting
$m_{ph}=0$, we get
\begin{equation}\label{f2}
F_2(q^2)=\frac{\alpha}{4\pi}\int_0^1 dy
   \frac{2m^2}{m^2-q^2y(1-y)}.
\end{equation}
For  $q^2=0$ it can be easily calculated and equals
\begin{equation}\label{anommag}
  F_2(q^2=0)=\frac{\alpha}{2\pi},
\end{equation}
which is nothing else but the first correction to the g-factor, which is called the
anomalous magnetic moment of electron (muon).

As for the form factor $F_1$, it is IR divergent. We calculate its divergent part in the
limit $m_{ph}\to 0 $. It comes only from the second integral in (\ref{for}). To simplify
the integration, we notice that the divergence is defined by the region of the parameter
$x\sim 0$. Therefore, we put $x=0$ everywhere in the numerator and in the coefficient of
 $m_{ph}$ in the denominator. Then one gets
\begin{equation}\label{f10}
 F_1(q^2) \simeq \frac{e^2}{16\pi^2}\int_0^1dy \int_0^1 x dx
   \frac{2(2m^2-q^2)}{[-m^2+q^2y(1-y)]x^2-m_{ph}^2}.
\end{equation}
The integral over $x$ is now easily evaluated
\begin{equation}\label{f11}
 F_1(q^2) \simeq \frac{\alpha}{4\pi}\int_0^1dy
   \frac{2m^2-q^2}{[-m^2+q^2y(1-y)]}\log
   \left(\frac{-m^2+q^2y(1-y)-m_{ph}^2}{-m_{ph}^2}\right).
\end{equation}
The remaining integral over $y$ is also simple. We calculate it in the limit
$-q^2\to\infty$. Then it takes the form
\begin{equation}\label{f12}
 F_1(q^2) \simeq -\frac{\alpha}{4\pi}\int_0^1\!\!dy
   \frac{q^2}{[-\!m^2\!+\!q^2y(1\!-\!y)]}\log
   \left(\!\frac{-q^2}{m_{ph}^2}\!\right) \simeq \! -\!\frac{\alpha}{2\pi}
   \log\left(\frac{-q^2}{m^2}\right)
  \log\left(\frac{-q^2}{m_{ph}^2}\right).
\end{equation}
 The obtained double logarithmic behaviour of the form-factor is called {\it the
Sudakov} double logarithm. It contains the infrared cutoff in the form of the photon
mass. In the amplitude of creation of the muon pair there are two of such form factors for
the electron and the muon vertices, respectively. The corrections to the fermion and the
photon propagators do not contain the IR divergences. Thus, the cross-section of the
process $e^+e^-\to\mu^+\mu^-$ is logarithmically divergent. In order to understand the
reason of appearance of the IR divergence and to find the method of its elimination,
consider the process of creation of the muon pair from the point of view of an observer.

\subsection{The soft photon emission}

During the process of electron-positron annihilation the muon pair is created with
momenta that satisfy the conservation law and can be measured. However, they are
registered with some accuracy, and momentum smaller than some value which depends on a
particular detector is not registered.  Therefore, if besides the muon pair the photon
with momentum smaller than this value is created, then this process with emission of the
"soft" $\gamma$-quantum  $e^+e^-\to \mu^+\mu^-\gamma$ is experimentally indistinguishable
from the initial process  $e^+e^-\to \mu^+\mu^-$. The diagrams corresponding to the
process $e^+e^-\to \mu^+\mu^-\gamma$ are shown in Fig.\ref{mu} e). They contain an
additional vertex and hence additional coupling, but being squared give a correction to
the main process of the order of $\alpha$, exactly as the radiative corrections due to
the virtual photon.

Let us compare the differential cross-sections of the precess $e^+e^-\to \mu^+\mu^-$ in
the one-loop approximation and $e^+e^-\to \mu^+\mu^-\gamma$ in the tree approximation. We
have, respectively,
\begin{eqnarray}\label{xsec}
 &&\hspace{-1cm} \frac{d\sigma}{d\Omega}({e^+e^-\to \mu^+\mu^-})=
 \left(\frac{d\sigma}{d\Omega}\right)_0\!
  \left[1\!-\!\frac{\alpha}{\pi} \log\left(\!\frac{-q^2}{m^2_{e,\mu}}\!\right)
  \log\left(\!\frac{-q^2}{m_{ph}^2}\!\right)\!+\! ...\! +\! {\cal O}(\alpha^2)\right] \\
&&\hspace{-1.2cm}\frac{d\sigma}{d\Omega}({e^+e^-\to
\mu^+\mu^-\gamma})=\left(\frac{d\sigma}{d\Omega}\right)_0\!
  \left[\!+\!\frac{\alpha}{\pi} \log\left(\!\frac{-q^2}{m^2_{e,\mu}}\!\right)
  \log\left(\!\frac{-q^2}{m_{ph}^2}\!\right)\!+\! ...\!+\! {\cal O}(\alpha^2)\right]\label{xsec2}
\end{eqnarray}
where the second cross-section is written down without derivation which we will perform
later. As follows from eqs.(\ref{xsec},\ref{xsec2}), each of these cross-sections is IR
divergent, but in the sum the divergences cancel and one gets the finite answer.

What is observable after all? In fact, neither the first nor the second process is
observable separately. In a real detector with limited sensitivity one observes the
process of creation of the muon pair plus an arbitrary number of  soft photons with the
total energy below the sensitivity threshold. In a given order of perturbation theory we
have to sum the cross-sections of the two processes in order to get the observed
cross-section
\begin{equation}\label{obs}
 \left(\frac{d\sigma}{d\Omega}\right)_{observable}=
 \left(\frac{d\sigma}{d\Omega}\right)({e^+e^-\to \mu^+\mu^-})+
  \left(\frac{d\sigma}{d\Omega}\right)({e^+e^-\to
  \mu^+\mu^-\gamma},E<E_{min}).
\end{equation}
The latter cross-section is given by the same formula (\ref{xsec2}) with the replacement
in the second logarithm of the photon energy by $E_{min}$. Thus, we get
\begin{equation}\label{obs2}
\left(\frac{d\sigma}{d\Omega}\right)_{observable}= \left(\frac{d\sigma}{d\Omega}\right)_0
  \left[1-\frac{\alpha}{\pi} \log\left(\frac{-q^2}{m^2_{e,\mu}}\right)
  \log\left(\frac{-q^2}{E_{min}^2}\right)+ ... + {\cal O}(\alpha^2)\right].
\end{equation}

As one can see, for the proper statement of the problem the cross-section of the
observable process is finite and does not depend on the IR regulator. At the same time,
it depends on the sensitivity of the detector $E_{min}$ and for improved sensitivity
tends to infinity. However, this infinity also is not physical and is the artefact of
perturbation theory: when the logarithm becomes large we go beyond the scope of
applicability of perturbation theory and it is necessary to perform the summation of
these corrections by analogy with what happens with the ultraviolet logarithms within the
renormalization group method.

\begin{figure}[ht]
 \begin{center}
 \leavevmode
  \epsfxsize=5cm
 \epsffile{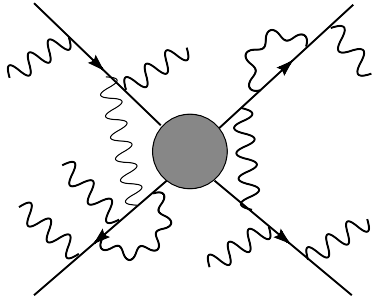}
 \end{center}\vspace{-0.3cm}
 \caption{The hard process with creation of the soft photons}
 \label{hard}
 \end{figure}

Thus, the IR divergences appear due to the contributions of the photons with "soft"
momenta: real with the energy smaller than $E_{min}$ and virtual with momenta
$k^2<E^2_{min}$. What is important is that the momenta of fermions are on mass shell,
otherwise the singularities in the propagator do not arise. The typical diagram of higher
order contains a big amount of real and virtual photon lines (see Fig.\ref{hard}).

Let us try to sum up the contributions of these soft photons. Consider first the external
fermion line with the outgoing photons (real and virtual).
\begin{figure}[ht]
 \begin{center}
 \leavevmode
  \epsfxsize=5cm
 \epsffile{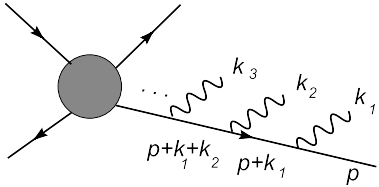}
 \end{center}\vspace{-0.5cm}
 \caption{The emission of the soft photons from the fermion line}\label{lin}
 \end{figure}

It corresponds to the following expression:
\begin{eqnarray}\label{emit}
  &&\bar u(p)\ (-ie\gamma^{\mu_1})\ \frac{i(\hat p+\hat k_1+m)}{2pk_1}\ (-ie\gamma^{\mu_2})\
  \frac{i(\hat p+\hat k_1+\hat k_2+m)}{2p(k_1+k_2)+O(k^2)}\ \cdots \\
 &&\hspace{2cm} \cdots\ (-ie\gamma^{\mu_n})\ \frac{i(\hat p+\hat k_1+\cdots
 +\hat k_n+m)}{2p(k_1+\cdots k_n)+O(k^2)}
  \ \ iM_{hard}\nonumber .
\end{eqnarray}
We use now the fact that the operator $\bar u(p)$ obeys the Dirac equation $\bar
u(p)(\hat p-m)=0$ and omit the momenta $k_i\ll p$ in the numerator. Then we get
\begin{equation}\label{sim}
  \bar u(p)\gamma^{\mu_1}(\hat p+m)\gamma^{\mu_2}
  (\hat p+m)\cdots= \bar u(p)2p^{\mu_1}\gamma^{\mu_2}
  (\hat p+m)\cdots=\bar u(p)2p^{\mu_1}2p^{\mu_2}\cdots.
\end{equation}
Hence, eq.(\ref{emit}) takes the form
\begin{equation}\label{emit2}
\bar u(p)\ (e\frac{p^{\mu_1}}{pk_1}) (e\frac{p^{\mu_2}}{p(k_1+k_2)})\cdots
(e\frac{p^{\mu_n}}{p(k_1+\cdots +k_n)}).
\end{equation}
The next step is the summation over all the permutations of the photon lines and the
permutations of momenta $k_i$. (So far we have not distinguished between the real and
virtual photons, we will do it later.)  This operation is non-trivial but leads to the
simple result. One has
\begin{equation}\label{perm}
  \sum\limits_{permutations}^{}\ \ \frac{1}{pk_1}\ \frac{1}{p(k_1+k_2)}\ \cdots\
  \frac{1}{p(k_1+ k_2+\cdots +k_n)}=\frac{1}{pk_1}\ \frac{1}{pk_2}\ \cdots\
  \frac{1}{pk_n}.
\end{equation}

The same procedure can be applied to the incoming fermion line. The difference is that
the fermion momentum has the opposite direction which leads to the replacement of
$(p+k_i)^2$ to $(p-k_i)^2$ in the propagator, i.e., the change of the sign $p\to -p$ in
the denominator. Collecting both factors together we get the following expression for the
amplitude of emission of soft photons from arbitrary points of the incoming and the
outgoing line (Fig.\ref{proc}):
\begin{figure}[ht]
 \begin{center}
 \leavevmode
  \epsfxsize=5cm
 \epsffile{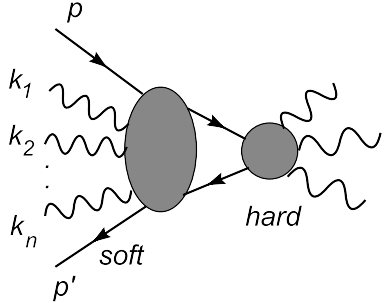}
 \end{center}\vspace{-0.5cm}
 \caption{The emission of soft photons from arbitrary points of the
incoming and the outgoing lines}\label{proc}
 \end{figure}
\begin{equation}\label{amp}
  {\cal M}=\bar u(p')\ i{\cal M}_{hard}\ u(p)\ e\left(\frac{p'^{\mu_1}}{p'k_1}-
  \frac{p^{\mu_1}}{pk_1}\right)e\left(\frac{p'^{\mu_2}}{p'k_2}-
  \frac{p^{\mu_2}}{pk_2}\right)\cdots e\left(\frac{p'^{\mu_n}}{p'k_n}-
  \frac{p^{\mu_n}}{pk_n}\right).
\end{equation}

Now we have to decide which photons are real and which are virtual. The virtual photon
can be obtained by joining the two photon momenta $k_i$ and $k_j$, taking $k_i=-k_j=k$,
multiplying by the photon propagator and integrating over $k$. In this way for any
virtual photon we get the expression:
\begin{equation}\label{virt}
  \frac{e^2}{2}\int\frac{d^4k}{(2\pi)^4}\frac{-i}{k^2}\left(\frac{p'}{p'k}-
  \frac{p}{pk}\right)\left(\frac{p'}{-p'k}-
  \frac{p}{-pk}\right),
\end{equation}
where the factor 1/2 compensates the double counting due to permutation of $k_i$ and
$k_j$. The obtained integral is nothing else but the vertex function in the one-loop
approximation, i.e., the form factor $F_1(q^2)$.

If the number of virtual photons equals $n$, one gets  the product of $n$ expressions
like (\ref{virt}) and the factor $1/n!$ taking into account the permutations which do not
change the result. The full answer is obtained with the help of summation over the soft
virtual photons, which gives\vspace{0.4cm}

\begin{equation}\label{virtpr}
\phantom{8888888888888}\times \sum\limits_{n=0}^{\infty}\frac{F_1^n}{n!}=\bar u(p')\
i{\cal M}_{hard}\ u(p)\exp(F_1).
 \end{equation}\vspace{-1.9cm}

 \epsfxsize=5cm
 \epsffile{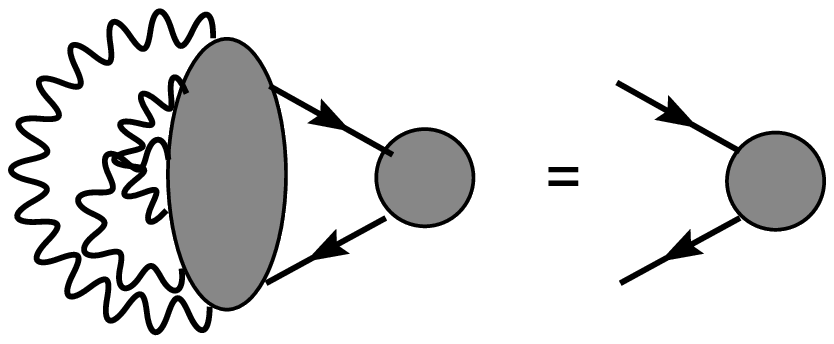}\vspace{0.2cm}

At the same time, if the real photon is emitted, then instead of the propagator one has
to multiply the amplitude by the polarization operator, sum up over all polarizations and
integrate the square of the matrix element  over the photon phase space. In this case,
one gets the following expression:
\begin{equation}\label{real}
I(q^2)=  e^2\int\frac{d^3k}{(2\pi)^3}\frac{-g^{\mu\nu}}{2|k|}\left(\frac{p'^\mu}{p'k}-
  \frac{p^\mu}{pk}\right)\left(\frac{p'^\nu}{p'k}-
  \frac{p^\nu}{pk}\right),
\end{equation}
which is the element of the cross-section of the process $e^+e^-\to  \mu^+\mu^-\gamma$.
The integration over the modulus of the three-vector $\vec k$ has to be performed within
the limits $(m_{ph},E_{min})$. Contracting the indices one gets
\begin{equation}\label{real2} I(q^2)=
-\frac{e^2}{(2\pi)^3}\int\frac{d^3k}{2|k|}\left(\frac{p'^2}{(p'k)^2}-
  2\frac{p'p}{(pk)(p'k)}+\frac{p^2}{(pk)^2}\right).
\end{equation}
The first and the last integrals are equal to each other. Let us consider the last one
and choose the frame where $\vec p=0$. This gives
\begin{equation}\label{real3}
  I_1=-\frac{e^2}{(2\pi)^3}4\pi\int_{m_{ph}}^{E_{min}}\frac{k^2dk}{2k}
  \frac{m^2}{(mk)^2}=-\frac{\alpha}{2\pi}\log(\frac{E_{min}^2}{m_{ph}^2}).
\end{equation}
As for the second integral, we proceed in the following way: first we also choose the
frame $\vec p=0$, and then we covariantize the answer. One has
\begin{eqnarray}\label{real4}
  I_2&=&\frac{e^2}{(2\pi)^3}2\pi\int_{m_{ph}}^{E_{min}}\frac{k^2dk}{k}\int_{-1}^{1}
  d\cos\theta \frac{m\sqrt{\vec{p^\prime}^2+m^2}}{(mk)(\sqrt{\vec{p^\prime}^2
  +m^2}k-|\vec{p^\prime}|k\cos\theta)}\nonumber\\
  &=&\frac{\alpha}{2\pi}\log(\frac{E_{min}^2}{m_{ph}^2})
  \frac{\sqrt{\vec{p^\prime}^2+m^2}}{|\vec{p^\prime}|}
  \log\left(\frac{\sqrt{\vec{p^\prime}^2+m^2}-|\vec{p^\prime}|}{\sqrt{\vec{p^\prime}^2
  +m^2}+|\vec{p^\prime}|}\right).
\end{eqnarray}
Covariantizing this answer and having in mind that $q=p-p',\ p^2=p'^2=m^2$ and, hence,
$q^2=2m^2-2m\sqrt{\vec{p^\prime}^2+m^2}$ one gets
\begin{equation}\label{cov}
I_2(q^2)=\frac{\alpha}{2\pi}\log(\frac{E_{min}^2}{m_{ph}^2})
\frac{2m^2-q^2}{\sqrt{-q^2(4m^2-q^2)}}
  \log\left(\frac{2m^2-q^2-\sqrt{-q^2(4m^2-q^2}}{2m^2-q^2+\sqrt{-q^2(4m^2-q^2}}\right).
\end{equation}
Thus,
\begin{equation}\label{realf}
I(q^2)=\frac{\alpha}{2\pi}\log(\frac{E_{min}^2}{m_{ph}^2})\left[
\frac{2m^2-q^2}{\sqrt{-q^2(4m^2-q^2)}}
  \log\left(\frac{2m^2\!-\!q^2\!-\!\sqrt{-\!q^2(4m^2\!-\!q^2}}{2m^2\!-\!q^2\!+
  \!\sqrt{-\!q^2(4m^2\!-\!q^2}}\right)
  \!-\!2\right].
\end{equation}
In the limit $-q^2\to\infty$ we get the desired answer
\begin{equation}\label{realff}
I(q^2)\to \frac{\alpha}{\pi}\log(\frac{E_{min}^2}{m_{ph}^2})
  \log(\frac{-q^2}{m^2}),
\end{equation}
coinciding with (\ref{xsec2}).

If there are $n$ real photons, there are $n$ such contributions and the symmetry factor
$1/n!$ taking into account the identity of the final particles. The cross-section of the
process with emission of an arbitrary number of photons with the energy smaller than
$E_{min}$ hence equals
\begin{equation}\label{xsec3}
  \sum\limits_{n=0}^{\infty}\frac{d\sigma}{d\Omega}(e^+e^-\!\!\to\!\mu^+\mu^-\!+n\gamma)=
\frac{d\sigma}{d\Omega}(e^+e^-\!\!\to\!\mu^+\mu^-)\times\sum\limits_{n=0}^{\infty}\frac{I^n}{n!}
=\frac{d\sigma}{d\Omega}(e^+e^-\!\!\to\!\mu^+\mu^-)e^I.
\end{equation}

Combining the results for the real and virtual photons one gets the final expression for
the observable cross-section with emission of an arbitrary number of photons with the
energy smaller than $E_{min}$
\begin{eqnarray}\label{obs3}
\left(\frac{d\sigma}{d\Omega}\right)_{observable}&=&
\left(\frac{d\sigma}{d\Omega}\right)_0 \times
  \exp{(2F_1)}\times \exp{(I)}\nonumber \\
  &&\hspace*{-4cm}=\left(\frac{d\sigma}{d\Omega}\right)_0
  \exp\left[-\frac{\alpha}{\pi} \log(\frac{-q^2}{m^2_{e,\mu}})
\log(\frac{-q^2}{m_{ph}^2})\right]
\exp\left[\frac{\alpha}{\pi}\log(\frac{-q^2}{m^2_{e,\mu}})
\log(\frac{E_{min}^2}{m_{ph}^2})\right]\nonumber\\
&&\hspace*{-4cm}=\left(\frac{d\sigma}{d\Omega}\right)_0
  \exp\left[-\frac{\alpha}{\pi} \log(\frac{-q^2}{m^2_{e,\mu}})
\log(\frac{-q^2}{E_{min}^2})\right].
\end{eqnarray}
The obtained expression is valid in all orders of perturbation theory. The exponential
factor does not depend on the IR cutoff but on the sensitivity of the detector. It is
called the {\it Sudakov form factor}. When $E_{min}$ tends to zero, the form factor
decreases and in the limit $E_{min}\to 0$ vanishes. This is the manifestation of the
statement that he amplitude of creation of the fermion pair without accompanying soft
photons indeed vanishes: the charged particle inevitably emits the low frequency
electromagnetic waves. This means that the cross-section of elastic electron scattering
without inclusion of emission of bremsstrahlung quanta should vanish, precisely as it
follows from eq.(\ref{obs3}).

Let us estimate the value of the Sudakov form factor for some real process. A good example
is the cross-section of $e^+e^-$ annihilation into hadrons which in the leading order in
the fine structure constant is described by one diagram with $Z$-boson exchange in the
$s$-channel. The cross-section has a maximum in the $Z$-boson peak where it is described
by the Breit-Wigner resonance formula. The energy is equal to the $Z$-boson mass $M_Z$
and the energy resolution is defined by the $Z$-boson width  $\Gamma_Z$. Substituting the
values $M_Z=91.187$ GeV, $\Gamma_Z=2.496 $ GeV, $m_e=0.5$ MeV, $\alpha=1/128$ into the
form factor (\ref{obs3}) we get
$$ \exp\left[-\frac{\alpha}{\pi} \log(\frac{M_Z^2}{m^2_{e}})
\log(\frac{M_Z^2}{\Gamma_Z^2})\right] \approx 0.648.$$ As one can see, the form factor,
despite the smallness of the fine structure constant, considerably departs from unity and
has to be taken into account when analysing the experimental data.

\subsection{The cancellation of the infrared divergences}

The considered example is typical of the QED and one can make the general statement
concerning the infrared divergences for the elements of the $S$-matrix.

{\it The infrared divergences in radiative corrections to the cross-section of any
physical process in QED are cancelled in every order of perturbation theory if to the
cross-section of the elastic process one adds the inelastic cross-section of the process
with emission of an arbitrary number of additional photons integrated over the phase
space with the requirement that the total photon energy does not exceed some value
$E_{min}$.}

This statement is also valid for the cross-sections of the processes in non-Abelian gauge
theories like the electroweak theory and some processes in QCD, though in this case, due
to the self-interaction of the non-Abelian gauge fields, there is no full factorization
with the exponentiation, and the proof of this statement presents some problem.
Nevertheless, for many processes the result has the same form. Thus, for example, the
electromagnetic form-factor in QCD has the same Sudakov form  (\ref{obs3}) but with the
replacement $\alpha\to C_F \alpha_s$.

Thus, one can say that the problem of obtaining the ultraviolet and the infrared finite
radiative corrections to the cross-sections of the physical processes is solved in two
steps: first, with the help of the renormalization procedure one gets rid of the
ultraviolet divergences, which is under full control in renormalizable theories; second,
defining the correct physical process including the emission of the soft quanta, the
cancellation of the infrared divergences takes place.

As we will see below, this is not sufficient in non-Abelian gauge theories with massless
gauge fields. They contain additional divergences which require some ads-inn to the
described procedure. We will consider this question in the last lecture.

\newpage
\vspace*{1cm}
\section{Lecture X: Collinear Divergences}
\setcounter{equation}{0}

\subsection{The collinear divergences in massless theory}

The obtained result  (\ref{obs3}) for the cross-section of creation of the muon pair in
the process of $e^+e^-$-annihilation with emission of additional soft photons is typical
of the theories with a massive fermion and massless photons. It can be generalized to
non-Abelian theories with massless gluon, though the gluon interactions cause some
problems in proving the cancellation of the IR divergences. Note, however, that eq.
(\ref{obs3}) contains the logarithmic singularity with respect to the fermion mass, and
if the latter tends to zero, one has the new divergence. This would not cause any problem
since all the fermions are massive but the masses of the electron and the light quarks
are so small compared to the characteristic energies of the scattering process that with
good precision it is reasonable to neglect them.  As for the QCD, considering the
processes with gluons in initial states due to the self-interaction of the gluons we face
this problem for the gluon amplitudes.

Let us analyse what is the reason for the appearance of the new divergence after the IR
divergence at small photon momenta if regularized by introducing the photon mass.
Consider for this purpose eq. (\ref{amp}) for the contribution of the real or virtual
photons. The difference is that in  one case the integration goes over the four-momentum
of the virtual photon; and in the other case, over the three-momentum of the real photon,
but what is essential that for the massless electron its propagator takes the form
\begin{equation}\label{col}
  \frac{1}{2pk}=\frac{1}{2(p^0k^0-\vec p\vec k )}\simeq \frac{1}{2(|\vec p||\vec k|-
  |\vec p||\vec k|\cos\theta)}=\frac{1}{2|\vec p||\vec k|(1-\cos\theta)},
\end{equation}
where $\theta$ is the angle between the electron and photon momenta. (In the case of a
virtual photon we use the fact that the contribution to the singularity comes from the
region of photon momentum close to the mass shell.)

Thus, the divergence appearing in the massless case comes from the integration over the
angles and not over the modulus, as in the case of the IR divergence, and is related to
the collinearity of momenta of two particles. For this reason it is called {\it the
collinear divergence}. To get rid of these divergences, one can  introduce the angular
sensitivity of the detector analogously to the IR divergence. This would reflect the fact
that two massless particles having almost parallel momenta are not distinguishable from a
single particle with the same total momentum. Hence, the observed cross-section should
include besides the main process the process of emission of the soft photons and the
process of emission of the collinear photons with the kinematically allowed absolute
values of momenta.

However, in real life the quarks and leptons are massive though their masses are very
small; therefore, the problem of collinear divergences occurs for the processes with the
gluon fields. Since the gluons are not free particles but exist inside  hadrons, any
process with the gluons has  a similar process with quarks and it is reasonable to
consider them together. For this reason, one usually speaks about  {\it the  inclusive}
processes where besides the particles of the main process one includes the creation of
all kinematically allowed particles, in particular the gluons. In this case, we do not
impose any restriction on the gluon energy, we do not introduce any detector sensitivity
to the energy or the angle, but sum over all the possibilities. It happens, however, that
this is not  sufficient to get the finite answer. It is necessary to take into account
the possibility of existence of collinear gluons in the initial state, and only after
this one can get the finite answer for the cross-section of the observable process.

The multiloop analysis in this case is much more complicated and is the subject of the
Kinoshita-Lee-Nauenberg theorem which states:\\

{\it The infrared and collinear divergences in a massless theory are cancelled in the
cross-section of any process if one takes into account the existence in the {\it initial}
and {\it final} states of an arbitrary number of the soft quanta as well as the particles
having the parallel momenta with the same total momentum. The probabilities of these
processes integrated over the phase space of these additional soft (collinear) quanta in
the initial and final states should be added to the probability of the initial process.}\\

\begin{figure}[ht]\vspace{0.2cm}
 \begin{center}
 \leavevmode
  \epsfxsize=15cm
 \epsffile{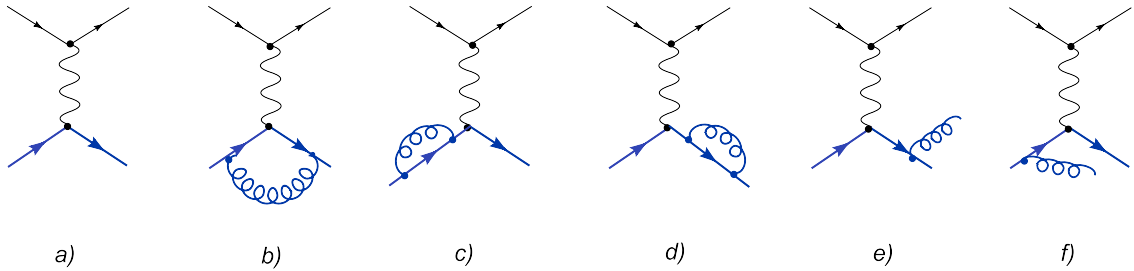}
 \end{center}\vspace{-0.5cm}
 \caption{The process of electron-quark scattering in the first order in  $\alpha_s$: а)
 the Born diagram, b)-d) the corrections due to the virtual gluons,
 e)-f) the corrections due to the real gluons
 }\label{toy}
 \end{figure}

As an illustration we consider the model example of the electron-proton (quark)
scattering and put all the masses equal to zero. We will be interested in the radiative
corrections in the first order with respect to the strong coupling $\alpha_s$. The
corresponding diagrams are shown in Fig.\ref{toy}.

We have already calculated the matrix elements corresponding to these diagrams, but now
we proceed in a different way. Since the ultraviolet divergences which appear in the
diagrams b)-d) are compensated due to the Ward identity in  QED ($Z_1=Z_2$), all the
arising divergences are solely infrared and collinear. To extract them we will use the
dimensional regularization. Then both the divergences are manifested in the form of the
poles over  $\varepsilon$ and, since we have both of them, there will be poles of the
first and the second order.

We start with the virtual corrections. The diagrams of self-energy c) and d) in the
massless case are identically zero due to the above-mentioned property of a massless
integral depending on one argument equal to zero ($p^2=0$ on the mass shell). As we
explained, here one has the cancellation of the UV and the IR divergences. Therefore, all
divergences in the vertex diagram b) may be considered as infrared. (The UV divergences
should  cancel with the UV ones from the self-energy diagrams and the latter in their
turn cancel with the IR).  The integral for the vertex part is defined by two form factors
$F_1(q^2)$ and $F_2(q^2)$ (\ref{form}). Taking  the expression for the vertex function
(\ref{ver3}) as the starting point, we put $m=0$ and go to the mass shell. The result is
\begin{eqnarray}
F_1(q^2)&=&-C_F\frac{\alpha_s}{4\pi}\left(\frac{\mu^2}{-q^2}\right)^\varepsilon
(\frac{2}{\varepsilon^2}+\frac{3}{\varepsilon}+8),\\
F_2(q^2)&=&0,
\end{eqnarray}
where instead of the logarithm of the photon mass as the IR regulator we have the pole
over $\varepsilon$. In order to avoid the transcendental numbers, we used the helpful
definition of the angular measure in the space of $4-2\varepsilon$ dimensions and
multiplied the standard expression by $\Gamma(1-\varepsilon)/(4\pi)^\varepsilon$. Then
the constants like $\gamma_E,\ log(4\pi)$ and $\zeta(2)$ disappear from the intermediate
expressions. Due to the cancellation of divergences in the final expressions, this
redefinition does not influence the answer.

Thus, the cross-section for the diagrams with virtual gluon has the form
\begin{equation}\label{virsec}
  \left(\frac{d\sigma}{d\Omega}\right)_{virt}=
\left(\frac{d\sigma}{d\Omega}\right)_0
\left[1-2C_F\frac{\alpha_s}{4\pi}\left(\frac{\mu^2}{-t}\right)^\varepsilon
(\frac{2}{\varepsilon^2}+\frac{3}{\varepsilon}+8)\right],
\end{equation}
where the differential cross-section in the Born approximation is given by
\begin{equation}\label{secborn}
\left(\frac{d\sigma}{d\Omega}\right)_0=\frac{\alpha^2}{2E^2}\left(\frac{s^2+u^2-\varepsilon
t^2 }{t^2}\right)\left(\frac{\mu^2}{s}\right)^\varepsilon.
\end{equation}
In the c.m. frame  $s=E^2, t=-E^2/2(1-\cos\theta), u=-E^2/2(1+\cos\theta)$, where the
angle $\theta$ is the electron scattering angle.

Consider now the diagrams with the emission of the real gluons e) and f). Besides the
squares of each of the diagrams one should also take into account the interference term.
The calculation in fact repeats that in QED but instead of the photon mass we again use
the dimensional regularization and do not restrict the integration region over the
momentum of additional gluon. The calculation is a bit tedious, after contracting all the
indices the phase integral takes the form
\begin{eqnarray}
d\sigma_{2 \rightarrow 3}& =&\frac{1}{2\pi E^2}\int\! d^{D}p_{3} \delta^{+}(p_{3}^{2})
\int\!\! \frac{d^{D}k}{(2\pi)^D}
\delta^{+}(k^{2}) \delta^{+}((p_{4}\!-\!k)^{2}) | M |^2_{p_{4}=p_{1}\!+\!p_{2}\!-\!p_{3}}\\
&&\hspace*{-2cm} |M|^2=\frac{e^4g^2}{4} 8 \frac{ M_{0} + \epsilon M_{1}+ \epsilon^2 M_{2}
}{t (s+t+u)},
\nonumber\\
&&\hspace*{-2cm} M_{0}= 4 s - 8  p_{1} k - 4 p_{2} k + \frac{- 8 (p_{1} k)^2 +4 (2 s + t)
p_{1} k - (3
s^2  + t^2 + u^2+ 2 s t)}{p_{2} k},\nonumber\\
&& \hspace*{-2cm}M_{1}=  \!-\! 4 (s\!+\!u)\! +\! 8  p_{1} k \!+\! 8 p_{2} k \!+\! \frac{
8 (p_{1} k)^2 \!-\! 4 (s \!+\! t\! +\!u) p_{1}
k\! +\! 2 ( s \!+\! t \!+\! u)^2 \!-\!2(u\!+\!s) t }{p_{2} k}\nonumber \\
&&\hspace*{-2cm}M_{2}= 4 (s+t+u)  - 4 p_{2} k - \frac{(s+t+u)^2}{ p_{2} k} = -
\frac{(s+t+u +2 p_{2} k)^2}{p_{2} k}.\nonumber
\end{eqnarray}
It is useful to pass to the spherical coordinates and use the c.m. frame. After the
integration over the phase volume the result can be represented in the form
\begin{equation}\label{realsec}
  \left(\frac{d\sigma}{d\Omega}\right)_{real}=
\left(\frac{d\sigma}{d\Omega}\right)_0\!\!
\left[2C_F\frac{\alpha_s}{4\pi}\left(\frac{\mu^2}{-t}\right)^\varepsilon\!\!\!
(\frac{2}{\varepsilon^2}\!+\!\frac{3}{\varepsilon}\!+\!8)\right]
+C_F\frac{\alpha^2}{E^2}\frac{\alpha_s}{4\pi}
\left(\!\frac{\mu^2}{s}\!\right)^\varepsilon\!\!\left(\!\frac{\mu^2}{-t}\!\right)^\varepsilon
(\frac{f_1}{\varepsilon}\!+\!f_2),
\end{equation}
where the functions $f_1$ and $f_2$  in the c.m. frame are ($x=\cos\theta$)
\begin{eqnarray}
f_1&=&-\!2 \frac{(1\!-\!x)(x^3\!+\!5x^2\!-\!3x\!+\!5)\log(\frac{1\!-\!x}{2})\!-\!
(x\!-\!1)^2(x\!+\!1)(x\!-\!11)/4}{(1-x)^2(1+x)^2},\label{f1} \\
f_2&=& -\frac{1}{(1-x)^2(1+x)^2}\left[(1-x)(x^3+5x^2-3x+5)
\log^2(\frac{1-x}{2})\right.\nonumber \\
&&+\left. \frac 12
(1-x)(3x^3\!+\!15x^2\!+\!77x\!-\!31)\log(\frac{1-x}{2})\!+\!(1+x)^2(x^2\!+\!5x\!+\!3)\pi^2\right.
\nonumber
\\&&\left.-12(9x^2\!+\!2x\!+\!5)Li_2(\frac{1+x}{2}) \!+\!\frac 12(1-x)(1+x)(5x^2\!-\!42x\!-\!23)\right].
\end{eqnarray}

As one can see from the comparison of the cross-sections of the processes with the
virtual (\ref{virsec}) and the real gluons (\ref{realsec}), in the sum the second order
poles cancel. However, the total cancellation of divergences does not happen. The
remaining divergences in the form of a single pole have a collinear nature. As was
already mentioned, for their cancellation one has to define properly the initial states.
The point is that the massless quark can emit the collinear gluon which will carry part
of the initial momentum and in this case, it is impossible to distinguish one particle
propagating with the speed of light from the two flying parallel.

\subsection{The quark distributions and the splitting functions}
\begin{figure}[ht]
 \begin{center}
 \leavevmode
  \epsfxsize=3cm
 \epsffile{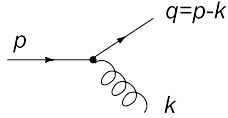}
 \end{center}\vspace{-0.5cm}
 \caption{The diagram corresponding to the splitting of the quark into the quark and the gluon
  }\label{split}
 \end{figure}

To take into account this possibility, let us come back to the scattering process and
assume that the initial quark has emitted  the parallel gluon (see Fig.\ref{split}). The
two particles can be  almost parallel with small relative transverse momentum.
The three four-momenta can be chosen in the form:
$$
p=(p;0,0,p),\ \ \ q\approx (zp;p_\perp,0,zp),\ \ \  k\approx ((1-z)p;-p_\perp,0,(1-z)p),
$$
so that all of them obey the condition $p^2=q^2=k^2=0$ with the accuracy up to
$p_\perp^2$. It is helpful, however, to use another method, namely to choose the momenta
in such a way that they obey the mass shell condition with the accuracy up to
$p_\perp^4$, but to give up the energy conservation in the order of $p_\perp^2$. The
advantage of this approach consists in the use of formulas for the spinors and the
polarization vectors on mass shell. Therefore, we choose the momenta as follows:
$$
p=(p;0,0,p),\  q\approx (zp+\frac{p_\perp^2}{2zp};p_\perp,0,zp),\  k\approx
((1\!-\!z)p+\frac{p_\perp^2}{2(1\!-\!z)p};-p_\perp,0,(1\!-\!z)p).
$$
The square of the matrix element corresponding to the process of splitting on mass shell
in this case can be written in the standard form
\begin{equation}\label{sp}
  |M(q\to qG)|^2=\frac{g^2}{2}С_FTr(\gamma^\mu \hat p \gamma^\nu \hat q)\sum\limits_{pol}
  {\epsilon^*}^\mu\epsilon^\nu,
\end{equation}
where the factor $1/2$ comes from the averaging over the spin states. Here we must take
into account the physical polarizations of the gluon only, i.e.
$$ \sum\limits_{pol}{\epsilon^*}^\mu\epsilon^\nu\to \delta^{ij}-\frac{k^ik^j}{(\vec k)^2},$$
which gives
\begin{equation}\label{sp2}
  |M(q\to qG)|^2=4g^2С_F \left[p^0q^0-\frac{(\vec p\vec k)(\vec q\vec k)}{(\vec
  k)^2}\right],
\end{equation}
or, substituting the values of momenta,
\begin{equation}\label{sp3}
  |M(q\to qG)|^2=C_F\frac{2g^2p_\perp^2}{z(1-z)} \frac{1+z^2}{1-z}, \ \ \ \  z<1.
\end{equation}
The obtained expression does not depend on the choice of momenta and has a universal
character.

Now one can calculate the cross-section of the process of interest. Graphically, it will
be the same diagram Fig.\ref{toy} е); however, the additional gluon will be referred not
to the final state but to the initial one. Here we use the standard Feynman rules when
the energy conservation law is not violated, but the massless particle is slightly off
shell. Since in the case of interest the quark with momentum $q$ is  virtual, it is
useful to choose the momenta like
$$p=(p;0,0,p),  q\approx (zp-\!\frac{p_\perp^2}{2(1\!\!-\!\!z)p};p_\perp,0,zp), k\approx
((1\!-\!z)p+\!\frac{p_\perp^2}{2(1\!\!-\!\!z)p};\!-\!p_\perp,0,(1\!-\!z)p).$$
 In this case,
\begin{equation}\label{q2}
q^2=-\frac{p_\perp^2}{1-z}.
\end{equation}
Then the cross-section of the process can be written in the factorized
form
\begin{equation}\label{factor}
  d\sigma(p)= \frac{1}{(2\pi)^3}\int \frac{d^3k}{2k^0}|M_{q\to qG}|^2
  (\frac{1}{q^2})^2 (\frac{p^0z}{p^0})d\sigma(pz),
\end{equation}
where the factor $(\frac{p^0z}{p^0})$ is due to fact that the cross-section is normalized
to the energy of initial particles, and we have replaced the quark with the energy $p^0$
by the quark with the energy $zp^0$.

Rewriting the differential $d^3k$ in terms of the new variables
$$d^3k=pdz d^2p_\perp=pdz \pi dp^2_\perp,$$
and substituting the value of the matrix element (\ref{sp3}) and $q^2$ from (\ref{q2}), we
get
\begin{eqnarray}\label{fact2}
  d\sigma(p)&=& C_F\frac{е^2}{16\pi^2}\int \frac{pdz dp^2_\perp}{(1-z)p}
  \frac{(1-z)^2}{p_\perp^4}\frac{2p^2_\perp}{z(1-z)} \frac{1+z^2}{1-z}z
  d\sigma(pz)\nonumber \\
  &=&C_F\frac{\alpha_s}{2\pi}\int \frac{dz dp^2_\perp}
  {p_\perp^2} \frac{1+z^2}{1-z}
  d\sigma(pz).
\end{eqnarray}
The integral over the transverse momentum is divergent at zero and this is nothing else
but  the manifestation of the collinear divergence. The upper limit is not of great
importance, it is restricted by kinematic considerations. We assume that the integration
over $p_\perp^2$ goes from zero to some scale $Q^2$. Later, we will see that one can
change this scale analogously to the change of the ultraviolet scale  $\mu^2$.

To extract the divergence we use the dimensional regularization. Changing the dimension
of transverse integration from $2$ to $2-2\varepsilon$ one gets
\begin{eqnarray}\label{spl}
d\sigma(p)&=&C_F\frac{\alpha_s}{2\pi}\int_0^1 dz \frac{1+z^2}{1-z} \int_0^{Q^2}
\frac{(p_\perp^2)^{-\varepsilon}(-\mu^2)^\varepsilon dp^2_\perp} {p_\perp^2} d\sigma(pz)
\nonumber\\
&=&C_F\frac{\alpha_s}{2\pi}\int_0^1 dz \frac{1+z^2}{1-z}\ \frac 1\varepsilon
\left(-\frac{\mu^2}{Q^2}\right)^\varepsilon d\sigma(pz).
\end{eqnarray}
At first sight the obtained expression still contains the pole in the integrand as $z\to
1$. However, it only looks like a singularity. It came from the matrix element
(\ref{sp3}), which we have calculated only for $z<1$ and it needs to be redefined for
$z\to 1$. We will come back to this question below and, at first, discuss the
interpretation of relation (\ref{spl}).

Let us introduce the notion of distribution of the initial quark with respect to the
fraction of the carried momentum $z$: $q(z)$. Then the initial distribution corresponds
to $q(z)=\delta(1-z)$, and the emission of a gluon leads to the splitting: the quark
carries the fraction of momentum equal $z$, while the gluon - $(1-z)$. The probability of
this event is given by the so-called {\it splitting functions} $P_{qq}(z)$ and
$P_{qG}(1-z)$. In the lowest order of perturbation theory in $\alpha_s$ the quark and
gluon distributions can be written in the form
\begin{eqnarray}\label{q}
q(z,Q^2)&=&\delta(1-z)+\frac{\alpha_s}{2\pi}\frac{1}{\varepsilon}
\left(\frac{\mu^2}{Q^2}\right)^\varepsilon P_{qq}(z), \\
G(z,Q^2)&=&\frac{\alpha_s}{2\pi}\frac{1}{\varepsilon}
\left(\frac{\mu^2}{Q^2}\right)^\varepsilon P_{qG}(1-z),
\end{eqnarray}
where the splitting functions are defined by the corresponding matrix elements one of
which for $P_{qq}(z)$ has been calculated in the leading order in $\alpha_s$ earlier (see
(\ref{sp3})). The result has the following form:
\begin{eqnarray}\label{qq}
P_{qq}(z)&=&C_F\left(\frac{1+z^2}{(1-z)_+}+\frac 32\delta(1-z)\right), \\
P_{qG}(z)&=&\frac{z^2+(1-z)^2}{2}.
\end{eqnarray}
Note that eq. (\ref{qq}) contains the redefinition of the function $P_{qq}(z)$ at the
point $z=1$ mentioned above, namely the sign $"+"$ should be understood as the following
integration rule:
$$ \int_0^1 dz \frac{f(z)}{(1-z)_+}\equiv \int_0^1 dz \frac{f(z)-f(1)}{(1-z)}, $$
and the coefficient of the  $\delta$-function is defined from the requirement of
conservation of the number of quarks
$$ \int_0^1 q(x,Q^2) dz=1 \ \ \Rightarrow \ \int_0^1 P_{qq}(z)dz =0.$$

Thus,  eq. (\ref{spl}) together with the Born diagram can be written as
\begin{equation}\label{init}
  d\sigma(p)=\int_0^1 dz \ q(z,Q^2)\ d\sigma(pz),
\end{equation}
where the quark distribution  $q(z,Q^2)$ is given by (\ref{q}).

It seems strange at  first sight that the answer depends on the scale $Q^2$ which defines
the quark distribution. However, it has the physical interpretation. This is the measure
of collinearity of the emitted gluons that can be distinguished, i.e., it refers to the
definition of the initial state. In fact, in the massless case one cannot define the
initial state that contains just the quark, it exists together with the set of collinear
gluons. (The same is true for the massless electron with collinear photons.) This scale
is sometimes called the factorization scale, at this scale the scattering cross-section
(\ref{init}) takes the factorized form. The factorization scale can be varied. The
dependence of the quark and the gluon distributions on the scale is governed by the
so-called DGLAP equations well known in QCD.

\subsection{The finite answers}

Thus, besides the two contributions to the cross-section from the virtual and the real
gluons there is one more contribution related to the splitted initial state (\ref{spl}).
In the lowest order of perturbation theory in $\alpha_s$ it can be written as
\begin{equation}\label{sp4} \left(\frac{d\sigma}{d\Omega}\right)_{split}=
\frac{1}{\varepsilon}\frac{\alpha_s}{2\pi}\int_0^1 dz
\left(\frac{\mu^2}{Q_f^2}\right)^\varepsilon P_{qq}(z) \frac{d\sigma_0}{d\Omega}(pz),
\end{equation}
where the Born cross-section is given by (\ref{secborn}) with the replacement of the
initial quark momentum $p$ by $pz$, and the factorization scale $Q_f^2$ is an arbitrary
quantity associated with the quark distribution function. Note that the scale $Q_f^2$ may
depend on $z$. It is quite natural to choose the factorization scale equal to the
characteristic scale of the process of interest. Thus, in our case this choice
corresponds to $Q_f^2=-\hat t$, where $\hat t$ is the Mandelstam parameter $t$ for the
process where $p$ is replaced by  $pz$. One has $\hat t = t\frac{2z}{(z+1)+(z-1)x}$. This
leads to the following result:
\begin{equation}\label{secsplit}
\left(\frac{d\sigma}{d\Omega}\right)_{split}=C_F\frac{\alpha^2}{2E^2}
\frac{\alpha_s}{2\pi} \left(\frac{\mu^2}{s}\right)^\varepsilon
\left(\frac{\mu^2}{-t}\right)^\varepsilon (-\frac{f_1}{\varepsilon}+f_3),
\end{equation}
where  $f_1$ is given by (\ref{f1}) and
\begin{eqnarray}\label{f3}
f_3&=& -\frac{1}{(1-x)^2(1+x)^2}\left[ 2(1-x)(x^3+x^2-33x+7)\log(\frac{1-x}{2})\right.
\nonumber
\\&&\left.+12(9x^2+2x+5)Li_2(\frac{1+x}{2}) -(1+x)^2(x^2+5x+3)\pi^2 \right. \nonumber \\
&&\left. -\frac 12(1-x)(1+x)(11x^2-19)\right].
\end{eqnarray}

Comparing the obtained expression with  (\ref{virsec}) and (\ref{realsec}) we see that
the last divergence cancels and the final expression for the cross-section of the
electron-quark scattering with account of possible creation of the gluon in the initial
and final states takes the form  $(x=\cos\theta)$
\begin{eqnarray}\label{fres}
  \left(\frac{d\sigma}{d\Omega}\right)_{набл}&=&\left(\frac{d\sigma}{d\Omega}\right)_{virt}+
\left(\frac{d\sigma}{d\Omega}\right)_{real}+\left(\frac{d\sigma}{d\Omega}\right)_{split}\\
&&\hspace*{-3cm}=\frac{\alpha^2}{2E^2}\left\{\frac{x^2+2x+5}{(1-x)^2}-\frac{\alpha_s}{2\pi}\frac{C_F}{(
1-x)(1+x)^2}\left[(x^3+5x^2-3x+5)\log^2\frac{1-x}{2}\right.\right.
\nonumber\\
&&\hspace*{-3cm}\left.\left.+\frac 12 (7x^3+19x^2-55x-3)\log\frac{1-x}{2}-
(1+x)(3x^2+21x+2)\right]\right\}. \nonumber
\end{eqnarray}

This expression is our final answer for the cross-section of the physical process of
electron-quark scattering where the initial and the final state include the soft and
collinear gluons. It includes also the definition of the initial state and can be
recalculated for the alternative choice of the factorization scale similar to what
happens to the ultraviolet scale which defines the coupling constant. Thus, we
practically deal with the scattering not of individual particles but rather with coherent
states with a fixed total momentum. Only this process has a physical meaning.

\begin{figure}[ht]
 %\begin{center}
 \leavevmode
  \epsfxsize=9.5cm
 \epsffile{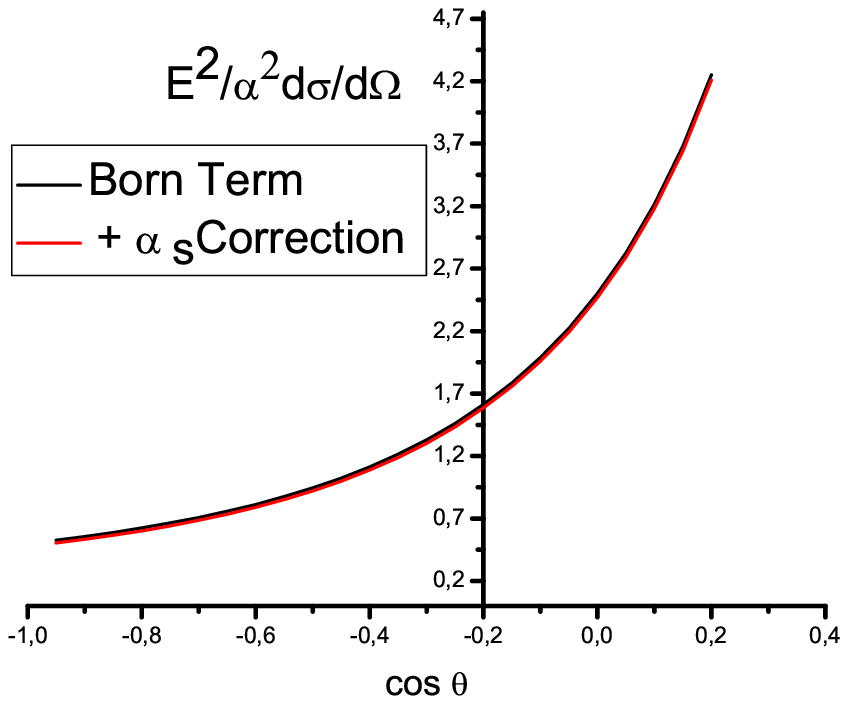}\vspace{-7.45cm}

\epsfxsize=8.0cm \hspace{9cm} \epsffile{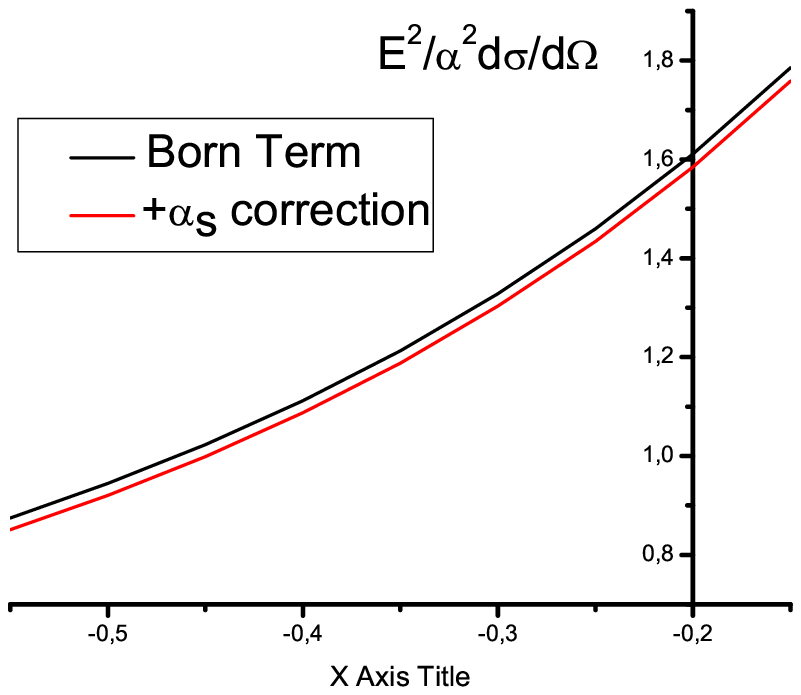}
% \end{center}\vspace{-0.5cm}
 \caption{The differential cross-section of  $eq$ scattering in the Born approximation
 and with allowance for the  $\alpha_s$ correction. On the right plane the same plot is shown in
 the bigger scale}\label{cr}
 \end{figure}

In Fig.\ref{cr}, we show the differential cross-section of this process as a function of
the electron scattering angle:  $\frac{E^2}{\alpha^2}
\frac{d\sigma}{d\Omega}(\cos\theta)$. We have chosen here the strong coupling
$\alpha_s=0.2$, and $C_F=4/3$.
As one can see, the inclusion of the radiative correction  $\sim \alpha_s$ practically
does not change the result, the difference from the Born approximation is less than a per
cent, that justifies the use of perturbation theory.

Let us stress once more that the obtained answer for the cross-section of the observable
process depends on: a) the ultraviolet subtraction scheme that manifests itself, in
particular, in the appearance of the ultraviolet scale $\mu^2$ (canceled in our case in
the lowest order of perturbation theory) and b) the definition of the initial coherent
state, which manifests itself in the appearance of the factorization scale $Q^2_f$. The
universality in the description of the physical processes is based on the fact that
choosing the UV and the IR scale one way or another and fitting the experimental data of
some process, one can then recalculate the obtained values of the running coupling and of
the quark (lepton) distribution for any other choice of the scales. This way the result
for the observable quantities does not depend on a particular choice of these scales and
is universal.

\newpage
\vspace*{1cm}
\section{Afterword}

Local quantum field theory, being the mathematical basis of elementary particle physics,
is the logical continuation of quantum mechanics. It exploits the same basic ideas, but
describing the system with an infinite number of degrees of freedom permits the creation
and annihilation of particles in the course of the interaction. The modern formulation is
based on the interaction representation which assumes the existence of the asymptotic
states of the free fields. In the S-matric approach we presume that these fields interact
in a local way in the space-time, and calculating the S-matrix elements one can find the
probabilities of various processes. The most developed and reliable method of these
calculations is the perturbation theory in the coupling constant which is similar to the
one in quantum mechanics. However, due to a much more complicated structure of the field
theory, the methods of perturbation theory encounter problems which have no analogy in
quantum mechanics, namely the divergence of the appearing integrals for the radiative
corrections. We have shown in these lectures how one can deal with these divergences
which have the ultraviolet and the infrared nature and how to get the finite answers for
the probabilities of the physical processes. We did not aim to prove the main theorems
like the Bogoliubov-Parasiuk or the Kinoshita-Lee-Nauenberg theorem, but have exemplified
how they work. The explicit calculations allow one to convince himself in the validity of
the final conclusions.

It should be noted that the formalism of quantum field theory contains the physical
principles which we have to follow sometimes not realizing it.  Thus, for example, the
ultraviolet divergences restrict the type of the interaction and, contrary to quantum
mechanics, there are only a few types of allowed Lagrangians. Not without reason the
renormalizability played such an important role in the formation of the Standard Model.
The other example is the notion of the asymptotic states. Even starting with the free
fields within the perturbation theory, from the requirement of the cancellation of the
infrared divergences we come to the definition of the physical initial and final states
which are essentially the coherent states.

The very fact that the gravitational interaction does not fit to the general scheme
probably means that local quantum field theory has a limited applicability and should be
replaced by a more general construction. It might be nonlocal like in the string theory,
or multidimensional one like in the brane-world theory. However, in any case, in the low
energy limit one has the local quantum field theory though possibly going beyond the
Standard Model that we considered here.

\section*{Acknowledgments}
The author is grateful to  A.V.Bednyakov, L.V.Bork, A.G.Grozin, S.V.Mikhailov,
N.G.Stefanis, G.S.Vartanov, A.A.Vladimirov, M.I.Vysotsky, and A.V.Zhiboedov  for numerous
helpful discussions. The work has been done with partial support from the RFBR grant №
08-02-00856 and the grant of the Ministry of Science and Education of RF for support of
the scientific schools № НШ-1027.2008.2

\end{document}